# The structure of aqueous lithium chloride solutions at high concentrations as revealed by a comparison of classical interatomic potential models

Ildikó Pethes

Wigner Research Centre for Physics, Hungarian Academy of Sciences, H-1525 Budapest, POB 49, Hungary

E-mail address: pethes.ildiko@wigner.mta.hu

Abstract

Highly concentrated aqueous lithium chloride solutions were investigated by classical molecular dynamics (MD) and reverse Monte Carlo (RMC) simulations. At first MD calculations were carried out applying twenty-nine combinations of ion-water interaction models at four salt concentrations. The structural predictions of the different models were compared, the contributions of different structural motifs to the partial pair correlation functions (PPCF) were determined. Particle configurations obtained from MD simulations were further refined using the RMC method to get better agreement with experimental X-ray and neutron diffraction data. The PPCFs calculated from MD simulations were fitted together with the experimental structure factors to construct structural models that are as consistent as possible with both the experimental results and the results of the MD simulations. The MD models were validated according to the quality of the fits. Although none of the tested MD models can describe the structure perfectly at the highest investigated concentration, their comparison made it possible to determine the main structural properties of that solution as well. It was found that four nearest neighbors (oxygen atoms and chloride ions together) are around a lithium ion at each concentration, while in the surroundings of the chloride ion hydrogen atom pairs are replaced by one lithium ion as the concentration increases. While in pure liquid water four water molecules can be found around a central water molecule, near the solubility limit nearly all water molecules are connected to two chloride ions (via their hydrogen atoms) and one lithium ion (by their oxygen atoms).

## 1. Introduction

Aqueous electrolyte solutions are in the centre of scientific interest being the media for biochemical reactions in living organisms. Besides that they have great importance in the fields of physical chemistry, medicine, electrochemistry and geochemistry, as well as in the industry. Due to its outstanding solubility in water, lithium chloride is an excellent prototype to investigate the properties of ionic complexes in highly concentrated aqueous solutions [1].

The structure of aqueous lithium chloride solutions has been investigated in numerous papers in the last 50 years. The most frequently applied experimental methods for structure determination are the diffraction techniques: X-ray diffraction (XRD) [2 - 11] and neutron diffraction (ND) mostly with isotopic substitution [4, 11 - 29]. Several other techniques have also been used during the decades, such as Raman spectroscopy [30 - 32], attenuated total reflection infrared spectra [33], X-ray Raman scattering spectroscopy [34], X-ray Compton scattering [35], dielectric relaxation spectroscopy [36] or nuclear magnetic resonance [37 - 39].

Several open questions still remain both for dilute and highly concentrated solutions, despite the huge experimental effort. The structural reordering of water molecules, the extent of ion-pairing and the hydration number of the chloride and lithium ions are crucial problems. (For the latter see e.g. a very recent paper considering dilute solutions [29].)

The complete description of the structure of an aqueous solution at the two-particle correlation level requires the combination of ten independent experimental data sets, since a four component system (H (D), O, anion, cation) has ten partial structure factors (or ten partial pair correlation functions (PPCF) in real space). Usually the number of experimental data sets is significantly lower than ten, e.g. only one ND or XRD data set is available. The low partial weight of the ion-related partials in the measured data can be another problem (mostly in dilute solutions).

Simulation methods are also often applied, sometimes in combination with the experimental techniques, because they are helpful to reduce the non-uniqueness of the interpretation of experimental data. The first molecular dynamics (MD) simulation on this system was carried out more than 30 years ago [40], and they are applied more frequently nowadays. Beside classical MD simulations [41 - 49], ab initio quantum mechanical MD [50 - 58], Monte Carlo [59, 60], molecular mechanics [50] and reverse Monte Carlo (RMC) [11, 61 - 64] calculations have also been used to investigate LiCl solutions. Ab initio MD simulations are increasingly important nowadays. Unfortunately, for highly concentrated LiCl solutions they are almost absent, and the comparison with experimental data sets (e.g. XRD or ND data) is completely missing.

Several interaction models (force fields, FF) have been developed and used for MD simulations in the context of electrolyte solutions. However, for concentrated aqueous LiCl solutions their applicability is very limited, as was shown recently [65]. In that work 29 different models were investigated; it turns out that their predictions about density, dielectric constant and self-diffusion constants scatter widely. Neutron and X-ray weighted structure factors were also calculated and compared with experimental data: it was found that the agreement between the experimental and simulation curves was poor, especially for the highest investigated concentration (19.55 mol/kg, near the solubility limit).

The reverse Monte Carlo method [66] is a useful tool to find three dimensional configurations consistent with experimental data sets. However the number of experimental data sets (mostly one or two structure factors from neutron and/or X-ray diffraction measurements) is low compared to the number of the independent partial pair correlation functions (10). This lack of information results in a highly underdetermined system (see the 'pure' RMC results in Refs. [11, 61, 62, 67]).

The combination of the experimental data driven RMC type simulations with the considerations of the interaction potential models (MD type simulations) can be successfully used to obtain reliable results. Such investigations were performed using hybrid RMC [67] or the empirical potential structure refinement (EPSR) method [26]. Unfortunately, both of these approaches were applied to dilute or moderately concentrated solutions; furthermore on the experimental side only neutron diffraction data were fitted during the simulations, which is sensitive mostly for the H-H (or D-D) and O-H (O-D) partials. Thus their [26, 67] results about the ion-ion correlations are less reliable. In highly concentrated solutions the effect of the ion-pairing and the weight of the ion-ion partials are more important. Data from X-ray diffraction can be helpful, since it is more sensitive to the O-O, Cl-O and Cl-Cl partials (in the case of highly concentrated solutions).

Another method is to combine experimental structure factors and radial distribution functions from MD simulations in one single structural model, generated by RMC modeling. This method was introduced in Ref. [68] for RbBr solutions. The method is also useful to determine whether various input data sets are consistent with each other, in which case they can be fitted simultaneously (see e.g. Ref. [69] on liquid water). Two papers used this technique to investigate aqueous LiCl solutions, Refs. [63, 64]. The first of them [63] fitted previously known experimental curves from the literature: ND and XRD data from Ref. [4] and 9 PPCFs from MD simulations of Ref. [46] (the Li-Li PPCF was not considered). The authors found that the Cl-O and O-O PPCFs from the MD simulations and the XRD data were not consistent with each other. They suggested that the ion-ion potential parameters need significant improvement. In the second paper [64] new ND and XRD data [11] were fitted together with 8 PPCFs from MD simulations (the Li-Li and Cl-Cl PPCFs were left out) at the same concentrations as were used in the experiments. Three different interaction potential models of water with the same ion potential parameters were investigated. It was found again that none of the investigated water models describe the structure properly. The RMC refined models fitted the experimental data better than that obtained from purely MD calculations, but the method resulted poor Cl-O, O-O and Li-Cl PPCFs fits, as in the previous paper. However the effect of the ion potential parameters on the quality of the fits was not investigated.

In the present study the 29 ion-water potential parameter sets collected in Ref. [65] are further investigated to establish their compatibility with experimental ND and XRD data sets. The structural properties calculated from MD simulations together with the results obtained from RMC refinements are discussed. Unlike in previous studies all the 10 PPCFs calculated by MD simulations are fitted simultaneously during these RMC simulations. It was possible to identify the structural motifs contributing to the PPCFs. The investigations revealed the probable structure of the LiCl solutions, which is compatible not only with the experimental data sets but with the better interatomic potential models as well.

## 2. Methods

*2.1 Experimental data sets and investigated concentrations*

Aqueous lithium chloride solutions were investigated at four different concentrations, all of them from the concentrated electrolyte solution region. Their molality values are 3.74 mol/kg, 8.30 mol/kg, 11.37 mol/kg, and 19.55 mol/kg. They were chosen because ND and XRD measurements were reported previously on all of these concentrations [11]. The exact number of ions per water molecules and the densities (at room temperature) are taken from Ref. [11] and shown in Table 1. The simulations were performed using the same cubic simulation boxes during the MD and RMC simulations. The boxes contained about 10,000 atoms. The number densities, number of atoms in the simulation boxes and the box sizes are presented in Table 1. The four investigated LiCl solutions will be denoted throughout this work according to their concentration as the 3.74m, 8.3m, 11.37m, and 19.55m samples.

*2.2 Interatomic potentials*

Pairwise additive non-polarizable intermolecular potentials were applied in the MD simulations, which consist of the Coulomb potential for electrostatics and the 12-6 Lennard-Jones (LJ) potential for the van der Waals interactions:

$$V_{ij}(r_{ij}) = \frac{1}{4\pi\varepsilon_0} \frac{q_i q_j}{r_{ij}} + 4\varepsilon_{ij} \left[ \left(\frac{\sigma_{ij}}{r_{ij}}\right)^{12} - \left(\frac{\sigma_{ij}}{r_{ij}}\right)^{6} \right]. \qquad (1)$$

Here $r_{ij}$ is the distance between particles $i$ and $j$, $q_i$ and $q_j$ are the point charges of the two particles, $\varepsilon_0$ is the vacuum permittivity, $\varepsilon_{ij}$ and $\sigma_{ij}$ are the 12-6 LJ potential parameters. For the FFs investigated here the $q_i$, $\varepsilon_{ii}$ and $\sigma_{ii}$ parameters are given in Table 2, the parameters between unlike atoms are calculated according to the combination rules, which are also given in the table. Two types of combination rules were used: the geometric (geom), where both the $\varepsilon_{ij}$ and $\sigma_{ij}$ are calculated as the geometric average of the homoatomic parameters, and the Lorentz-Berthelot type (LB), in which the $\varepsilon_{ij}$ is calculated as geometric, while $\sigma_{ij}$ is the arithmetic average of the relevant parameters. (The $\varepsilon_{ij}$ and $\sigma_{ij}$ parameters are collected also on Fig. S.1 of the Supplementary material, for easier comparison.) All of the FFs tested here apply one of the following water models: SPC/E [70], TIP4P [71] and TIP4PEw [72]. These models are amongst the simplest water models, which are rigid and non-polarizable. The SPC/E model is a 3-site model (3 point-like atoms representing the oxygen and the two hydrogen atoms), while the other two models use a fourth, virtual site as well (the 'partial charge of the oxygen atom' is located on this fourth site). The $\varepsilon_{HH}$ and $\sigma_{HH}$ values are equal to 0 for all models. The other parameters of these water models are collected in Table 3.

The investigated FFs are detailed in Ref. [65], so they are only briefly summarized here.

- Chandrasekhar FF (Ch) [73]. The earliest 12-6 LJ parameter set.
- Dang-Smith parameter set (DS) [74, 75]. It is one of the most frequently used FFs.
- Jensen-Jorgensen set (JJ) [76]. Developed with the purpose to create a consistent set for OPLS-AA [77] FF.
- Joung-Cheatham III sets (JC-S, JC-T) [78]. Parameters were determined for three different water models, two of them are tested here: SPC/E (JC-S) and TIP4PEw (JC-T).

- Horinek-Mamatkulov-Netz parameter sets (HS-g, HM-g, HL-g, HS-LB, HM-LB, HL-LB) [79]. The ion-oxygen parameters ($\varepsilon_{iO}$ and $\sigma_{iO}$) are determined in Ref. [79], leaving the choice of the combination rule free, thus both possibilities are tested here. 3 different $\varepsilon$ parameters (small (S), medium (M) and large (L)) are presented in the paper, all of them are investigated.

- Gee et al. FF (Gee) [80]. This FF uses geometric combination rule with a modification for the calculation of the $\varepsilon_{LiO}$ parameter: $\varepsilon_{LiO} = 0.4\ (\varepsilon_{LiLi}\ \varepsilon_{OO})^{1/2}$ (denoted as mgeom).

- Reif-Hünenberger FFs (RH, RM, RL, RL-sLB) [81]. Three different hydration free energy values have been used for the determination of these FF parameters (high (H), medium (M) and low (L)). Although the parameters were calculated for both the SPC and SPC/E water models, only the later is tested here. For RH, RM and RL FFs the geometric combination rule is applied, the L set is studied with a modified combination rule as well (denoted with sLB), in which the $\sigma_{LiO}$ and $\sigma_{ClO}$ values are calculated according to the geometric combination rule, but $\sigma_{LiCl}$ is obtained using the LB combination rule.

- Mao-Pappu parameter sets (MP-S, MP-T) [82] A solvent independent ionic potential set, which is tested here using SPC/E (MP-S) and TIP4PEw (MP-T) water models.

- Deublein-Vrabec-Hasse (DVH) [83] and Reiser-Deublein-Vrabec-Hasse (RDVH) [84] parameter sets. The two parameter sets differ only in their $\varepsilon$ parameters, which parameters are the same for all ions, and two times higher in the RDVH than in the DVH FF.

- Li-Song-Merz FF (Li-HFE-S, Li-HFE-T, Li-IOD-S, Li-IOD-T) [85]. Two 12-6 LJ parameter sets are available in Ref. [85], they were fitted to hydration free energies (Li-HFE) and ion oxygen distance (Li-IOD). Both of them were optimized for three water models, from which two are tested here: SPC/E (Li-HFE-S and Li-IOD-S) and TIP4PEw (Li-HFE-T, Li-IOD-T).

- Åqvist+Chandrasekhar parameter combination (AqCh). [73, 86] This FF is a combination of two parameter sets: the parameters of the Cl$^-$ ion are taken from the paper of Chandrasekhar [73] (as in the Ch FF above), the Li$^+$ ion parameters are from the paper of Åqvist [86]. Although this combination seems to be somewhat arbitrary (e.g. during the parametrization the two developers applied different water models) this combination is widely used in simulations, with different water models. This is because this combination is provided by default for the OPLS-AA [77] force field in the frequently applied

GROMACS software package [87]. This combination was studied in Ref. [64] with 3 different water models. This combination with SPC/E water will be investigated here.

- Pluharova et al. parameters (Pl) [27, 88]. This is the only FF investigated here in which the charge of the ions is not equal to ±1, but ±0.75. The advantageous effects of using reduced charges (which originates from the electronic continuum model [89]) are examined in several papers (see e.g. [65, 90]).

- Aragones et al. parameters (Ar) [48] This FF is a modification of the JC-T FF. The anion-cation parameters are calculated according to a modified LB rule (mLB): $\varepsilon_{LiCl} = 1.88(\varepsilon_{LiLi}\varepsilon_{ClCl})^{1/2}$ and $\sigma_{LiCl} = 0.932 (\sigma_{LiLi} + \sigma_{ClCl})/2$. These modified parameters were chosen to obtain the best possible agreement of the simulated Li-Cl PPCF of a 1.38 mol/kg solution with the so called 'experimental' PPCF of Ref. [26]. (Unfortunately the uncertainty of the fitted 'experimental' PPCF was rather high, since this partial is nearly invisible (the weight of the Li-Cl partial structure factor is very small) in the ND experiment applied. The average number of direct ion contact pairs is 0.2-0.3 ±0.4 according to Ref. [26].)

- Singh-Dalvi-Gaikar parameter sets (SDG-S, SDG-T) [91]. These models are combinations, in which the $Li^+$ ion parameters are obtained from the DS FF, and the $Cl^-$ ion parameters are equal to the JJ $Cl^-$ parameters. The combination was investigated here with SPC/E (SDG-S) and TIP4P (SDG-T) water models.

*2.3 MD simulations*

The classical molecular dynamics simulations were performed by the GROMACS software (version 5.1.1) [87]. The initial particle configurations were obtained by placing the ions and water molecules randomly into the simulation boxes. Energy minimization was carried out using the steepest descent method. After that the calculations were performed applying the leap-frog algorithm at constant volume and temperature (NVT ensemble) at $T = 300$ K. The cutoff distance for the Coulomb and van der Waals forces (which are determined according to the parameters detailed in the previous section) was 10 Å. The total simulation time was 10 ns, with a time step of 2 fs. After a 2 ns equilibration period, 50 particle configurations (sampled every 160 ps) were collected to calculate the PPCFs. For a more detailed description of the MD simulation method see Ref. [65].

*2.4 Reverse Monte Carlo simulations*

RMC modeling is described in detail in Refs. [66, 92, 93]. RMC is an inverse method to obtain three dimensional particle configurations that are consistent with the supplied experimental and/or theoretical data sets: in this case the neutron and X-ray weighted structure factors from the ND and XRD experiments and the PPCFs from the MD simulations. The total scattering structure factors ($S(Q)$) and the PPCFs ($g_{ij}(r)$) are connected through the partial structure factors ($S_{ij}(Q)$) by Eqs. (2-3):

$$S_{ij}(Q) - 1 = \frac{4\pi\rho_0}{Q}\int_0^\infty r(g_{ij}(r) - 1)\sin(Qr)\mathrm{d}r \tag{2}$$

$$S(Q) = \sum_{i \leq j} w_{ij}^{X,N}(Q) S_{ij}(Q). \tag{3}$$

Here $Q$ is the amplitude of the scattering vector, and $\rho_0$ is the average number density. The scattering weights for neutron ($w_{ij}^N$) and X-ray ($w_{ij}^X$) are given by Eqs. (4-5):

$$w_{ij}^N = (2 - \delta_{ij})\frac{c_i c_j b_i b_j}{\sum_{ij} c_i c_j b_i b_j} \tag{4}$$

$$w_{ij}^X(Q) = (2 - \delta_{ij})\frac{c_i c_j f_i(Q) f_j(Q)}{\sum_{ij} c_i c_j f_i(Q) f_j(Q)}. \tag{5}$$

Here $\delta_{ij}$ is the Kronecker delta, $c_i$ denotes the atomic concentration and $b_i$ is the coherent neutron scattering length, and $f_i(Q)$ is the atomic form factor. The weighting factors for the four investigated concentrations are shown in Fig. S.2.

During RMC calculation particles are moved randomly in the simulation box and the differences between the experimental and model curves are minimized (the PPCFs from MD simulations are taken into account in the same way as the experimental curves). In the case where all of the input data are consistent, particle configurations are obtained which are compatible with all input data and the differences between the model and experimental curves are small (the qualities of all the fits are adequate). To measure the quality of the fit (and to make them comparable) 'goodness-of-fit' values, (denoted here as *R*-factors) are calculated:

$$R = \frac{\sqrt{\sum_i (A_{\mathrm{mod}}(x_i) - A_{\mathrm{exp}}(x_i))^2}}{\sqrt{\sum_i (A_{\mathrm{exp}}(x_i) - 1)^2}}. \tag{6}$$

Here $A_{\mathrm{mod}}$ and $A_{\mathrm{exp}}$ are the calculated and experimental data (structure factors or PPCFs), for $S(Q)$ fitting $x_i = Q_i$ and for $g(r)$ fitting $x_i = r_i$ and the summation is over the $i$ experimental data points.

The RMC++ code [94] was used here for fitting the neutron and X-ray weighted $S(Q)$ functions from the experiments (from Ref. [11]) and the 10 intermolecular PPCF curves from MD simulations simultaneously. To keep the atoms in the water molecules together during the calculations the 'fixed neighbor constraints' (FNC) option was applied [93]. This algorithm connects two hydrogen atoms and their central oxygen atom permanently via their identity numbers. The O-H and H-H intramolecular distances were kept between minimum and maximum values; these values for the different water models are shown in Table 4. It was found earlier [69, 95] that the RMC with FNC method unfortunately can lead to unrealistic intramolecular H-H curves even for pure water with the deformation of the water molecule (e.g. the H-H intramolecular distances tend to be shorter). To avoid this effect and to keep the shape of the water molecules similar to their shape in the MD simulations, intramolecular $g_{HH}(r)$ PPCFs were also fitted (different for the different water models). These curves were obtained from RMC_POT simulations [95, 96] of pure water using the corresponding water model. (They are mainly Gaussian curves around the expected intramolecular H-H distance of the applied water model. The width of these curves was set according to the minimum and maximum distances of the FNC constraints.) By applying this method it is ensured that the shape of the water molecules is the same in all models, which are using the same water model. Thus the intramolecular contributions of these models to the $S(Q)$ curves are also the same and the differences in the quality of the fits originate only from the ion-water and ion-ion interactions.

The PPCF curves were fitted (in $r$) from the appropriate cutoff distances up to 10 Å, the intramolecular H-H curves were fitted from 0 – 1.65 Å for TIP4P and TIP4PEw water and 0 - 1.70 Å for SPC/E water. The intermolecular Li-Li and Cl-Cl $g(r)$ curves obtained from the MD simulations were smoothed before putting them into the RMC calculations. The cutoff distances are collected in Table 5.

The experimental $S^N(Q)$ and $S^X(Q)$ structure factors [11] were fitted in the 0.5 – 16 Å$^{-1}$ range.

The particle configurations obtained from MD simulations were used as initial configurations in the RMC calculations. The bin size (the resolution in $r$-space) was 0.05 Å, the maximum move of atoms was set to 0.1 Å. The relative weights of the input data sets (control parameters or σ values that influence the tightness-of-fit) were the same for all concentrations and all models. It was set to 0.01 for the 11 $g(r)$ curves, 0.001 and 0.002 for $S^N(Q)$, and for $S^X(Q)$, respectively. These parameter values were chosen to get appropriate fits for the best models.

All the RMC simulations were carried out under the same circumstances: the fitted curves contained the same number of points, the number of the generated moves were the same for all tested FFs. The number of accepted moves was around $10^6$ in each calculation.

## 3. Results and discussion

*3.1 Main structural units in the MD models: the first coordination shells*

The basic units in the structure of aqueous LiCl solutions are the H-bonded water molecules, the hydrated ions (through Li-O and Cl-H pairs) and the (contact) ion pairs. The proportion of these units depends of course on the concentration, but it was also found to be different in the various FFs studied. These units are crucial to the understanding of the structure of aqueous LiCl solutions: not only the first coordination shells of the ions and water molecules can be described through them, but the medium range structure (second or third nearest neighbors) is also determined by the ratio of these basic pairs.

*3.1.1. The nearest neighbors of $Li^+$ ions*

The first coordination shell of lithium ions consists of water molecules and chloride ions at each investigated concentrations for each FFs. The Li-O and Li-Cl PPCFs are shown in Figs. 1 and 2 for some selected FFs. The $g_{LiO}(r)$ and $g_{LiCl}(r)$ curves have well-defined first peaks (around $r_{max,LiO}$ and $r_{max,LiCl}$) followed by a region, where the neighboring particles are completely missing for most FFs. Thus the $N_{LiO}$ and $N_{LiCl}$ coordination numbers (the number of the particles in the first coordination shell, a.k.a. the number of the nearest neighbors) can be obviously determined, by counting particles up to the first minimum of the curves ($r_{min,LiO}$ and $r_{min,LiCl}$). The $N_{LiCl}$ and $N_{LiO}$ values are shown in Tables 6 and 7, while the respective $r_{min}$ and $r_{max}$ positions are collected in Tables S.1 and S.2. The number of oxygens in the first coordination shell of Li is decreasing and the number of $Cl^-$ ions is increasing as the concentration increases. The sum of the two values, $N_{Li}=N_{LiO}+N_{LiCl}$ (the total coordination number of $Li^+$) is around 4 at all investigated concentration for nearly all FFs (see Table 7). The exceptions are RH, DVH and RDVH, they will be discussed later.

The cosine distributions of the O-Li-O, O-Li-Cl and Cl-Li-Cl angles confirm the tetrahedral arrangement around $Li^+$ (see Fig. 3); all three functions have peaks around 100-120°. In the configurations, in which $N_{LiCl}$ is small (in certain models, at low concentrations) and at most one $Cl^-$ ion is present around a $Li^+$ ion, all three peaks are around 109°. In the configurations, in which more neighboring $Cl^-$ ions can be found around a $Li^+$ ion, the two adjacent $Cl^-$ ion cause a slight torsion of the tetrahedral layout: the typical Cl-Li-Cl angle is higher than 109° (the peak in the distribution function is around 114°), while the O-Li-O angle is smaller (the peak of the curve is around 105°).

At low and medium concentrations, where the number of water molecules is high enough to satisfy the coordination requirements of the Li$^+$ ions, the hydration number of Li$^+$ is around 4, as was shown by neutron diffraction with isotopic substitution (see e.g. [25, 28]). According to the results of the FFs investigated here, as the concentration is increased the water molecules are replaced by chloride ions, to keep the number of nearest neighbors of Li$^+$ around 4. The rate of this replacement is different in the investigated models. At the smallest concentration (3.74m sample) ion pairing is not necessary for the fourfold coordinated Li$^+$, since there are 14.84 water molecules per ion pair. Some of the models result $N_{LiCl} \approx 0$ in the 3.74m sample, such as JC-S, HL-LB, JC-T, HM-LB, Gee (etc.), however in some models (e. g. RH, RM, JJ or Li-IOD-T) $N_{LiCl} \gg 0$ ($N_{LiCl} > 1$) even for this concentration (see Table 6). The tendency of the FFs to produce ion pairing is shown in Fig. 4 and in Table 6. This ion pairing tendency will be denoted below as IPT, where lower IPT means that the model results a lower $N_{LiCl}$ value.

*3.1.2. The first coordination shell of Cl$^-$ ions*

The first coordination shell of Cl$^-$ ions contains water molecules and lithium ions. The water molecules turn towards the ion by their positively charged H atoms. These Cl$^-$-H pairs result in the first peak of $g_{ClH}(r)$ around 2.1-2.4 Å (see Fig. 5). The height of this peak is inversely proportional to the IPT of the model, and the number of Cl$^-$-H pairs is smaller at higher concentrations, in accordance with the hypothesis that hydrating water molecules are replaced by Li$^+$ ions. The boundary of the first shell is not as well defined as in the case of the Li-O and Li-Cl partials, because $g_{ClH}(r)$ is not equal to zero at any $r$ distance. Thus the uncertainty of the $N_{ClH}$ coordination number is higher. The positions of the first peak ($r_{max,ClH}$) and the upper limits of the integral in the calculation of the $N_{ClH}$ coordination number ($r_{min,ClH}$) are shown in Table S.3, and the $N_{ClH}$ values are presented in Table 8. The $r_{max,ClH}$ values of the different FFs are more strongly scattered than the $r_{max,LiO}$ values. The shortest ones can be observed in the RH (1.98 Å) and RM (2.06 Å) models, the longest ones in the RDVH (2.44 Å), Li-HFE-T (2.4 Å) models (and they are above 2.3 Å in the Li-HFE-S, SDG-T, DVH and JJ models as well).

The $N_{ClH}$ coordination number is decreasing about twice as fast as $N_{ClLi}(=N_{LiCl})$ is increasing with the concentration, suggesting that each counter ion substitutes two water molecules in the first coordination sphere of a Cl⁻ ion. The 'weighted total coordination number', $N_{Cl}=N_{ClLi}+1/2N_{ClH}$ value is around 3.5 at each investigated concentrations in most models, see Table 8. (The exceptions are the same as was found for $N_{Li}$: RH, DVH and RDVH models.)

The cosine distributions of the H-Cl-H, Li-Cl-Li and H-Cl-Li angles are shown in Fig. 6. While the most likely value of the H-Cl-H angle is around 75°, the Li⁺ ions prefer tetrahedral arrangement around the Cl⁻ ions. The small angles in H-Cl-H distributions are almost completely absent, which suggest that most water molecules turn toward the Cl⁻ ion by only one of their H atoms. The cosine distribution of the Cl-H-O angle is shown in Fig. 7 (here H-O denotes the intramolecular H-O bond and H...O will be used for the intermolecular, H-bonded H-O pairs). The most preferred arrangement of the Cl, H and O atoms is the straight one, similarly to the regular hydrogen bond. This cosine distribution confirms that almost every water molecules turn toward the anion by only one of their hydrogen atoms.

It follows from the previous considerations that essentially every Cl-H nearest neighbor pairs belong to different Cl-water pairs, thus the number of water molecules around a Cl⁻ ion is about equal to the number of Cl-H pairs. As the $N_{ClLi}+1/2N_{ClH}$ value is around 3.5 for all concentrations (see Table 8), thus the average hydration number of Cl⁻ is around 7 at lower concentrations, where the number of the contact ion pairs is negligibly small. The hydration number of Cl⁻ in low and medium concentrated solutions was found to be around 6 experimentally (see e.g. Refs. [6, 8, 21, 25, 26]).

*3.1.3. The first coordination sphere of the water molecules*

The environment of the water molecules can be deduced from the Li-O, Cl-H and O-H PPCFs (Figs. 1, 5, 8). The ion-water nearest neighbors were discussed in the previous sections. The first intermolecular peak in $g_{OH}(r)$ (Fig. 8) at 1.8 Å, belongs to the H-bonded water molecules (in pure water this distance is 1.77 Å for the SPC/E water model and 1.82 Å for the TIP4P and TIP4PEw models). The positions of this peak ($r_{max,OH}$) and the upper limits of the integral in the calculation of the $N_{OH}$ coordination number ($r_{min,OH}$) are shown in Table S.4. The $N_{OH}$ coordination numbers are collected in Table 9. The number of H-bonded water molecules is higher for the high IPT models, even in the 3.74m sample, but this difference between the FFs is more pronounced at higher concentrations. In the 19.55m sample the average number of O...H pairs is around 1.07 in the RM model and only 0.25 in the JC-S model.

The value of the $N_{OH}$ intermolecular coordination number decreases with increasing concentrations, about twice as fast as the $N_{OLi}$ coordination number increases. For pure water there are 2 intra and 2 intermolecular hydrogen neighbors of an oxygen atom, in these solutions they have – beside the two intramolecular ones – either two intermolecular hydrogen atoms or one $Li^+$ ion neighbor. Two O…H bonds are replaced by one O-$Li^+$ pair. The 'weighted total coordination number' $N_O = N_{OLi} + N_{OH}/2$ value (see Table 9) is around 1 for every FFs and only slightly increases as the concentration is raised. (The exceptions are again the DVH and RDVH models: oxygen atoms, which are surrounded by two $Li^+$ ions or one $Li^+$ ion and one intermolecular H atom, can also be found in the configurations obtained in simulations using these FFs.)

*3.1.4. The RH, DVH and RDVH models*

There are three models which strongly differ from all the others: these are the RH, DVH and RDVH FFs. It was shown earlier [65] that the RH model is not able to describe the structure of concentrated LiCl solutions, since precipitation can be observed in the configurations obtained in simulations with this FF, even in the lowest concentration (3.74m sample). Many properties of the DVH and RDVH models deviate from the properties of the other FFs. The typical $Li^+$-O and $Li^+$-$Cl^-$ pair distances are significantly higher in the DVH and RDVH models (see Tables S.1 and S.2), and the $g_{LiO}(r_{min,LiO})$ and $g_{LiCl}(r_{min,LiCl})$ values are far from zero. The $N_{LiCl}$ coordination numbers in these models are in the same range as the other models, but the $N_{LiO}$ values are significantly higher. While in the other models the oxygen atoms have at most one $Li^+$ nearest neighbors, in the configurations obtained with DVH and RDVH FFs a significant number of oxygen atoms can be found which are connected to two $Li^+$ ions. Moreover in these configurations $Li^+$-O...H-O motifs are frequently present, which units are rare in the other models.

*3.2 Beyond the first coordination shell: detailed analysis of the PPCFs obtained from MD simulations*

The first peaks of the Li-Cl, Li-O, Cl-H and O-H PPCFs were discussed in the previous sections. In this section the other features of these partials and the remaining 6 PPCFs will be analyzed. The $g_{ij}(r)$ curves obtained from MD simulations at the highest and lowest investigated concentrations are shown in Figs. 1, 2, 5, 8, 10-15 for some selected FFs. (The entire concentration dependence can be seen in Figs. S.3-S.12, for two selected FFs.) It can be seen that there is a strong difference between the curves from various models: not only the positions and heights of the peaks vary, but for several *i-j* pairs the overall shape of the curves seems to be dissimilar at first sight.

As it was discussed in section 3.1, the number of the contact ion pairs strongly depends on the FFs. It will be shown below that this is the origin of the strong difference between the PPCFs from various models. It was found that the PPCFs obtained in simulations with different FFs can be characterized according to the IPT of their models. Five FFs, with different $N_{LiCl}$ values were selected in Figs. 1, 2, 5, 8, 10-15 to present the main properties of the PPCFs: JC-S, MP-S, RL-sLB, RL and RM (from the lowest to the highest IPT respectively; for the 3.74m sample $N_{LiCl} = 0$ in the configuration obtained with JC-S, and $N_{LiCl} = 1.95$ for RM). Each of these FFs uses the SPC/E water model, thus the differences shown are due to the varying ion-ion and ion-water parameters. The shape of the curves are similar for the FFs applying other water models, the main differences are in the intramolecular part of the O-H and H-H PPCFs.

The comparison of the $g_{ij}(r)$ functions obtained with different FFs helps to identify the contributions of the different structural motifs (second, third or fourth neighbors). This will be detailed below. The motifs so identified can also be discovered in the snapshots of the configurations: some of them are presented in Fig. 9. The most probable pair distances were calculated using the nearest neighbor distances and cosine distributions presented in section 3.1. The distance distributions of second and (some) third neighbors were calculated directly from the configurations as well, which calculations confirm the results presented below. (In what follows Li-Cl, Li-O and Cl-H nearest neighbor pairs will be referred to as "bonds", for simplicity.)

*3.2.1 The $g_{LiCl}(r)$ curve (Fig.2)*

The first peak in the $g_{LiCl}(r)$ curves (see Figs. 2 and S.3), which is around 2.3-2.4 Å, belongs to the $Li^+$-$Cl^-$ contact ion pairs; the height of this peak shows the IPT of the FF. A second peak can be seen around 4.5 Å, which is higher for models with lower IPT. This peak can be attributed to the distance of $Li^+$ and $Cl^-$ ions in $Li^+$-O-H-$Cl^-$ "chains". (Such a motif can be seen in Fig. 9a, marked with blue connecting lines.) The most probable $Li^+$-$Cl^-$ distance in this motif (4.5 Å) can be calculated taking into account the typical $Cl^-$-H (2.2 Å) and $Li^+$-O (2 Å) nearest neighbor distances (see section 3.1), the Li$^+$-O-H angle (which is around 120°, see Fig. 7a), and the O-H-Cl$^-$ angle (which is nearly 180°, see Fig. 7b). The third peak in the $g_{LiCl}(r)$ curves around 5.8 Å is more significant for higher IPT and for higher concentrations, and it belongs to the Li$^+$-Cl$^-$-Li$^+$-Cl$^-$ chains. Such a motif can be seen in Fig 9b, marked with magenta lines.

*3.2.2. The $g_{LiO}(r)$ curve (Fig. 1)*

The first peak of the $g_{LiO}(r)$ curve (Figs. 1 and S.4) represents the Li$^+$-O "bonds", with an average "bond" distance 1.9-2.0 Å. There are some FFs, in which this distance is significantly higher: RH (2.18 Å), RM (2.08 Å), DVH (2.2 Å), RDVH (2.25 Å) and Li-IOD-T (2.08 Å). This peak is universally higher for lower IPT, showing the higher number of Li$^+$-O "bonds" in those models. Two smaller peaks can be observed at higher distances: around 4.2 Å and 5.4 Å. The smaller one can be attributed mostly to Li$^+$-O-H...O chains. (The O-H...O distance is around 2.8 Å, the Li$^+$-O distance is about 2 Å and the typical Li$^+$-O-H angles are around 120°; which give a Li$^+$-O distance distribution around 4.2 Å.) In the 19.55m sample these motifs are more frequent in models with higher IPT, because the number of H-bonded water molecules is higher there. In the 3.74m sample the number of Li$^+$-O pairs determines the frequency of this motif. The Li$^+$-Cl$^-$-H-O chains can result in similar Li$^+$-O distances (around 3.8 – 4.9 Å) but with a broader distribution. The number of these chains are rather small at low concentrations (3.74m sample) in the configurations obtained with low IPT FFs, however the peak around 4.2 Å is also present in these models, thus the contributions of the Li$^+$-Cl$^-$-H-O motifs to this peak must be small. The small third peak around 5.4 Å can be the sign of the Li$^+$-Cl$^-$-Li$^+$-O chains, which are completely missing for low IPT models at low concentrations, where this peak is also missing.

*3.2.3. The $g_{LiLi}(r)$ curve (Fig. 10)*

The first peak in $g_{LiLi}(r)$ curves (Figs. 10 and S.5) is around 3.8-4.1 Å, it belongs to the Li$^+$-Li$^+$ distances from Li$^+$-Cl$^-$-Li$^+$ chains (second neighbors). For low IPT FFs these pairs are not present in the 3.74m sample, and their number is small even in the 19.55m sample. However in the case of the high IPT FFs, such ion pairs can be found even in the 3.74m sample. The first peak is followed by a broad distribution of distances, which can be due to the contributions of fourth neighbors (such as Li$^+$-Cl$^-$-Li$^+$-Cl$^-$-Li$^+$ chains).

*3.2.4. The $g_{LiH}(r)$ curve (Fig. 11)*

The height of the first peak of the $g_{LiH}(r)$ (at 2.6 - 2.7 Å) increases together with the height of the first peak in the $g_{LiO}(r)$ (see Figs. 11 and S.6). This peak belongs to the Li$^+$-O-H motifs. The shoulder in the 3 – 3.9 Å region shows the contribution of the $Li^+$-$Cl^-$-H chains, these motifs have greater significance in the high IPT models and at higher concentrations. The smaller second peak around 4.7 Å is the contribution of the $Li^+$-O-H...O-H chains and it is significant mostly in lower concentrations, where the number of the H-bonded water molecules and the $Li^+$-O pairs are both high.

*3.2.5. The $g_{ClH}(r)$ curve (Fig.5)*

The first peak of the $g_{ClH}(r)$ curve (Fig. 5 and S.7) belongs to the nearest neighbor $Cl^-$-H pairs and was discussed in section 3.1.2. The $g_{ClH}(r)$ curve has a second peak around 3.6 Å and a third peak around 4.3 Å. The shorter one can be identified as the distance between the ion and the distant hydrogen atom of the neighboring water molecule. (The $Cl^-$-H-O angle is nearly 180° and the $Cl^-$-(H-)O distance is around 3-3.1 Å (see section 3.2.7).) This peak is higher in the lower IPT models, which is in agreement with the higher number of $Cl^-$-H "bonds" in these models. The peak around 4.3 Å can be attributed to $Cl^-$-$Li^+$-O-H chains, in accordance with that its height is proportional to the IPT.

*3.2.6. The $g_{ClCl}(r)$ partial (Fig. 12)*

The first peak of $g_{ClCl}(r)$ (Figs. 12 and S.8) can be found around 3.9 - 4 Å, its height is proportional to IPT and this peak stems from $Cl^-$-$Li^+$-$Cl^-$ chains. Such a motif is shown in Fig. 9b, marked with green. This peak is completely missing from the PPCF obtained with low IPT models at the lowest concentration (3.74m sample). The second peak around 5 Å is stronger for low IPT models and can be identified as the contribution of $Cl^-$-$Cl^-$ pairs, which are sharing a water molecule ($Cl^-$-H-O-H-$Cl^-$ motifs, see Fig. 9a, marked with green).

*3.2.7.The $g_{ClO}(r)$ curve (Fig. 13)*

The $g_{ClO}(r)$ curve is maybe the most interesting one, it changes the most with concentration and with IPT, see Figs. 13 and S.9. It has two characteristic distances close to each-other in the 2.8 - 4 Å region, their 'contest' can be recognized in the varying shape of the curve in this region: for some models and concentrations it has a single narrow peak at lower *r* distances (such as in the JC-S model at 3.74m, around 3.1 Å), it can have a low *r* peak with a shoulder on its right hand side (such as in the DS or JC-S models at the highest concentration), a peak at higher *r* and a shoulder on the left hand side (such as in the 19.55m sample by the RL model), or two distinct peaks (as in the 19.55m sample obtained with the RM model). The shorter of the two distances around 3.05-3.1 Å can be identified as the $Cl^-$-water contribution: the $Cl^-$-H-O unit. The longer one, around 3.4 Å, originates from $Cl^-$-$Li^+$-O chains (two corners of the tetrahedron around a $Li^+$ ion, see Fig. 9b, marked with yellow). The first one dominates at lower concentrations: the second one is completely missing from the $g_{ClO}(r)$ curves of the 3.74m sample obtained with low IPT models. As the concentration increases, the number of ion pairs increases and the longer $Cl^-$-O distance becomes significant. This behavior explains the different $Cl^-$-O 'first neighbor' distances of the models, presented in Fig.12. of Ref. [65]. The third characteristic $Cl^-$-O distance around 4.9 Å can be seen only in the curves of the less concentrated sample and mostly for models with low IPT. This peak can be associated with the second hydration shell of $Cl^-$, the $Cl^-$-H-O-H...O or $Cl^-$-H-O...H-O chains. (The Cl-H-O and the O-H...O angles are about 180°, the $Cl^-$-O first neighbor distance is around 3.1 Å, the O(-H)...O distance in H-bonded water molecules is about 2.8 Å, and the hydrogens around the oxygen are in tetrahedral arrangement.)

*3.2.8. The O-O PPCF (Fig. 14)*

The shortest characteristic distance in the $g_{OO}(r)$ curve (Figs. 14 and S.10) can be attributed to the H-bonded water molecules (see Fig. 9a, marked with orange): the O-H...O distance is around 2.75 Å. It can be recognized clearly in the curves of the 3.74m sample, but this peak is almost missing at the highest concentration for some FFs. The first peak in the $g_{OO}(r)$ curve of the 19.55m sample obtained with the JC-S model (around 3.15 Å) can be associated with the neighboring oxygens around the same $Li^+$. (The oxygens around the $Li^+$ ion are in tetrahedral arrangement (O-$Li^+$-O angles are 109°), the typical O-$Li^+$ "bond" length is around 1.9-2 Å, which leads to O-$Li^+$-O distances in the 3.1-3.25 Å region.) These pairs can be observed in the curves of the 3.74m sample as well, but their number is much smaller than the number of the H-bonded water molecules. In the models where the IPT is high, the number of the ion-water units is rather small compared to the number of ion-ion and water-water pairs, as it can be seen in Fig. 14. The distance of O-O pairs from oxygen atoms around the same $Cl^-$ ions (O-H-$Cl^-$-H-O chains) has a broad distribution in the 3.5-4.4 Å region (it can be estimated from the $Cl^-$-O distance (3.1 Å) and from the broad distribution of the H-$Cl^-$-H angle (see Fig. 6a).

The second neighbor water molecules (O-H...O-H...O chain) contribute to the peak around 4.4 - 4.5 Å in the $g_{OO}(r)$ curve of the 3.74m sample (it also can be observed in the 19.55m curve as a shoulder), while the peak around 6 Å is probably related to ion-water motifs again (maybe from O-H-$Cl^-$-$Li^+$-O chain), since it is stronger in low IPT models.

*3.2.9. The O-H partial (Fig. 8)*

The first intermolecular peak of the $g_{OH}(r)$ (Figs. 8 and S.11) is the contribution of the H-bonded water molecules (O...H), as discussed in section 3.1.3. The second peak, which is present in pure water as well, is around 3.25 Å. This peak shows the contribution of the O-H pairs, in which pairs the hydrogen is the non-bonding hydrogen from an H-bonded pair of water molecules (O...H-O-H or O-H...O-H motifs). The water-ion motifs, such as H-Cl$^-$-H-O chains and the O-Li$^+$-O-H structures, can lead to distances in the 3.3-5.4 Å and 3.4-4.4 Å regions, respectively. The shoulder around 4 Å is probably caused by these motifs.

*3.2.10. The $g_{HH}(r)$ curve (Fig. 15)*

The first peak of the $g_{HH}(r)$ curves (Figs. 15 and S.12) at 1.63 Å (models with SPC/E water) or 1.51 Å (models with TIP4P and TIP4PEw water) is the contribution of the intramolecular H-H distance. The second peak around 2.3-2.4 Å originates from the H-bonded water molecules (two hydrogens around the same oxygen atom: H-O...H or H...O...H unit). This peak, similarly to the first intermolecular O...H peak, is more significant for high IPT models, particularly at higher concentrations. The third, broader peak in the 3-5 Å region are present in all models and concentrations. A similar (but narrower) peak around 3.9 Å is present in pure water also, it is the contribution of those hydrogen pairs, which are from neighboring water molecules, but not participate in the H-bond (H-O...H-O-H). In LiCl solutions the hydrogens of the hydrating water molecules around the same ion, such as the H-Cl$^-$-H or H-O-Li$^+$-O-H motifs, give distances in the 2.6-5 Å region. Thus the third peak in the 3-5 Å region can be attributed to water-water and water-ion interactions as well, which explains why it can be observed in all FF similarly.

*3.3 Classification of the FFs by RMC refinement of the configurations*

The comparison of the performances of the FFs cannot be achieved directly from their PPCFs, since the PPCFs cannot be directly measured. The ND or XRD weighted total $S(Q)$ functions, which can be calculated from the PPCFs, can be measured however, and the experimental and simulated curves can be compared. This comparison was performed in Ref. [65]. It was found that the neutron weighted structure factor is reproduced reasonably well by the simulations, for the lowest concentration the best fits were obtained using those FFs which use the TIP4PEw water model. The discrepancy of the experimental and simulated X-ray weighted structure factors is significantly higher, and stronger differences can be observed between the FFs according to the X-ray measurements. The quality of the fits deteriorates by increasing concentration, and it is far from satisfactory at the highest concentration even for the best performing (JC-S and JC-T) models.

The configurations from MD simulations can be further refined by RMC simulations to obtain better agreement with experimental data. Fitting together the experimental $S(Q)$ functions and the PPCFs of the MD model can lead to more precise structural information. The method is also useful to investigate the compatibility of the MD models with experimental data: whether it is possible to find a configuration which fits properly all the input sets together. In case of an inconsistency the method can help to find its origin: the partials, which are problematic.

Simulations were performed to test the PPCF sets of the FFs, one by one, under the same conditions (the number of the fitted points, the relative weights of the input sets, the number of the generated moves etc. were the same). The $R$-factors of the fits are used here to compare the FFs. The results are collected in Figs. 16-18 and S.3-S.12.

The neutron and X-ray weights of the partials are different (see Fig. S.2). The quality of the fit of the ND structure factor is sensitive to the O-H (O-D), H-H (D-D) pairs mostly and also to the Cl-H (Cl-D) partial at higher concentrations. In the XRD total structure factor the contribution of the O-O and Cl-O partials are the highest, at lower concentrations the O-H and at higher concentrations the Cl-Cl partials are also significant. There are some partials, which have only small weight in both measured quantities, such as Li-Li, Li-Cl, Li-O or Li-H. The reliability of these PPCFs cannot be investigated directly. However, it is possible to test their reliability indirectly using the results presented in the previous section, because the contributions of the main structural units can be identified in several different PPCFs.

This invisibility of these partials also manifests itself in the quality of the fits of these PPCFs, which is remarkably good independently from the FF, although the shape of these curves is FF dependent (see the previous section). The $R$-factors of the fits of Li-Li, Li-Cl, Li-O and Li-H partials are collected in Fig. 16, and the quality of these fits are shown for two FFs (one with low and one with high IPT) at the 4 investigated concentrations (Figs. S.3-S.6). Most of the values of the $R$-factors are around or below 5 %, except for the lowest concentration, where the Li-Li and Li-Cl partials have higher $R$-factors. This latter is caused by the high noise due to the small number of these pairs in the configurations. It can be concluded that these invisible partials can always be fitted properly, and thus they are not suitable to compare the FFs.

The 'ND related input sets' are the $S^N(Q)$, $g_{OH}(r)$, $g_{HH}(r)$ and $g_{ClH}(r)$; their $R$-factors are shown in Fig. 17. Some examples of the fits (for the same two FFs as previously) are presented in Figs. 19, S.7, S.11, S.12. The $R$-factors of the Cl-H partials are usually significantly below 5 %, for the H-H partials they are around 5 %, for the O-H partials they are mostly between 5-8 %, and the $R$-factors of the $S^N(Q)$ are scattered between 5-10 %. There are small differences between the various results: The quality of the $g_{OH}(r)$ fits are better for higher concentrations; at the lowest concentration the FFs, which apply the TIP4PEw (or TIP4P) water models perform better than those that use the SPC/E model (this can be caused by the intramolecular contributions). The models, which according to these $R$-factors seem to be significantly worse than the others are: JJ, RH, and RDVH.

The highest differences between the performance of the FFs are in the quality of the fits of the 'XRD-related quantities': the $S^X(Q)$, $g_{ClCl}(r)$, $g_{ClO}(r)$, and $g_{OO}(r)$ functions. Some examples of the quality of their fits are presented in Figs. 20, S.8-S.10, the $R$-factors are shown in Fig. 18.

The quality of the fits of the $S^X(Q)$ deteriorates with increasing concentration for several models: their $R$-factors are around 4-5 % in the case of the 3.74m sample and around 7-10 % for the 19.55m sample. The best fits are obtained for DS, JC-S, JC-T, HM-LB, HL-LB, Gee, MP-S and Pl models, which all have low IPT. The worst fits belong to the JJ, RH, RM models, which have the highest IPT. The IPT dependence of the $R$-factors of the fit of the $S^X(Q)$ is presented in Fig. 21, which clearly shows, that the models with lower IPT perform better in this test.

The $R$-factors of the Cl-Cl PPCF are between 5 and 20 %. The worst fit of the 3.74m sample again comes from the high noise due to the small number of Cl-Cl pairs for the low IPT models. At higher concentrations the heights of the first peaks of the $g_{ClCl}(r)$ (proportional to the number of the Cl$^-$-Cl$^-$ (second) neighbors) are different in the MD and the RMC configurations, see Fig. S.8. This suggests that the RMC simulation – driven by the experimental data – try to force a given value (region). The best fits at the highest concentration are obtained for the JC-S, Gee, HL-LB, DS and HM-LB models (these all are low-IPT models).

The Cl-O $R$-factors are more scattered. For the 3.74m sample the quality of the fits are good enough, the $R$-factors are between 2 and 6 %. As the concentration increases the weight of this partial becomes higher, and the quality of the fits deteriorates. The highest $R$-factors can be found in this set: the curve obtained by the JJ for the 19.55m sample gives the worst result. Two tendencies can be observed in the RMC curves (see Fig. S.9). First they tend toward the low IPT curves: the shorter Cl-O distances are more typical for them. Secondly, in the curves obtained by the RMC refinement there is a dip around 4 Å, which shows that there is not enough $Cl^-$-O pairs in the RMC configurations to satisfy the requirements of the MD-obtained curve. A similar tendency can be seen in the O-O PPCFs around 4 Å (see Fig. S.10). As this distance is the typical length between the nearest $Cl^-$-$Cl^-$ pairs, and the $S^X(Q)$ is sensitive to the amount of the $Cl^-$-$Cl^-$, $Cl^-$-O and O-O pairs, it is possible that the too high contribution of $Cl^-$-$Cl^-$ to the total $S^X(Q)$ is compensated by the lack of $Cl^-$-O and O-O pairs. At the 8.30m and 11.37m samples the lowest $R$-factors are produced by the Ar, MP-T, JC-T, Pl, MP-S and Gee models. At the highest concentration (19.55m sample) the JC-S and JC-T models have the lowest $R$-factors (about 9 %). The changes between the $g_{OO}(r)$ curves obtained from MD and RMC (Fig. S.10) show again that at high concentrations the RMC refinement modifies the curves of high IPT models toward the shape of the curves of the low IPT models. The smallest (best) $R$-factors are around 5-6 % for the 3.74m (Ar, Pl, MP-T, RL-sLB, MP-S, Gee models), the 11.37m (JC-T, MP-S, MP-T, Ar models) and the 19.55m (JC-S, JC-T models) samples, and below 5 % for the 8.30m sample (Ar, MP-T models). Several of the high IPT models cannot be fitted properly, their $R$-factors are above 10 %, these are the JJ, RH, RM, DVH, RDVH, Li-HFE-S, Li-HFE-T, SDG-S, SDG-T, HL-g, HM-g, AqCh, Rm, Li-IOD-T, HS-g and HS-LB models.

In summary it can be concluded that the models, which have low IPT perform significantly better during the experiment based RMC refinement than those, which have high IPT. The most sensitive data sets are the 'X-ray related' data sets: the $S^X(Q)$ structure function, and the $g_{ClO}(r)$ and $g_{OO}(r)$ PPCFs. These PPCFs if obtained from high IPT FFs are not compatible with the experimental $S^X(Q)$ data. These PPCFs require the strongest modifications during the RMC refinement procedure, which suggests that high-IPT MD models predict them poorly. The question then arises, whether the 'invisible' and 'ND related' PPCFs, obtained with the high IPT FFs are trustworthy or not. The answer is probably no for two reasons: First, the method is highly insensitive to these PPCFs (almost any curve can be fitted properly). Secondly, the structural motifs make connections between the PPCFs, as it was shown in section 3.1 and 3.2, and thus modifications in the 'X-ray related' data sets should lead to variations of the other PPCFs as well.

*3.4 The structure of the aqueous LiCl solutions at high concentrations*

As it was seen in the previous sections there are huge differences between the structural features predicted by different FFs. Taking into account the quality of the fits, some conclusions can be obtained about the structure. At first it should be noted that the quality of the fits is worse for the three 'outlying models': RH, DVH and RDVH, which models have significantly different properties than the other FFs (see section 3.1.4.). It suggests that these models are inadequate in describing aqueous LiCl solutions, thus these models will not be discussed further.

The most important variation in the other models is the number of the ion pairs (IPT), which has significant effects on the fits of the $S^X(Q)$ function and also the XRD-related partials. From the quality of these fits it can be concluded that the models with lower IPT predict the structure better.

Furthermore it is common in all the models, that there are four neighboring atoms around the $Li^+$ ions. The weights of the $Li^+$ related partials are relatively small in the measured $S(Q)$ functions, but it can be concluded that the tetrahedral arrangement around $Li^+$ is consistent with the experiments. The proportion of the neighboring $Cl^-$ ions and water molecules are different in different models. At lower concentrations the experimental data sets can be fitted even if the number of the ion pairs is relatively high, but as the concentration is increasing, the quality of the fits is more sensitive to the ion pairing tendency. At the highest investigated concentration even the lowest $N_{LiCl}$ coordination number (=1.28, obtained using JC-S model) is too high to get proper fits. It suggests that to get the best model the number of ion pairs should be kept on a minimum value. It is easy to calculate the minimum of $N_{LiCl}$, which still satisfies the coordination requirement of $Li^+$. If one assumes that all water molecules are connected to one $Li^+$ ion and every $Li^+$ ions have 4 nearest neighbors, that requires an average of 1.16 Li-Cl pairs for the 19.55m sample, in which sample there are 2.84 water molecules per ion pairs. This minimum is lower than the $N_{LiCl}$ of the best fitting (lowest IPT) model, which points toward a possible direction of further refinement of the MD models of LiCl solutions.

The chloride ions are surrounded by an average of 7 hydrogen atoms at lower concentrations. These hydrogen atoms belong to 7 water molecules, as it can be concluded from the cosine distribution of the $Cl^-$-H-O (Fig. 7b) and H-$Cl^-$-H (Fig. 6a) angles. The hydrogen atoms (water molecules) can be replaced pairwise by one $Li^+$ ion as the concentration increases. Another concentration and model independent property of the structure is that the $N_{ClLi}+1/2N_{ClH}$ value is around 3.5 always.

The ions built in the H-bonded water structure are connected by the water molecules through ion-water motifs. In pure water the water molecules are in tetrahedral arrangement, every oxygen atoms are bonded to two intermolecular hydrogen atoms. In LiCl solutions for some of the water molecules these two H-bonds are replaced by one $Li^+$-O connection. The hydrogen atoms of some water molecules, which were connected (through H-bonds) to the neighboring water molecules in pure water, lose their oxygen pair and connect to $Cl^-$ ions in LiCl solutions. As the concentration increases, more and more water molecules are bonded to the ions, and the number of the water-water connections decreases. At the highest concentration studied here the water molecules on average are bonded to two $Cl^-$ ions (with their hydrogen atoms) and one $Li^+$ ion (through their oxygen) and some small number of H-bonded water molecules remain as well.

## 4. Conclusions

The structure of highly concentrated aqueous lithium chloride solutions was investigated by MD and RMC simulations. The predictions of 29 interaction models were compared. The contributions of the different structural units to the characteristic distances in the PPCF curves were identified. It was found that the PPCF curves obtained with different FFs can be ordered by their ion pairing tendency (the number of $Li^+$-$Cl^-$ pairs in the configurations obtained by the models). Compatibility of the model $g(r)$ curves obtained from MD simulations with experimental total structure factors (from neutron and X-ray diffraction experiments) were investigated by RMC technique. It was found that although none of the FFs describe the structure perfectly, those FFs, which have lower ion pairing tendency show better agreement with the experiments.

Several model independent properties of the structure of LiCl solutions were identified complemented with such structural properties, which are common to the best fitting low-IPT models.

The local environment of $Li^+$ ions consists 4 water molecules at the lowest concentration studied here (3.74 mol/kg), the experimental data can be fitted without any ion-pairing. As the concentration increases the number of cation-anion pairs increases. The results suggest that the $N_{LiCl}$ is not higher than 1.28 (the value of the JC-S model) at the highest concentration (19.55 mol/kg). The number of the surrounding water molecules decreases with increasing concentration. The sum of the O atoms and $Cl^-$ ions around a $Li^+$ ion is constant (four) in the entire investigated concentration range.

The chloride ions are surrounded by about 7 hydrogen atoms (from different water molecules) at the lowest concentration. As the concentration increases, two of these hydrogen atoms can be substituted by one $Li^+$ ion in the environment of some of the $Cl^-$ ions. Thus the number of the neighboring water molecules reduces twice as fast as the number of anion-cation pairs increases.

As the ion concentration is increasing, the network of the H-bonded water molecules is reordering. Two H-bonds of the oxygen atom can be replaced by one $Li^+$-O bond, while the released hydrogen atoms of the water molecule are bonded to the chloride ions.

According to these findings the JC-S and JC-T models give the best results for the structure of aqueous LiCl solutions near the solubility limit. The results suggest that even these models can be improved by lowering their ion pairing tendency.

Acknowledgments

The author is grateful to the National Research, Development and Innovation Office (NKFIH) of Hungary for financial support through Grants No. SNN 116198 and K 124885. The author would like to thank L. Pusztai (Wigner RCP) for the helpful discussions and suggestions.

**Table 1**

Investigated concentrations. The numbers of ion pairs / water molecules and the experimental densities are taken from Ref. [11].

| Molality [mol/kg] | 3.74 | 8.30 | 11.37 | 19.55 |
|---|---|---|---|---|
| $N_{LiCl}$ | 200 | 500 | 700 | 1000 |
| $N_{water}$ | 2968 | 3345 | 3416 | 2840 |
| Density [g/cm$^3$] | 1.076 | 1.1510 | 1.1950 | 1.2862 |
| Number density [Å$^{-3}$] | 0.09735 | 0.0939 | 0.0919 | 0.0871 |
| Box length [nm] | 4.5721 | 4.8982 | 5.0232 | 4.94102 |

**Table 2**

Force field parameters of the investigated models. The applied water models and combination rules are also shown. For the definitions of the combination rules see the text.

| Model | $q_{Li}/q_{Cl}[e]$ | $\sigma_{LiLi}$ [nm] | $\varepsilon_{LiLi}$ [kJ/mol] | $\sigma_{ClCl}$ [nm] | $\varepsilon_{ClCl}$ [kJ/mol] | comb. rule | water model | References |
|---|---|---|---|---|---|---|---|---|
| Ch | +1/-1 | 0.126 | 26.1495 | 0.4417 | 0.4928 | geom | TIP4P | [73] |
| DS | +1/-1 | 0.1506 | 0.6904 | 0.4400 | 0.4184 | LB | SPC/E | [74, 75] |
| JJ | +1/-1 | 0.2870 | 0.0021 | 0.4020 | 2.9706 | geom | TIP4P | [76] |
| JC-S | +1/-1 | 0.1409 | 1.4089 | 0.4830 | 0.0535 | LB | SPC/E | [78] |
| JC-T | +1/-1 | 0.1440 | 0.4351 | 0.4918 | 0.0488 | LB | TIP4PEw | [78] |
| HS-g | +1/-1 | 0.2880 | 0.0006 | 0.4520 | 0.4200 | geom | SPC/E | [79] |
| HM-g | +1/-1 | 0.1700 | 0.6500 | 0.4520 | 0.4200 | geom | SPC/E | [79] |
| HL-g | +1/-1 | 0.1630 | 1.5400 | 0.4520 | 0.4200 | geom | SPC/E | [79] |
| HS-LB | +1/-1 | 0.2870 | 0.0006 | 0.4400 | 0.4200 | LB | SPC/E | [79] |
| HM-LB | +1/-1 | 0.1470 | 0.6500 | 0.4400 | 0.4200 | LB | SPC/E | [79] |
| HL-LB | +1/-1 | 0.1370 | 1.5400 | 0.4400 | 0.4200 | LB | SPC/E | [79] |
| Gee | +1/-1 | 0.182 | 0.7 | 0.44 | 0.47 | mgeom | SPC/E | [80] |
| RH | +1/-1 | 0.3529 | 0.0007 | 0.3493 | 1.7625 | geom | SPC/E | [81] |
| RM | +1/-1 | 0.3078 | 0.0015 | 0.3771 | 1.1137 | geom | SPC/E | [81] |
| RL | +1/-1 | 0.2679 | 0.0035 | 0.4096 | 0.6785 | geom | SPC/E | [81] |
| RL-sLB | +1/-1 | 0.2679 | 0.0035 | 0.4096 | 0.6785 | sLB | SPC/E | [81] |
| MP-S | +1/-1 | 0.1715 | 0.2412 | 0.4612 | 0.1047 | LB | SPC/E | [82] |
| MP-T | +1/-1 | 0.1715 | 0.2412 | 0.4612 | 0.1047 | LB | TIP4PEw | [82] |
| DVH | +1/-1 | 0.1880 | 0.8314 | 0.4410 | 0.8314 | LB | SPC/E | [83] |
| RDVH | +1/-1 | 0.1880 | 1.6629 | 0.4410 | 1.6629 | LB | SPC/E | [84] |
| Li-HFE-S | +1/-1 | 0.2242 | 0.0115 | 0.4112 | 2.6931 | LB | SPC/E | [85] |
| Li-HFE-T | +1/-1 | 0.2184 | 0.0071 | 0.4136 | 2.7309 | LB | TIP4PEw | [85] |
| Li-IOD-S | +1/-1 | 0.2343 | 0.0249 | 0.3852 | 2.2240 | LB | SPC/E | [85] |
| Li-IOD-T | +1/-1 | 0.2343 | 0.0249 | 0.3852 | 2.2240 | LB | TIP4PEw | [85] |
| AqCh | +1/-1 | 0.2126 | 0.0765 | 0.4417 | 0.4928 | geom | SPC/E | [73, 86] |
| Pl | +0.75/-0.75 | 0.1800 | 0.0765 | 0.4100 | 0.4928 | LB | SPC/E | [27, 88] |
| Ar | +1/-1 | 0.1440 | 0.4351 | 0.4918 | 0.0488 | mLB | TIP4PEw | [48] |
| SDG-S | +1/-1 | 0.1506 | 0.6945 | 0.402 | 2.9706 | LB | SPC/E | [91] |
| SDG-T | +1/-1 | 0.1506 | 0.6945 | 0.402 | 2.9706 | LB | TIP4P | [91] |

**Table 3**

Parameters of the water-models. In the TIP4P and TIP4PEw models there is a fourth (virtual) site (M). It is situated along the bisector of the H-O-H angle and coplanar with the oxygen and the hydrogen atoms. The negative charge is placed in M.

|        | $\sigma_{OO}$ [nm] | $\varepsilon_{OO}$ [kJ/mol] | $q_H$ [e] | $d_{O-H}$ [nm] | $\theta_{H-O-H}$ [deg] | $d_{O-M}$ [nm] | Ref. |
|--------|--------|--------|--------|--------|--------|--------|--------|
| SPC/E   | 0.3166 | 0.6502 | +0.4238  | 0.1     | 109.47 | -      | [70] |
| TIP4P   | 0.3154 | 0.6485 | +0.52    | 0.09572 | 104.52 | 0.015  | [71] |
| TIP4PEw | 0.3164 | 0.6809 | +0.52422 | 0.09572 | 104.52 | 0.0125 | [72] |

**Table 4**

Minimum and maximum O-H and H-H intramolecular distances used in fixed neighbor constraints for the different water models (in Å).

| Water model | O-H       | H-H       |
|-------------|-----------|-----------|
| SPC/E       | 0.94-1.05 | 1.45-1.70 |
| TIP4P       | 0.94-1.05 | 1.45-1.65 |
| TIP4PEw     | 0.94-1.05 | 1.45-1.65 |

**Table 5**

Intermolecular closest approach distances (cutoff distances) between atoms (in Å). [a]The higher value was used for SPC/E water and the lower value for TIP4P and TIP4PEw water models.

| Li-Li | Li-Cl | Li-O | Li-H | Cl-Cl | Cl-O | Cl-H | O-O | O-H | H-H |
|-------|-------|------|------|-------|------|------|-----|-----|-----|
| 2.7   | 2.1   | 1.7  | 2.2  | 3.1   | 2.8  | 1.6  | 2.5 | 1.5 | 1.7/1.66[a] |

**Table 6** $N_{\text{LiCl}}$ coordination numbers (calculated up to the first minimum of the $g_{\text{LiCl}}(r)$ curves) obtained from the MD simulations.

| forcefield | 3.74m | 8.3m | 11.37m | 19.55m |
|---|---|---|---|---|
| Ch | 1.28 | 1.59 | 1.75 | 2.06 |
| DS | 0.12 | 0.42 | 0.71 | 1.5 |
| JJ | 2.05 | 2.26 | 2.35 | 2.49 |
| JC-S | 0.01 | 0.08 | 0.24 | 1.28 |
| JC-T | 0.08 | 0.31 | 0.59 | 1.47 |
| HS-g | 0.94 | 1.29 | 1.47 | 1.86 |
| HM-g | 1.04 | 1.35 | 1.51 | 1.92 |
| HL-g | 1.08 | 1.42 | 1.57 | 2.01 |
| HS-LB | 0.95 | 1.34 | 1.52 | 1.97 |
| HM-LB | 0.09 | 0.42 | 0.69 | 1.5 |
| HL-LB | 0.06 | 0.27 | 0.56 | 1.42 |
| Gee | 0.15 | 0.37 | 0.64 | 1.47 |
| RH | 3.06 | 3.49 | 3.61 | 3.44 |
| RM | 1.95 | 2.26 | 2.32 | 2.51 |
| RL | 1.28 | 1.71 | 1.89 | 2.18 |
| RL-LB | 0.99 | 1.37 | 1.55 | 2 |
| MP-S | 0.18 | 0.59 | 0.89 | 1.66 |
| MP-T | 0.46 | 0.93 | 1.27 | 1.86 |
| DVH | 0.08 | 0.48 | 0.97 | 2.24 |
| RDVH | 0.16 | 0.73 | 1.27 | 2.6 |
| Li-HFE-S | 0.57 | 1 | 1.22 | 1.81 |
| Li-HFE-T | 0.81 | 1.16 | 1.34 | 1.84 |
| Li-IOD-S | 1.15 | 1.61 | 1.82 | 2.19 |
| Li-IOD-T | 1.66 | 2.08 | 2.22 | 2.44 |
| AqCh | 1.28 | 1.67 | 1.83 | 2.15 |
| Pl | 0.22 | 0.62 | 0.9 | 1.56 |
| Ar | 0.47 | 0.92 | 1.22 | 1.82 |
| SDG-S | 0.45 | 0.92 | 1.19 | 1.8 |
| SDG-T | 1.16 | 1.58 | 1.76 | 2.15 |

**Table 7** $N_{LiO}$ coordination numbers (calculated up to the first minimum of the $g_{LiO}(r)$ curves) and the total coordination numbers of Li$^+$ ions ($N_{Li}=N_{LiCl}+N_{LiO}$) obtained from the MD simulations.

| forcefield | $N_{LiO}$ | | | | $N_{Li}=N_{LiO}+N_{LiCl}$ | | | |
|---|---|---|---|---|---|---|---|---|
| | 3.74m | 8.3m | 11.37m | 19.55m | 3.74m | 8.3m | 11.37m | 19.55m |
| Ch | 2.73 | 2.41 | 2.25 | 1.94 | 4.01 | 4 | 4 | 4 |
| DS | 4.01 | 3.7 | 3.38 | 2.56 | 4.13 | 4.12 | 4.09 | 4.06 |
| JJ | 2.06 | 1.83 | 1.73 | 1.63 | 4.11 | 4.09 | 4.08 | 4.12 |
| JC-S | 4.16 | 4.05 | 3.84 | 2.76 | 4.17 | 4.13 | 4.08 | 4.04 |
| JC-T | 3.95 | 3.72 | 3.42 | 2.53 | 4.03 | 4.03 | 4.01 | 4 |
| HS-g | 3.06 | 2.7 | 2.53 | 2.14 | 4 | 3.99 | 4 | 4 |
| HM-g | 2.97 | 2.66 | 2.49 | 2.08 | 4.01 | 4.01 | 4 | 4 |
| HL-g | 2.94 | 2.59 | 2.43 | 2 | 4.02 | 4.01 | 4 | 4.01 |
| HS-LB | 3.06 | 2.66 | 2.48 | 2.03 | 4.01 | 4 | 4 | 4 |
| HM-LB | 3.96 | 3.63 | 3.35 | 2.52 | 4.05 | 4.05 | 4.04 | 4.02 |
| HL-LB | 4.09 | 3.86 | 3.55 | 2.65 | 4.15 | 4.13 | 4.11 | 4.07 |
| Gee | 3.86 | 3.63 | 3.37 | 2.53 | 4.01 | 4 | 4.01 | 4 |
| RH | 1.92 | 1.55 | 1.43 | 1.3 | 4.98 | 5.04 | 5.04 | 4.74 |
| RM | 2.1 | 1.77 | 1.69 | 1.51 | 4.05 | 4.03 | 4.01 | 4.02 |
| RL | 2.72 | 2.29 | 2.11 | 1.82 | 4 | 4 | 4 | 4 |
| RL-LB | 3.05 | 2.65 | 2.46 | 2.01 | 4.04 | 4.02 | 4.01 | 4.01 |
| MP-S | 3.91 | 3.47 | 3.15 | 2.36 | 4.09 | 4.06 | 4.04 | 4.02 |
| MP-T | 3.88 | 3.28 | 2.89 | 2.24 | 4.34 | 4.21 | 4.16 | 4.1 |
| DVH | 5.37 | 5.01 | 4.55 | 3.42 | 5.45 | 5.49 | 5.52 | 5.66 |
| RDVH | 5.54 | 5.09 | 4.59 | 3.29 | 5.7 | 5.82 | 5.86 | 5.89 |
| Li-HFE-S | 3.44 | 3 | 2.79 | 2.2 | 4.01 | 4 | 4.01 | 4.01 |
| Li-HFE-T | 3.19 | 2.85 | 2.66 | 2.16 | 4 | 4.01 | 4 | 4 |
| Li-IOD-S | 3.03 | 2.51 | 2.29 | 1.89 | 4.18 | 4.12 | 4.11 | 4.08 |
| Li-IOD-T | 2.63 | 2.1 | 1.95 | 1.72 | 4.29 | 4.18 | 4.17 | 4.16 |
| AqCh | 2.76 | 2.34 | 2.19 | 1.86 | 4.04 | 4.01 | 4.02 | 4.01 |
| Pl | 3.78 | 3.39 | 3.1 | 2.45 | 4 | 4.01 | 4 | 4.01 |
| Ar | 3.54 | 3.09 | 2.79 | 2.18 | 4.01 | 4.01 | 4.01 | 4 |
| SDG-S | 3.65 | 3.17 | 2.9 | 2.28 | 4.1 | 4.09 | 4.09 | 4.08 |
| SDG-T | 3.17 | 2.67 | 2.48 | 2.09 | 4.33 | 4.25 | 4.24 | 4.24 |

**Table 8** $N_{ClH}$ coordination numbers (calculated up to the first minimum of the $g_{ClH}(r)$ curves) and the 'weighted total coordination number' of Cl$^-$ ions ($N_{Cl}=N_{ClLi}+N_{ClH}/2$) obtained from the MD simulations.

| | $N_{ClH}$ | | | | $N_{Cl}=N_{ClH}/2+N_{ClLi}$ | | | |
|---|---|---|---|---|---|---|---|---|
| forcefield | 3.74m | 8.3m | 11.37m | 19.55m | 3.74m | 8.3m | 11.37m | 19.55m |
| Ch | 4.04 | 3.48 | 3.22 | 2.75 | 3.3 | 3.33 | 3.36 | 3.435 |
| DS | 6.79 | 6.31 | 5.82 | 4.39 | 3.515 | 3.575 | 3.62 | 3.695 |
| JJ | 2.91 | 2.5 | 2.38 | 2.25 | 3.505 | 3.51 | 3.54 | 3.615 |
| JC-S | 6.86 | 6.83 | 6.56 | 4.72 | 3.44 | 3.495 | 3.52 | 3.64 |
| JC-T | 6.47 | 6.04 | 5.61 | 4 | 3.315 | 3.33 | 3.395 | 3.47 |
| HS-g | 5.04 | 4.39 | 4.08 | 3.41 | 3.46 | 3.485 | 3.51 | 3.565 |
| HM-g | 4.94 | 4.32 | 4.04 | 3.34 | 3.51 | 3.51 | 3.53 | 3.59 |
| HL-g | 4.82 | 4.2 | 3.97 | 3.21 | 3.49 | 3.52 | 3.555 | 3.615 |
| HS-LB | 5.06 | 4.33 | 4 | 3.27 | 3.48 | 3.505 | 3.52 | 3.605 |
| HM-LB | 6.88 | 6.29 | 5.82 | 4.31 | 3.53 | 3.565 | 3.6 | 3.655 |
| HL-LB | 6.94 | 6.63 | 6.09 | 4.56 | 3.53 | 3.585 | 3.605 | 3.7 |
| Gee | 6.72 | 6.35 | 5.9 | 4.39 | 3.51 | 3.545 | 3.59 | 3.665 |
| RH | 2.36 | 1.86 | 1.67 | 1.61 | 4.24 | 4.42 | 4.445 | 4.245 |
| RM | 3.25 | 2.69 | 2.58 | 2.29 | 3.575 | 3.605 | 3.61 | 3.655 |
| RL | 4.37 | 3.57 | 3.23 | 2.8 | 3.465 | 3.495 | 3.505 | 3.58 |
| RL-LB | 5 | 4.35 | 4.07 | 3.3 | 3.49 | 3.545 | 3.585 | 3.65 |
| MP-S | 6.61 | 5.85 | 5.34 | 3.99 | 3.485 | 3.515 | 3.56 | 3.655 |
| MP-T | 5.75 | 4.95 | 4.35 | 3.33 | 3.335 | 3.405 | 3.445 | 3.525 |
| DVH | 7.08 | 6.66 | 5.98 | 4.07 | 3.62 | 3.81 | 3.96 | 4.275 |
| RDVH | 7.23 | 6.52 | 5.67 | 3.75 | 3.775 | 3.99 | 4.105 | 4.475 |
| Li-HFE-S | 6.08 | 5.31 | 4.81 | 3.85 | 3.61 | 3.655 | 3.625 | 3.735 |
| Li-HFE-T | 5.27 | 4.55 | 4.25 | 3.35 | 3.445 | 3.435 | 3.465 | 3.515 |
| Li-IOD-S | 4.82 | 4.08 | 3.7 | 3.1 | 3.56 | 3.65 | 3.67 | 3.74 |
| Li-IOD-T | 3.69 | 2.94 | 2.72 | 2.38 | 3.505 | 3.55 | 3.58 | 3.63 |
| AqCh | 4.39 | 3.75 | 3.5 | 2.98 | 3.475 | 3.545 | 3.58 | 3.64 |
| Pl | 5.21 | 4.71 | 4.34 | 3.4 | 2.825 | 2.975 | 3.07 | 3.26 |
| Ar | 5.64 | 4.87 | 4.34 | 3.28 | 3.29 | 3.355 | 3.39 | 3.46 |
| SDG-S | 6.29 | 5.49 | 5.09 | 3.91 | 3.595 | 3.665 | 3.735 | 3.755 |
| SDG-T | 4.69 | 4 | 3.69 | 3.06 | 3.505 | 3.58 | 3.605 | 3.68 |

**Table 9** $N_{OH}$ intermolecular coordination and the 'weighted total coordination number' of the oxygen atoms ($N_O = N_{OLi} + N_{OH}/2$) obtained from the MD simulations.

| | $N_{OH}$ intermolecular | | | | $N_O = N_{OH}/2 + N_{OLi}$ | | | |
|---|---|---|---|---|---|---|---|---|
| forcefield | 3.74m | 8.3m | 11.37m | 19.55m | 3.74m | 8.3m | 11.37m | 19.55m |
| Ch | 1.58 | 1.35 | 1.22 | 0.95 | 0.97 | 1.04 | 1.07 | 1.16 |
| DS | 1.44 | 0.93 | 0.72 | 0.39 | 0.99 | 1.02 | 1.05 | 1.10 |
| JJ | 1.65 | 1.52 | 1.39 | 1.15 | 0.96 | 1.03 | 1.05 | 1.15 |
| JC-S | 1.4 | 0.88 | 0.57 | 0.25 | 0.98 | 1.05 | 1.07 | 1.10 |
| JC-T | 1.46 | 1.01 | 0.8 | 0.51 | 1.00 | 1.06 | 1.10 | 1.15 |
| HS-g | 1.56 | 1.21 | 1.05 | 0.68 | 0.99 | 1.01 | 1.04 | 1.09 |
| HM-g | 1.53 | 1.2 | 1.03 | 0.71 | 0.97 | 1.00 | 1.03 | 1.09 |
| HL-g | 1.59 | 1.26 | 1.08 | 0.77 | 0.99 | 1.02 | 1.04 | 1.09 |
| HS-LB | 1.55 | 1.23 | 1.03 | 0.74 | 0.98 | 1.01 | 1.02 | 1.08 |
| HM-LB | 1.44 | 0.94 | 0.72 | 0.37 | 0.99 | 1.01 | 1.05 | 1.07 |
| HL-LB | 1.39 | 0.91 | 0.67 | 0.32 | 0.97 | 1.03 | 1.06 | 1.09 |
| Gee | 1.45 | 0.96 | 0.69 | 0.38 | 0.99 | 1.02 | 1.04 | 1.08 |
| RH | 1.71 | 1.6 | 1.53 | 1.3 | 0.98 | 1.03 | 1.06 | 1.11 |
| RM | 1.66 | 1.47 | 1.33 | 1.07 | 0.97 | 1.00 | 1.01 | 1.07 |
| RL | 1.56 | 1.34 | 1.18 | 0.84 | 0.96 | 1.01 | 1.02 | 1.06 |
| RL-LB | 1.55 | 1.23 | 1.02 | 0.72 | 0.98 | 1.01 | 1.01 | 1.07 |
| MP-S | 1.45 | 1.01 | 0.79 | 0.49 | 0.99 | 1.02 | 1.04 | 1.08 |
| MP-T | 1.49 | 1.15 | 1 | 0.74 | 1.01 | 1.07 | 1.09 | 1.16 |
| DVH | 1.38 | 0.91 | 0.72 | 0.5 | 1.05 | 1.20 | 1.29 | 1.45 |
| RDVH | 1.38 | 0.97 | 0.79 | 0.63 | 1.06 | 1.25 | 1.34 | 1.47 |
| Li-HFE-S | 1.47 | 1.12 | 0.91 | 0.58 | 0.97 | 1.01 | 1.03 | 1.06 |
| Li-HFE-T | 1.55 | 1.24 | 1.06 | 0.8 | 0.99 | 1.05 | 1.08 | 1.16 |
| Li-IOD-S | 1.53 | 1.27 | 1.11 | 0.86 | 0.97 | 1.01 | 1.02 | 1.10 |
| Li-IOD-T | 1.63 | 1.46 | 1.36 | 1.08 | 0.99 | 1.04 | 1.08 | 1.15 |
| AqCh | 1.56 | 1.33 | 1.19 | 0.85 | 0.97 | 1.01 | 1.04 | 1.08 |
| Pl | 1.51 | 1.19 | 1.03 | 0.73 | 1.01 | 1.10 | 1.15 | 1.23 |
| Ar | 1.5 | 1.18 | 1.02 | 0.76 | 0.99 | 1.05 | 1.08 | 1.15 |
| SDG-S | 1.49 | 1.09 | 0.88 | 0.57 | 0.99 | 1.02 | 1.03 | 1.09 |
| SDG-T | 1.55 | 1.28 | 1.15 | 0.86 | 0.99 | 1.04 | 1.08 | 1.17 |

Figures

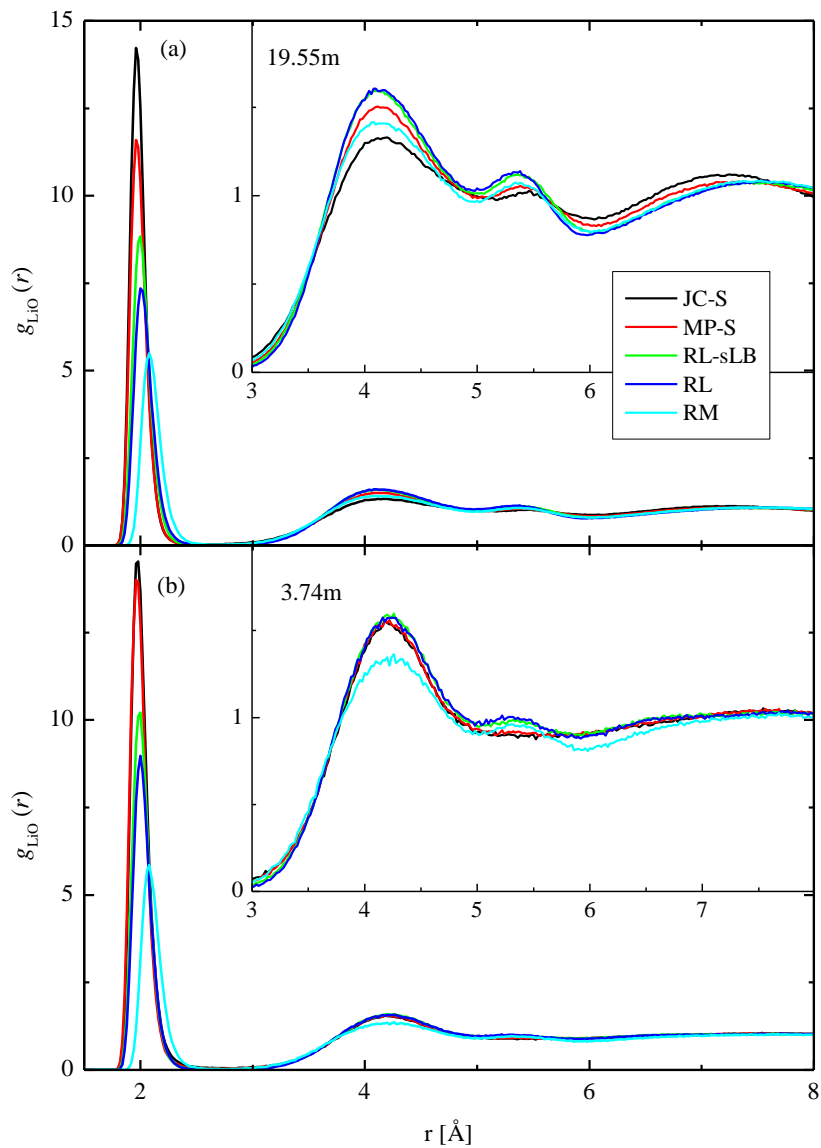

**Figure 1.** Li-O partial pair correlation functions as calculated from MD simulations with different interatomic potential models (black) JC-S, (red) MP-S, (light green) RL-sLB, (blue) RL, (cyan) RM. The curves are shown for (a) the 19.55m and (b) the 3.74m samples. The inset is an enlargement of the high *r* part of the curves.

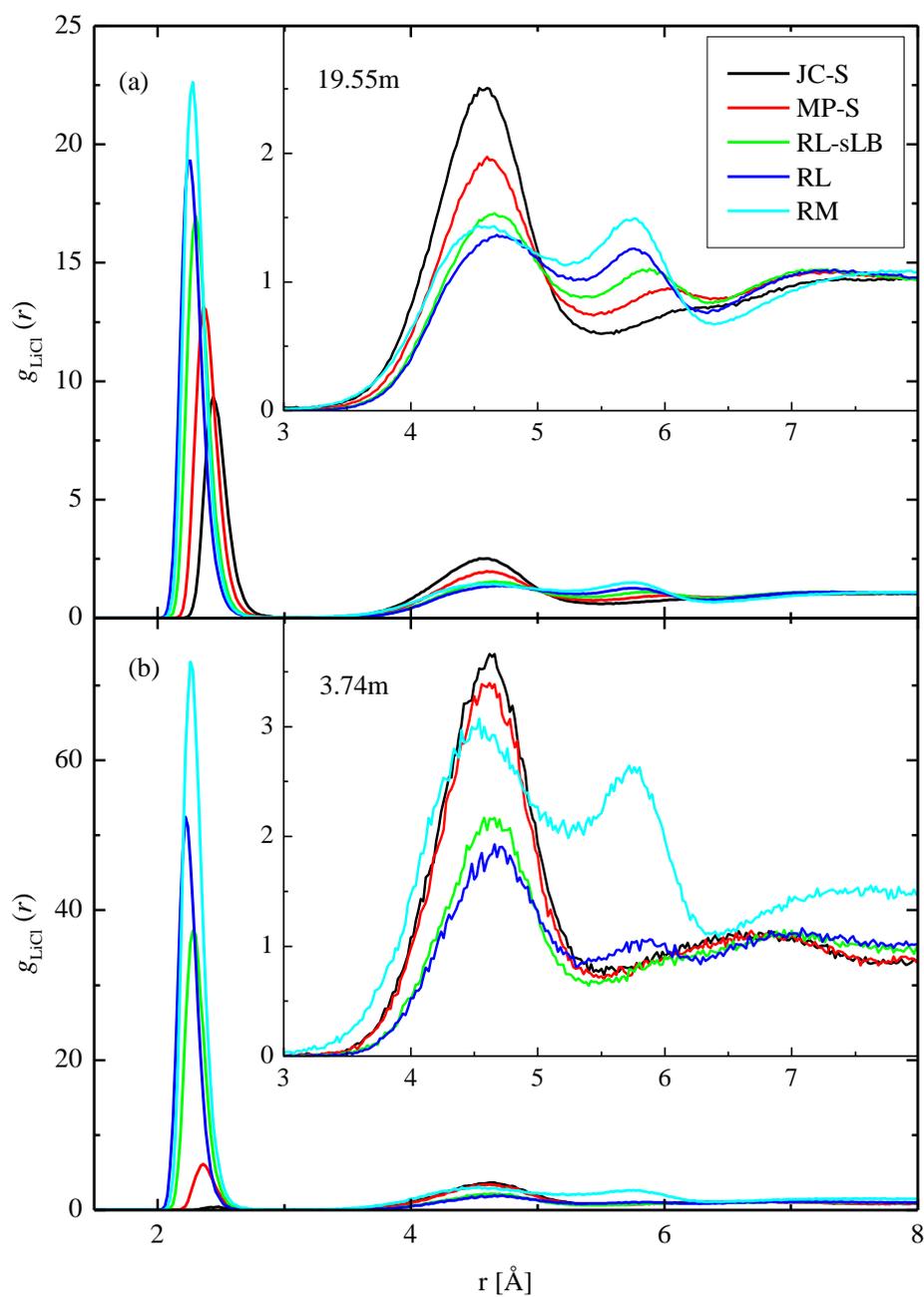

**Figure 2.** Li-Cl partial pair correlation functions as computed from MD simulations with different interatomic potential models (black) JC-S, (red) MP-S, (light green) RL-sLB, (blue) RL, (cyan) RM. The curves are shown for (a) the 19.55m and (b) the 3.74m samples. The inset is an enlargement of the high *r* part of the curves.

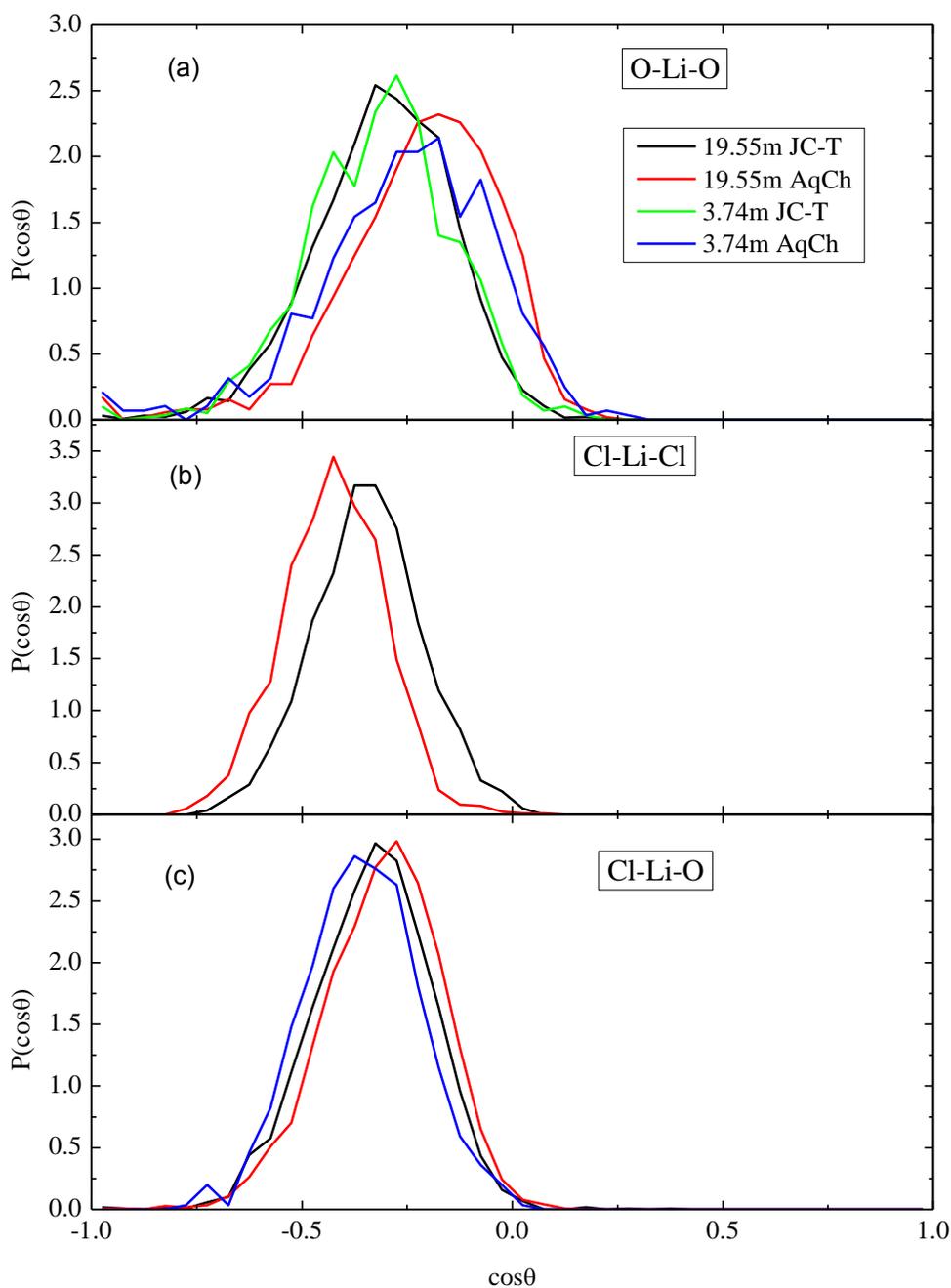

**Figure 3.** First coordination shell of Li$^+$ ions: (a) O-Li-O, (b) Cl-Li-Cl, and (c) Cl-Li-O cosine distributions obtained from MD simulations (the RMC simulations produced similar curves), for the 19.55m (black and red curves) and 3.74m (green and blue curves) samples. The distributions are shown for two FFs with different ion pairing tendency (see text): JC-T (low IPT, black and green) and AqCh (high IPT, red and blue). Some triplets are missing from the 3.74m sample.

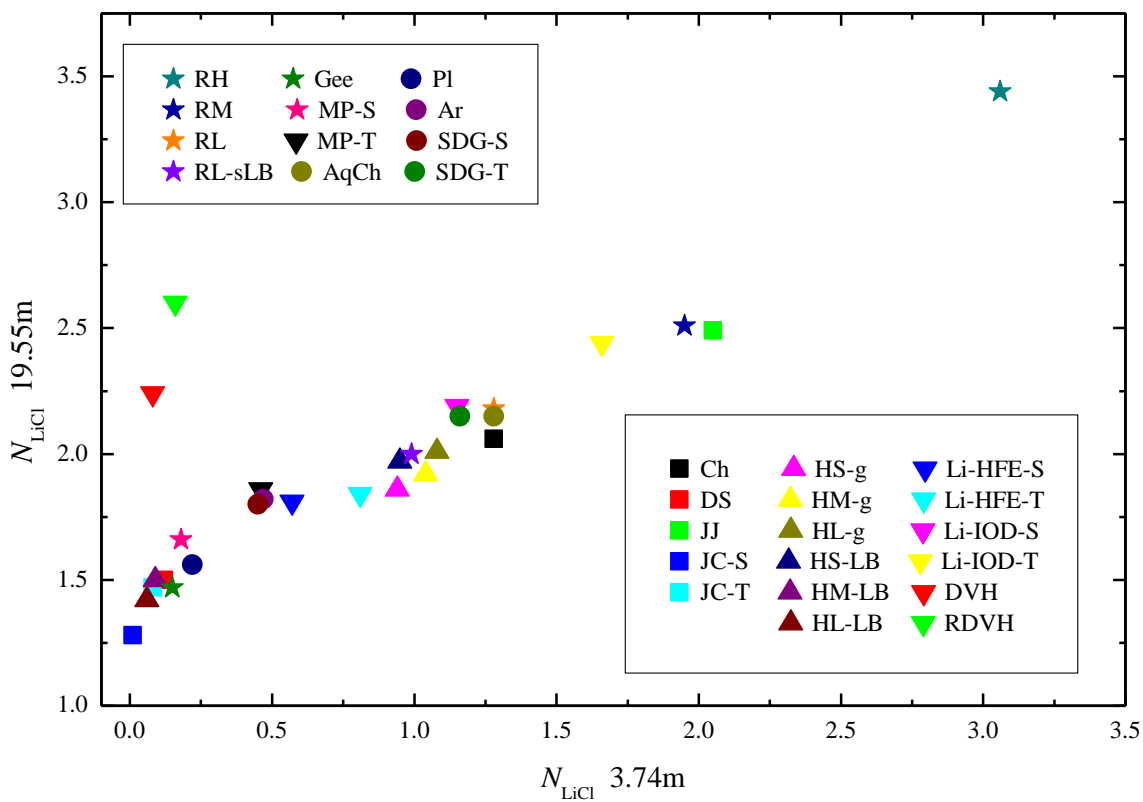

**Figure 4.** Ion pairing tendency of the investigated FFs: $N_{\text{LiCl}}$ coordination numbers of the 3.74m and 19.55m samples.

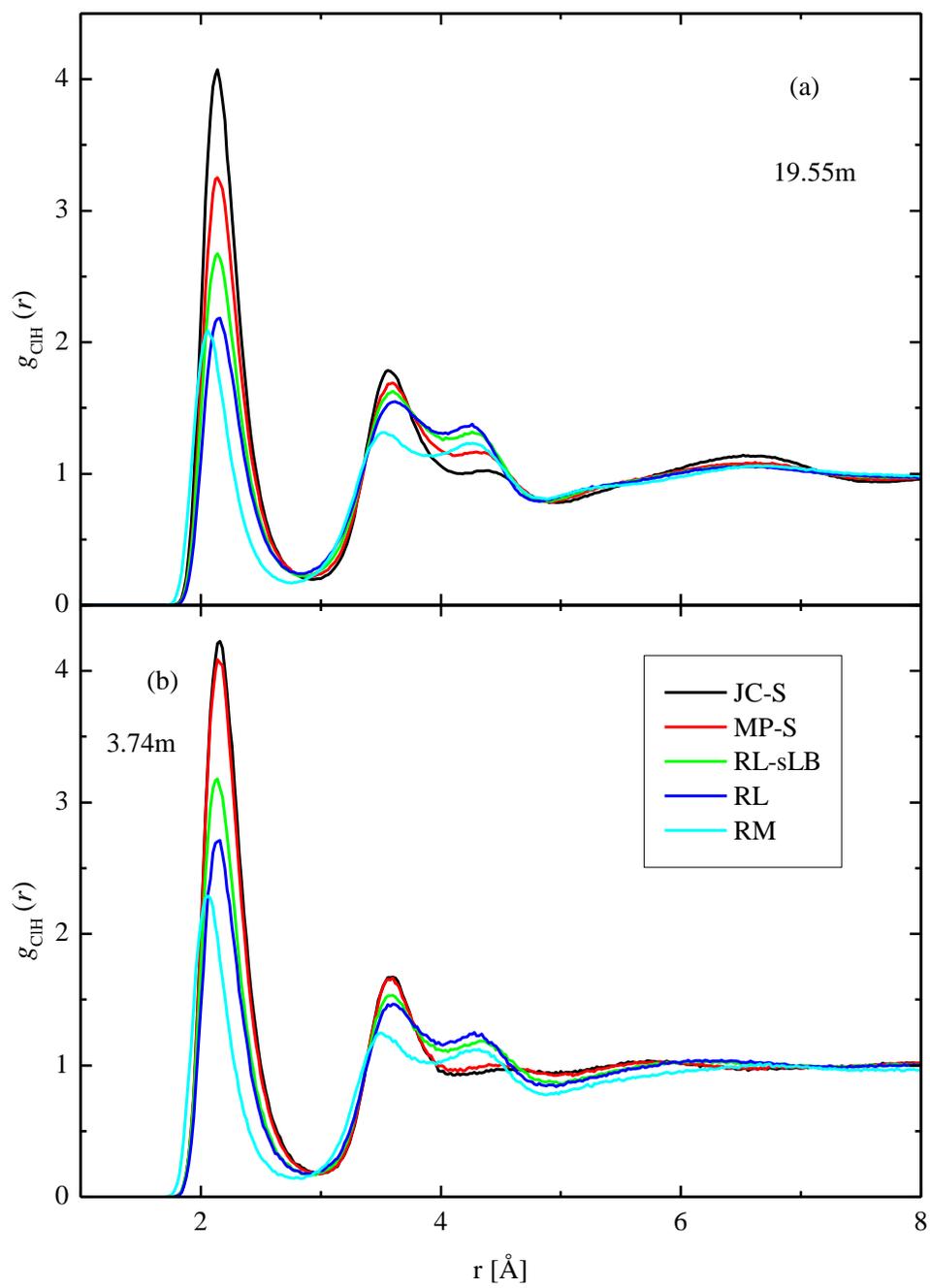

**Figure 5.** Cl-H partial pair correlation functions as computed from MD simulations with different interatomic potential models (black) JC-S, (red) MP-S, (light green) RL-sLB, (blue) RL, (cyan) RM. The curves are shown for (a) the 19.55m and (b) the 3.74m samples.

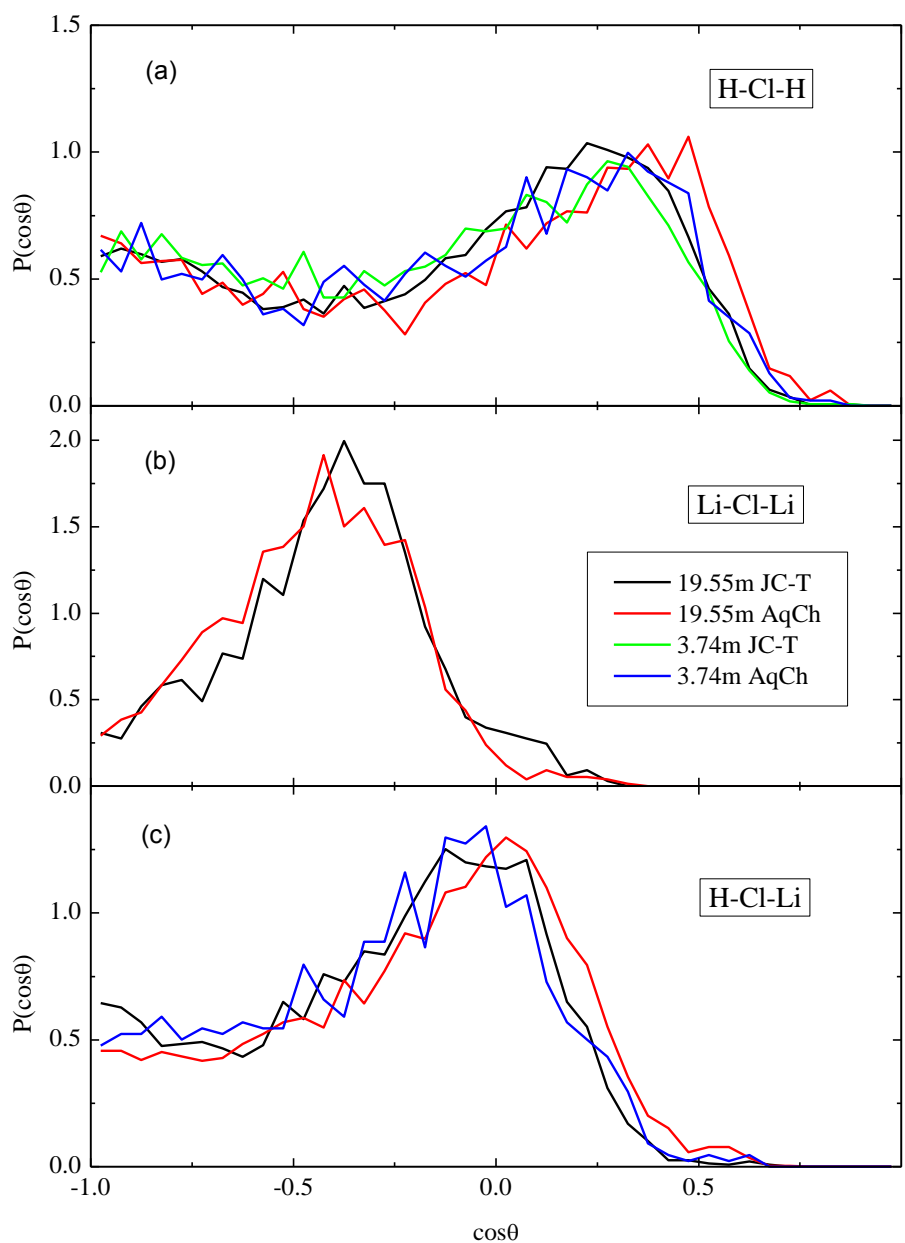

**Figure 6**. First coordination shell of Cl$^-$ ions: (a) H-Cl-H, (b) Li-Cl-Li, and (c) H-Cl-Li cosine distributions obtained from MD simulations (the RMC simulations led to similar curves), for the 19.55m (black and red curves) and 3.74m (green and blue curves) samples. The distributions are shown for two FFs with different ion pairing tendency (see text): JC-T (low IPT, black and green) and AqCh (high IPT, red and blue). Some triplets are missing from the 3.74m sample.

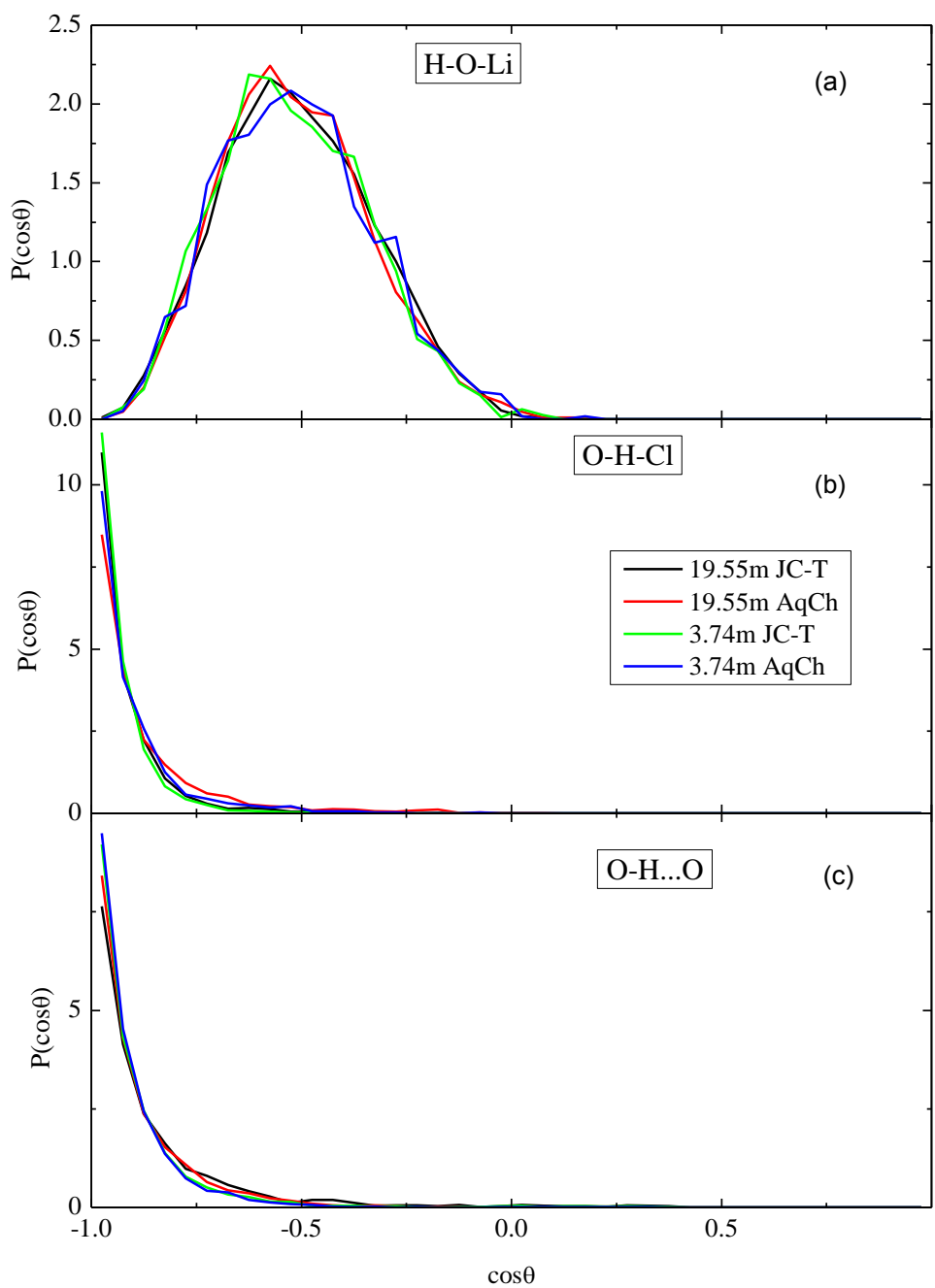

**Figure 7.** Cosine distributions of the (a) H-O-Li, (b) O-H-Cl, and (c) O-H...O angles, obtained from MD simulations (similar but less sharp curves were obtained from RMC simulations).

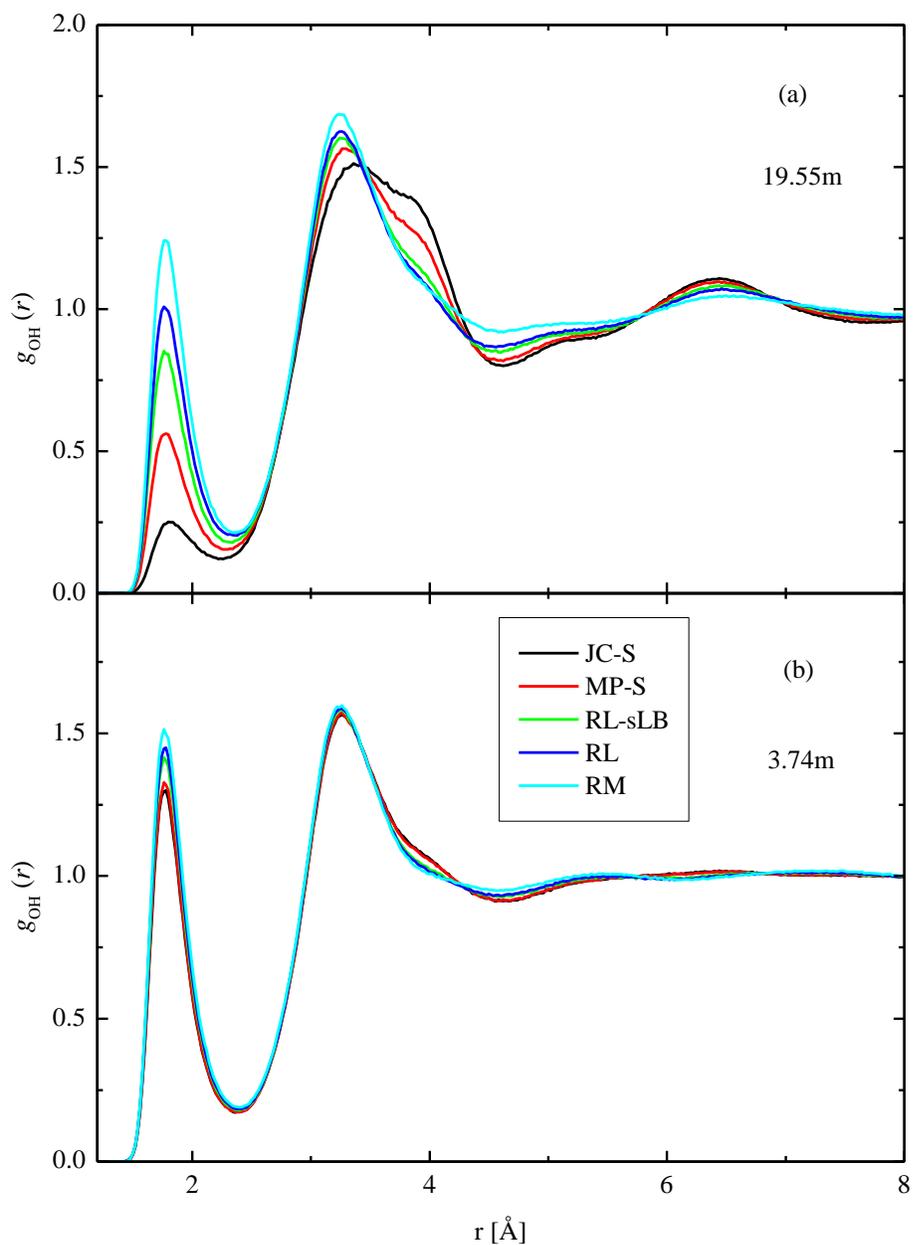

**Figure 8**. O-H partial pair correlation functions as derived from MD simulations with different interatomic potential models (black) JC-S, (red) MP-S, (light green) RL-sLB, (blue) RL, (cyan) RM. The curves are shown for (a) the 19.55m and (b) the 3.74m samples. (Only the intermolecular parts of the curves are shown.)

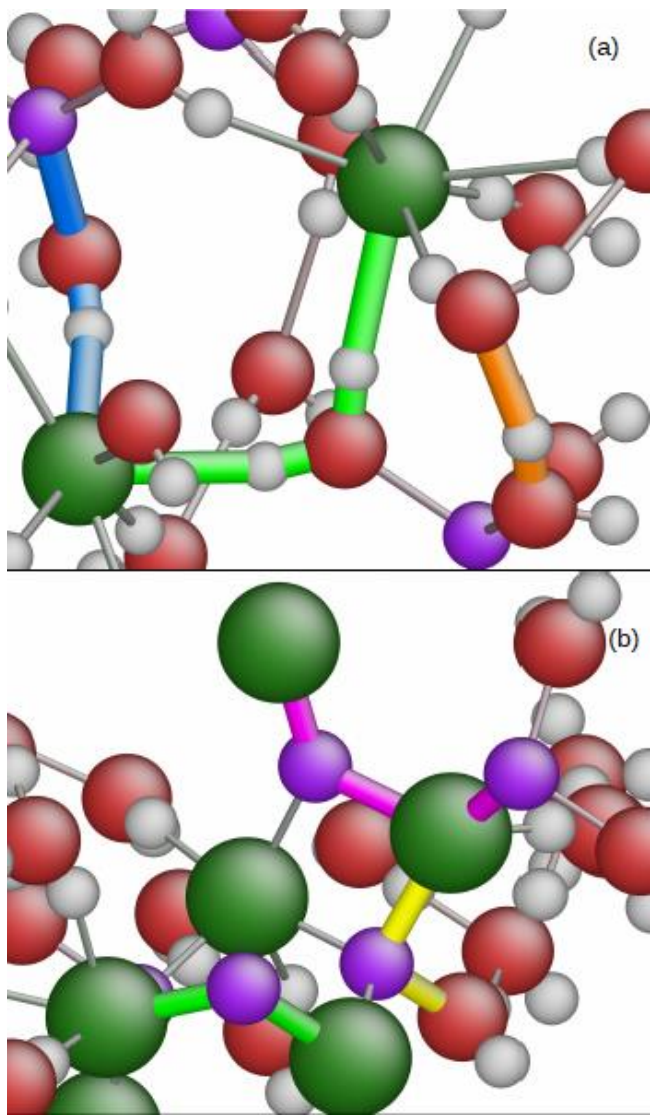

**Figure 9**. Selected part of the configuration of (a) the 3.74m sample obtained with the JC-S model and (b) the 19.55m sample obtained with the RM model. $Li^+$, $Cl^-$, O and H particles are represented by purple, dark green, red and grey balls, respectively. Nearest neighbor particles are connected with sticks. Some structural motifs are marked by colored connecting sticks: (a) green: $Cl^-$-H-O-H-$Cl^-$, blue: $Li^+$-O-H-$Cl^-$, orange O...H-O and (b) green: $Cl^-$-$Li^+$-$Cl^-$, magenta: $Li^+$-$Cl^-$-$Li^+$-$Cl^-$, yellow: $Cl^-$-$Li^+$O. (Atomic configurations are visualized by the AtomEye program[97].)

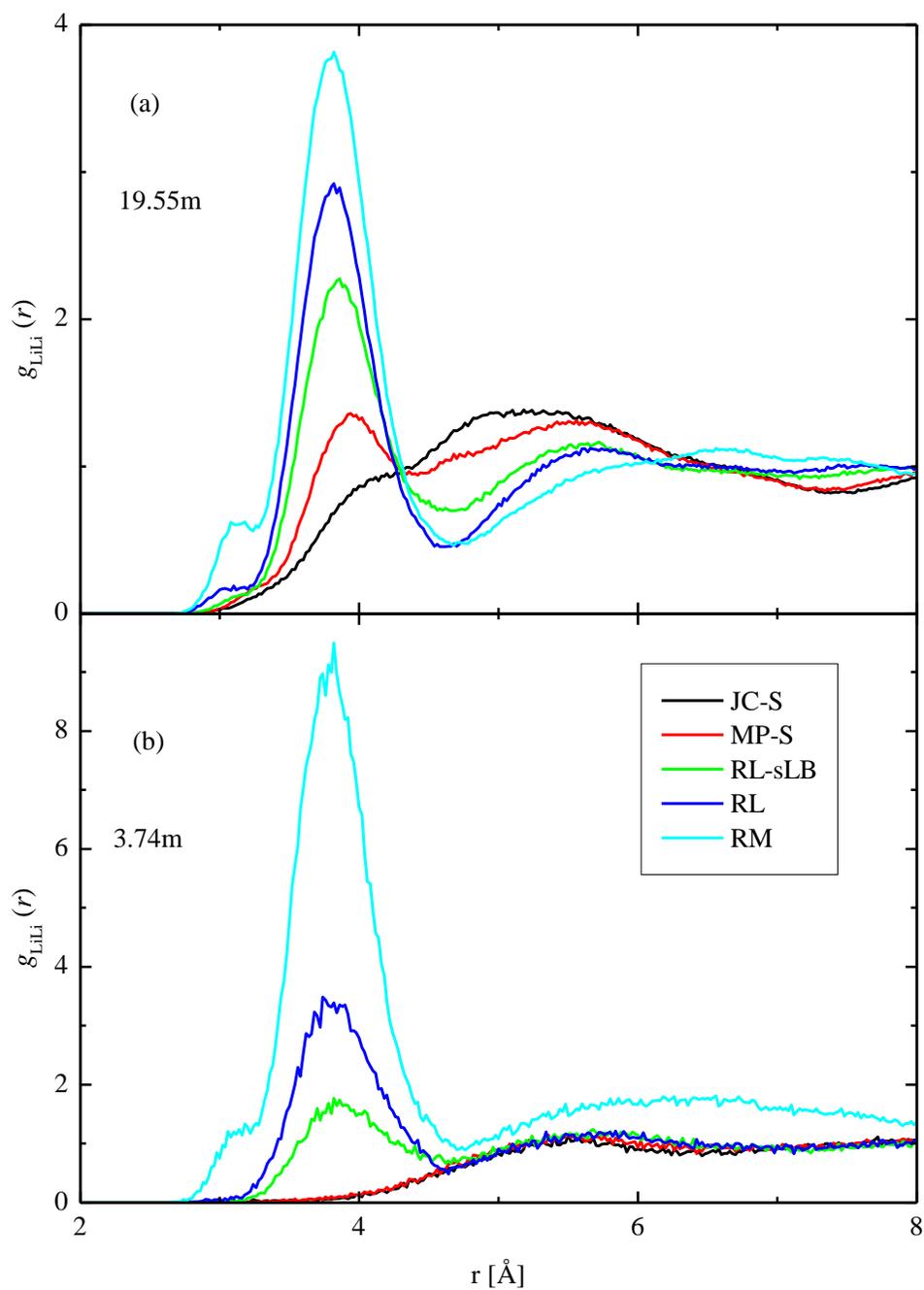

**Figure 10.** Li-Li partial pair correlation functions as derived from MD simulations with different interatomic potential models (black) JC-S, (red) MP-S, (light green) RL-sLB, (blue) RL, (cyan) RM. The curves are shown for (a) the 19.55m and (b) the 3.74m samples.

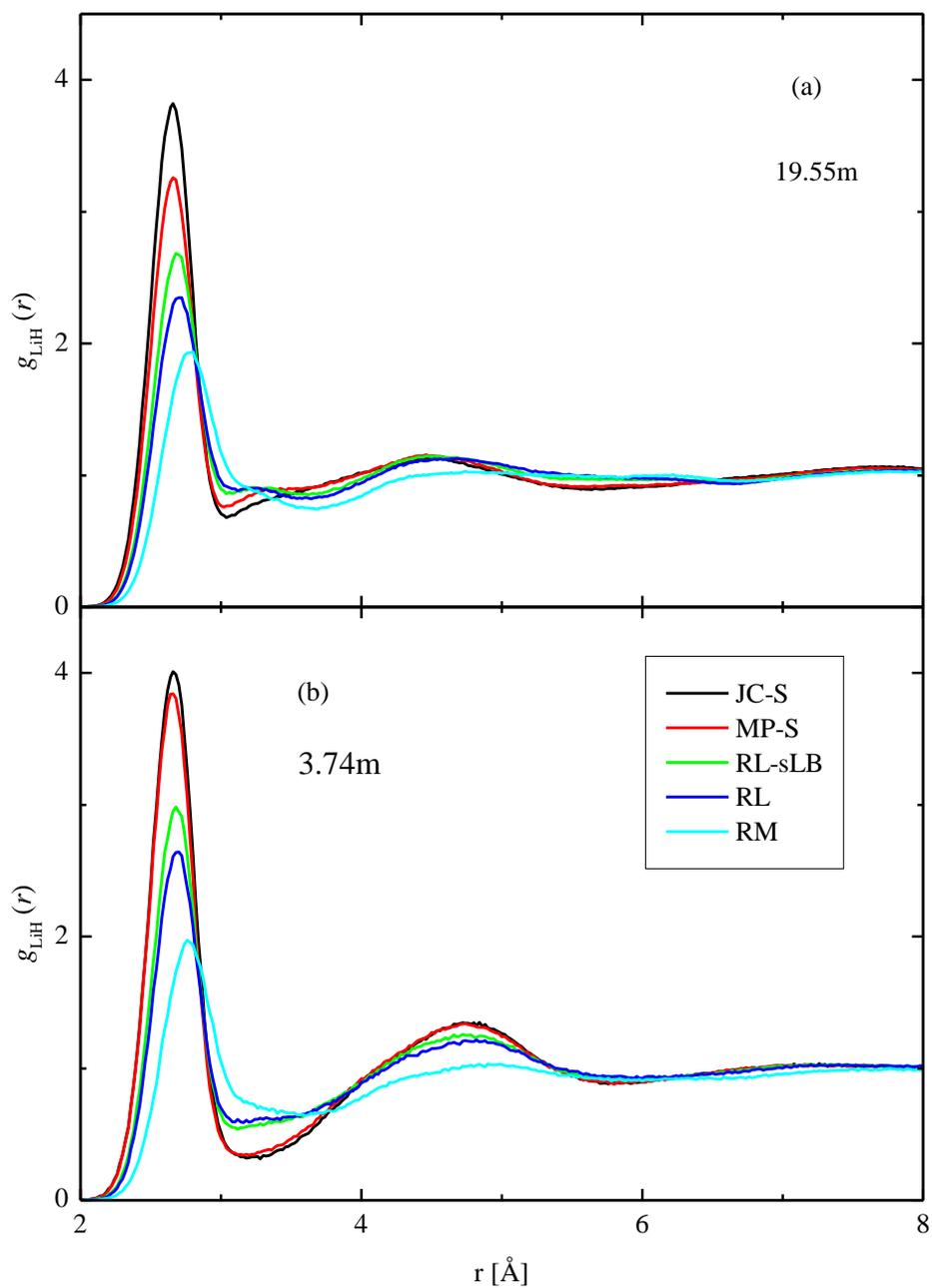

**Figure 11.** Li-H partial pair correlation functions as calculated from MD simulations with different interatomic potential models (black) JC-S, (red) MP-S, (light green) RL-sLB, (blue) RL, (cyan) RM. The curves are shown for (a) the 19.55m and (b) the 3.74m samples.

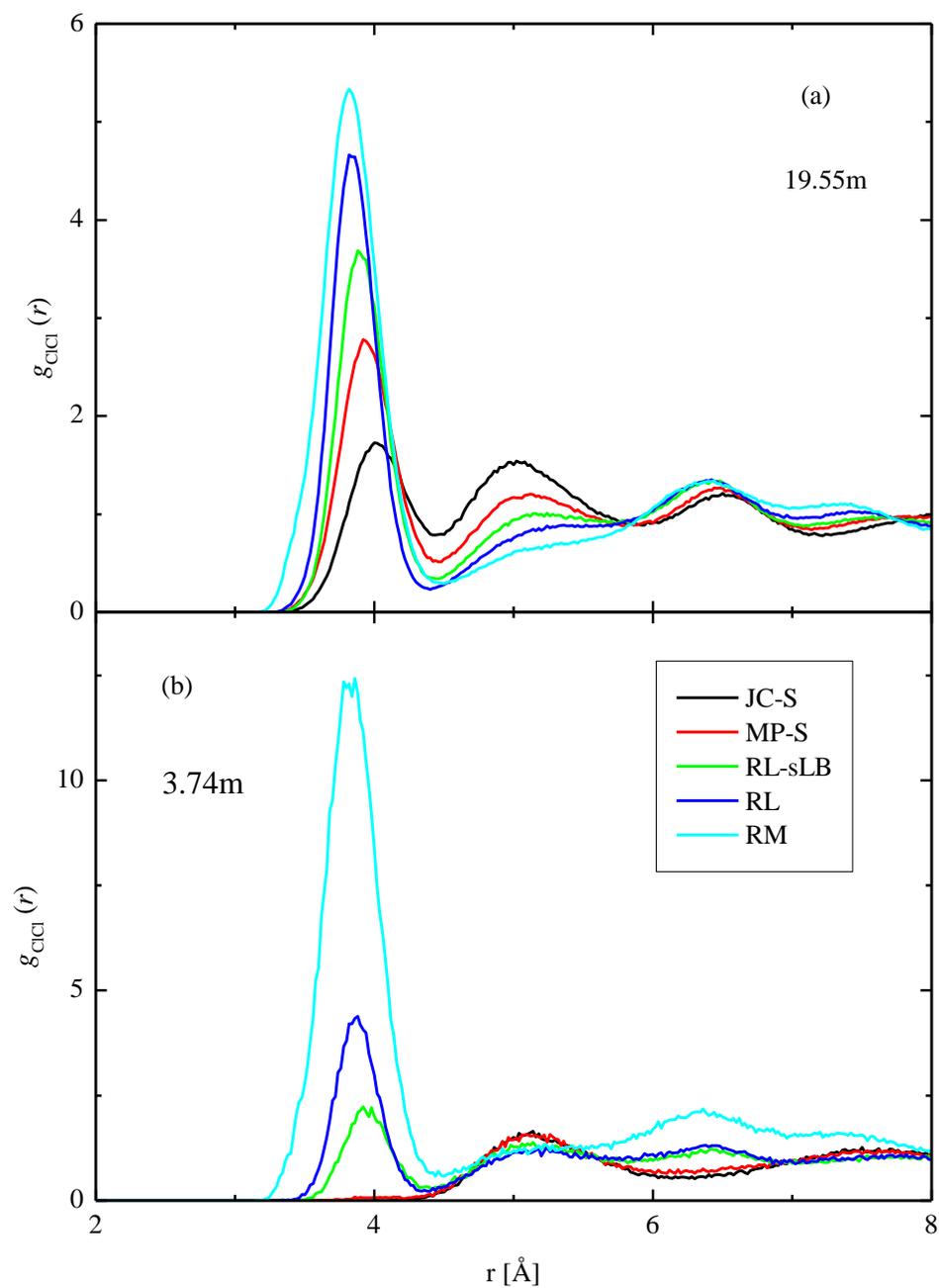

**Figure 12.** Cl-Cl partial pair correlation functions as computed from MD simulations with different interatomic potential models (black) JC-S, (red) MP-S, (light green) RL-sLB, (blue) RL, (cyan) RM. The curves are shown for (a) the 19.55m and (b) the 3.74m samples.

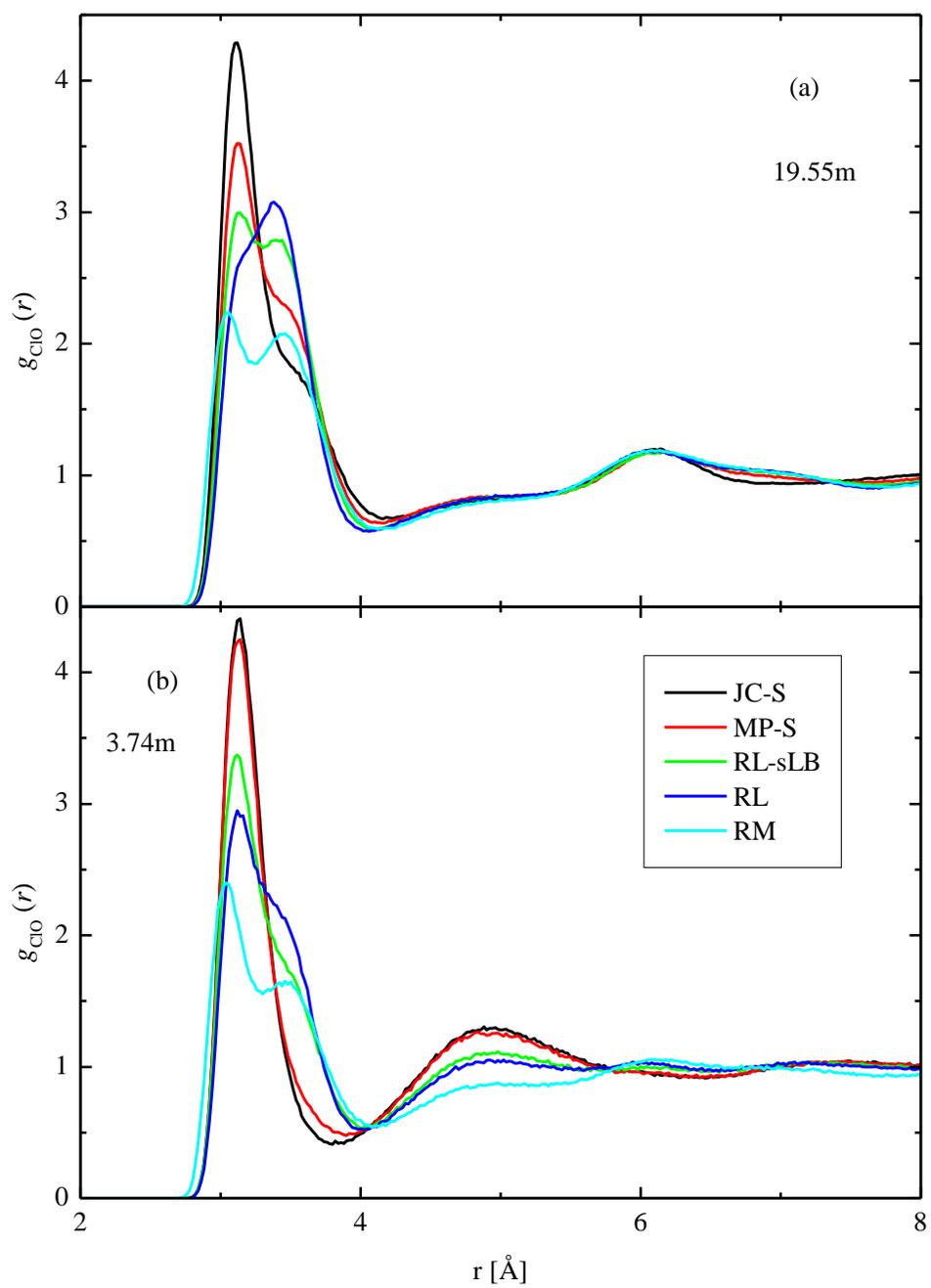

**Figure 13.** Cl-O partial pair correlation functions as derived from MD simulations with different interatomic potential models (black) JC-S, (red) MP-S, (light green) RL-sLB, (blue) RL, (cyan) RM. The curves are shown for (a) the 19.55m and (b) the 3.74m samples.

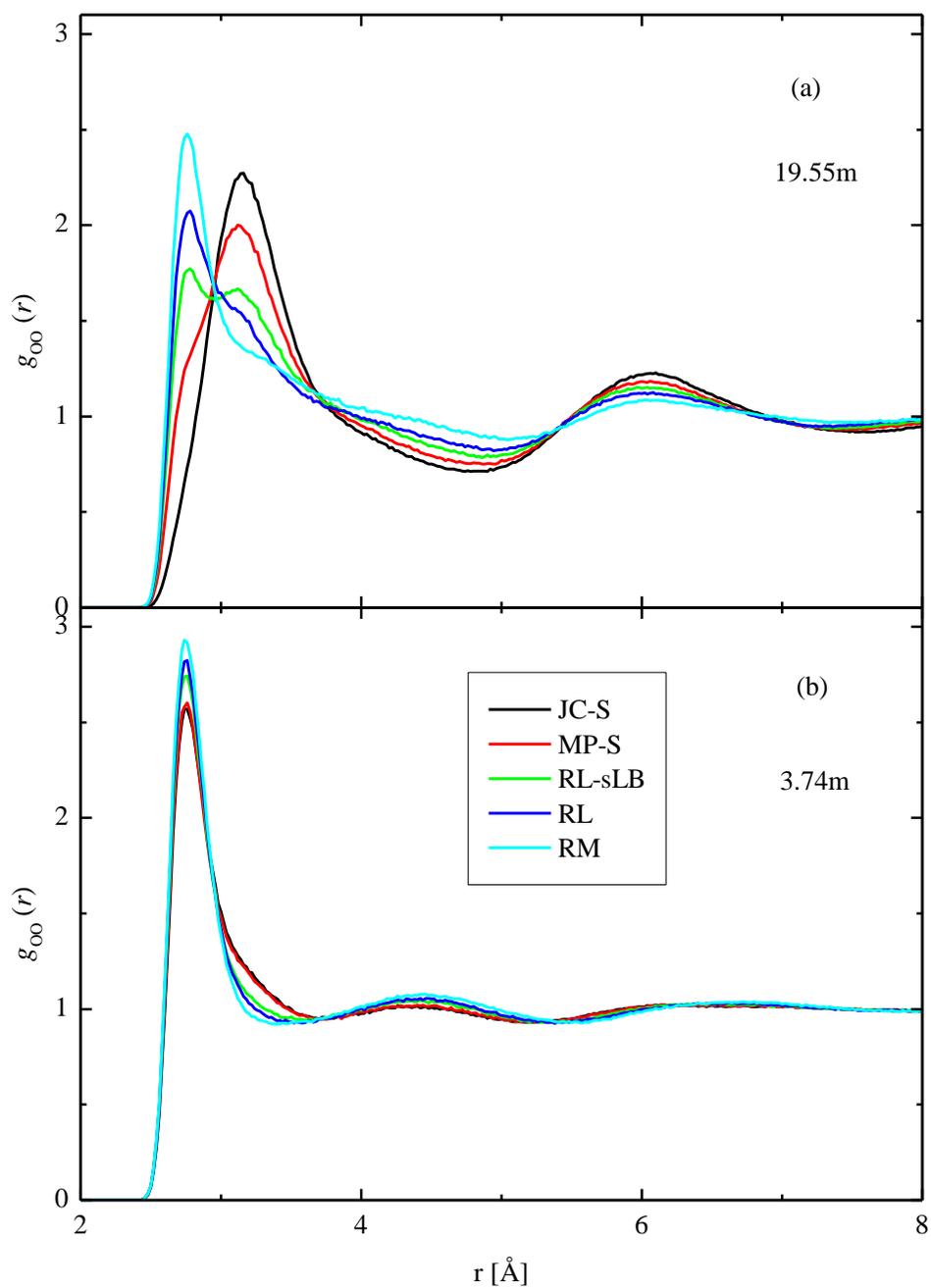

**Figure 14**. O-O partial pair correlation functions as calculated from MD simulations with different interatomic potential models (black) JC-S, (red) MP-S, (light green) RL-sLB, (blue) RL, (cyan) RM. The curves are shown for (a) the 19.55m and (b) the 3.74m samples.

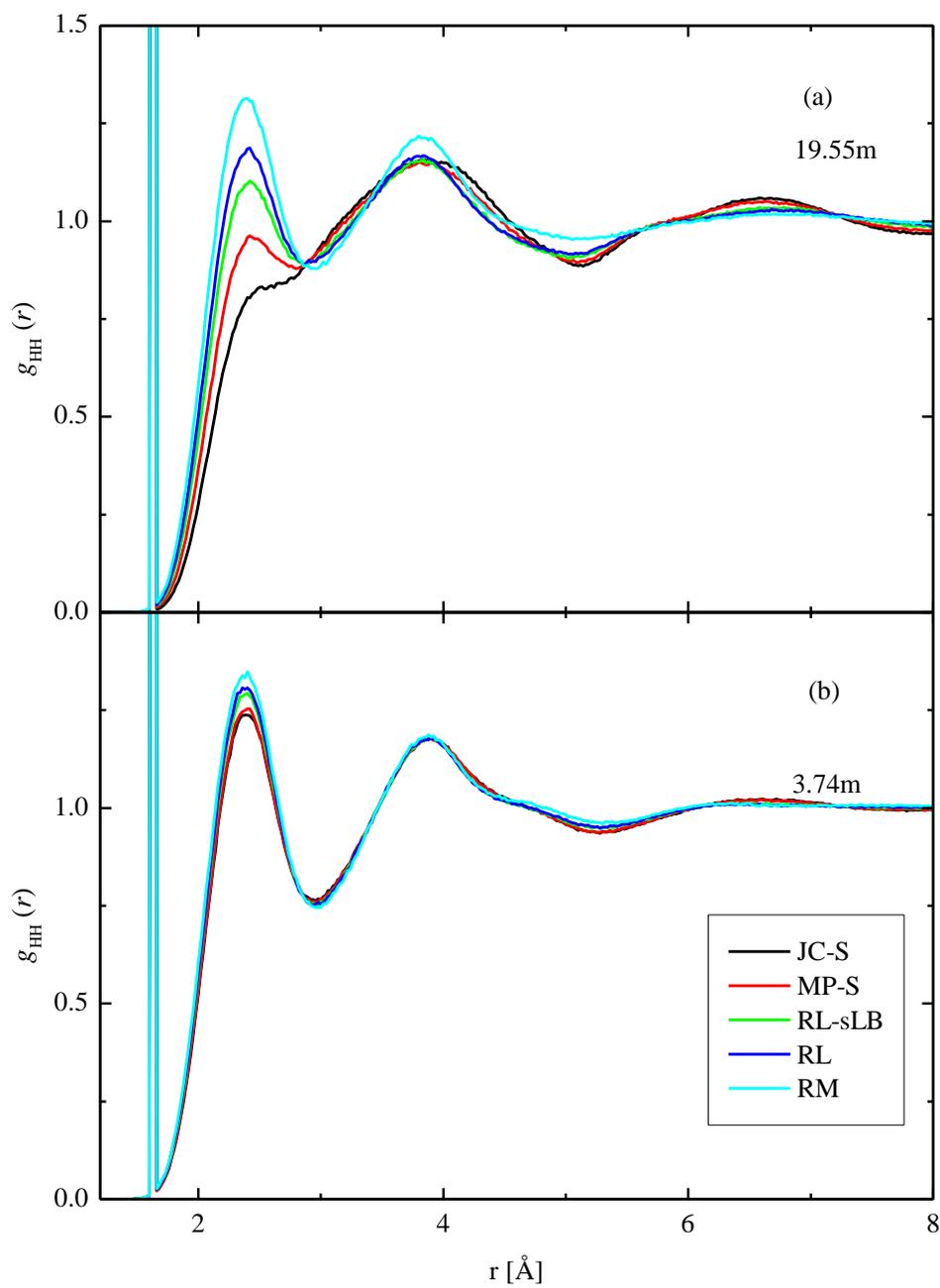

**Figure 15**. H-H partial pair correlation functions as computed from MD simulations with different interatomic potential models (black) JC-S, (red) MP-S, (light green) RL-sLB, (blue) RL, (cyan) RM. The curves are shown for (a) the 19.55m and (b) the 3.74m samples.

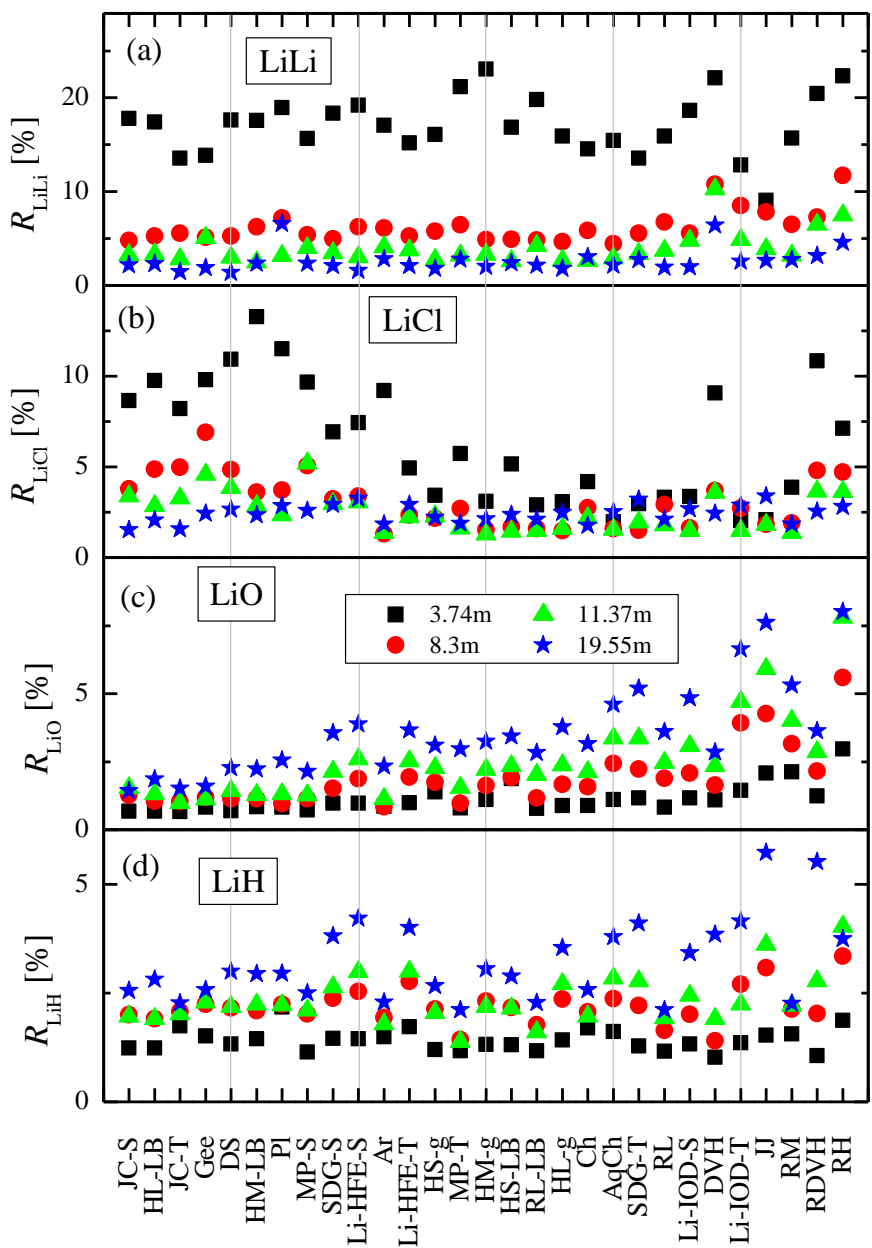

**Figure 16.** *R*-factors of (a) Li-Li PPCFs, (b) Li-Cl PPCFs, (c) Li-O PPCFs and (d) Li-H PPCFs for the different models at four concentrations, as calculated for the final stages of the RMC refinements. (Black squares 3.74m, red circles 8.3m, green triangles 11.37m, blue stars 19.55m samples.) The FFs are sorted according to their IPT (see text).

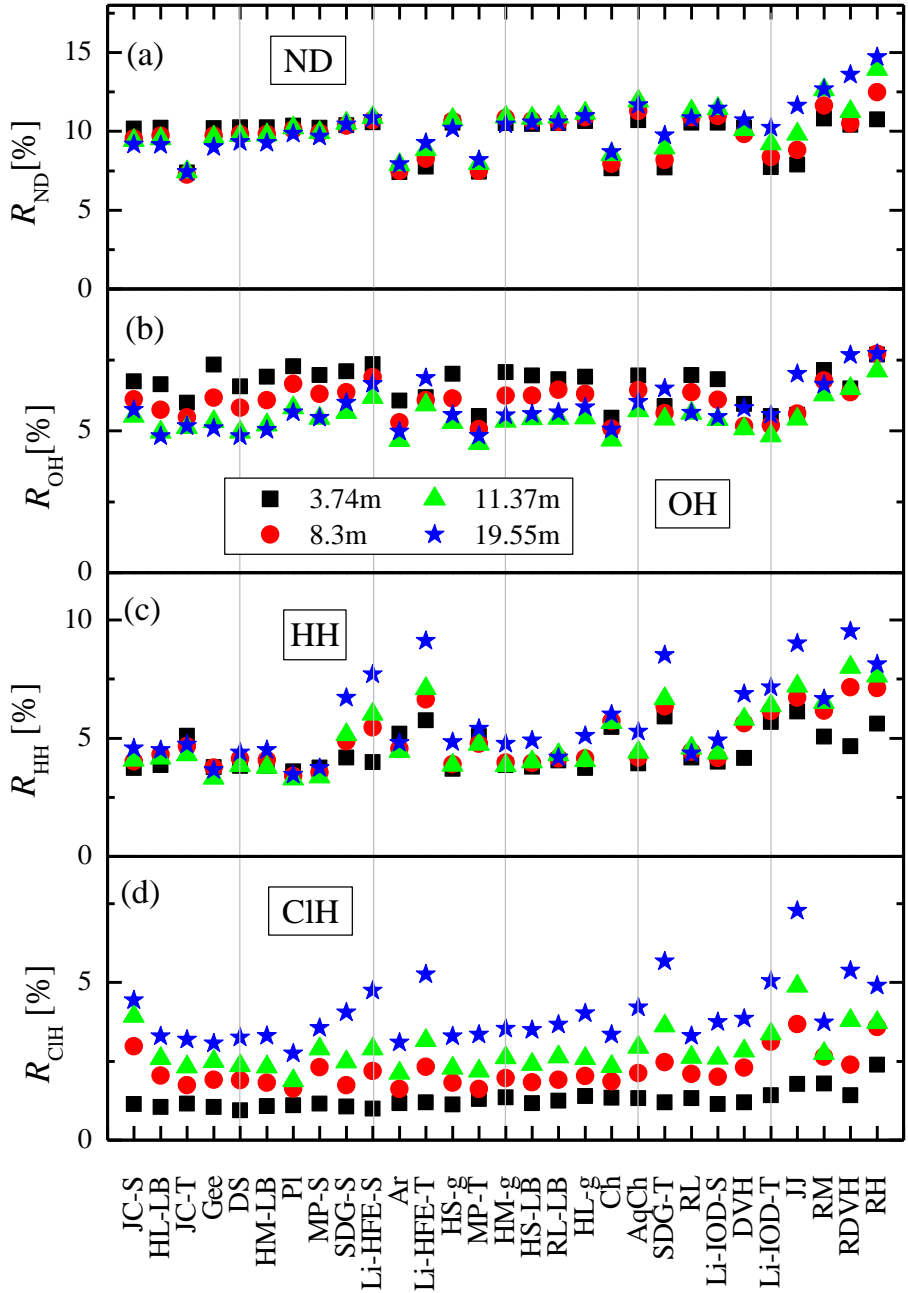

**Figure 17.** *R*-factors of (a) the neutron structure factors, (b) O-H PPCFs, (c) H-H PPCFs and (d) Cl-H PPCFs for the different models at four concentrations, as calculated for the final stages of the RMC refinements. (Black squares 3.74m, red circles 8.3m, green triangles 11.37m, blue stars 19.55m samples.) The FFs are sorted by their IPT (see text).

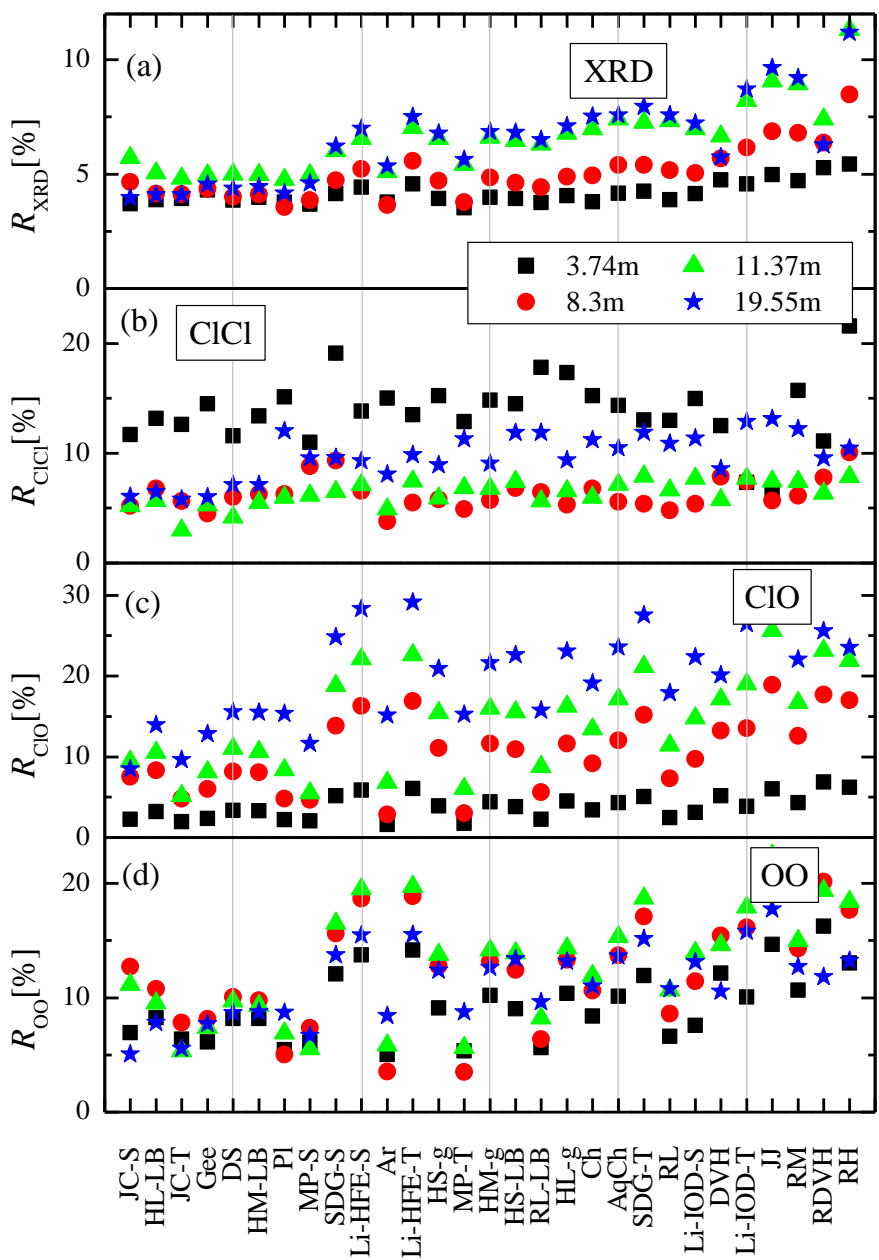

**Figure 18.** *R*-factors of (a) the X-ray structure factors, (b) Cl-Cl PPCFs, (c) Cl-O PPCFs and (d) O-O PPCFs for the different models at four concentrations, as calculated for the final stages of the RMC refinements. (Black squares 3.74m, red circles 8.3m, green triangles 11.37m, blue stars 19.55m samples.) The FFs are sorted according to their IPT (see text).

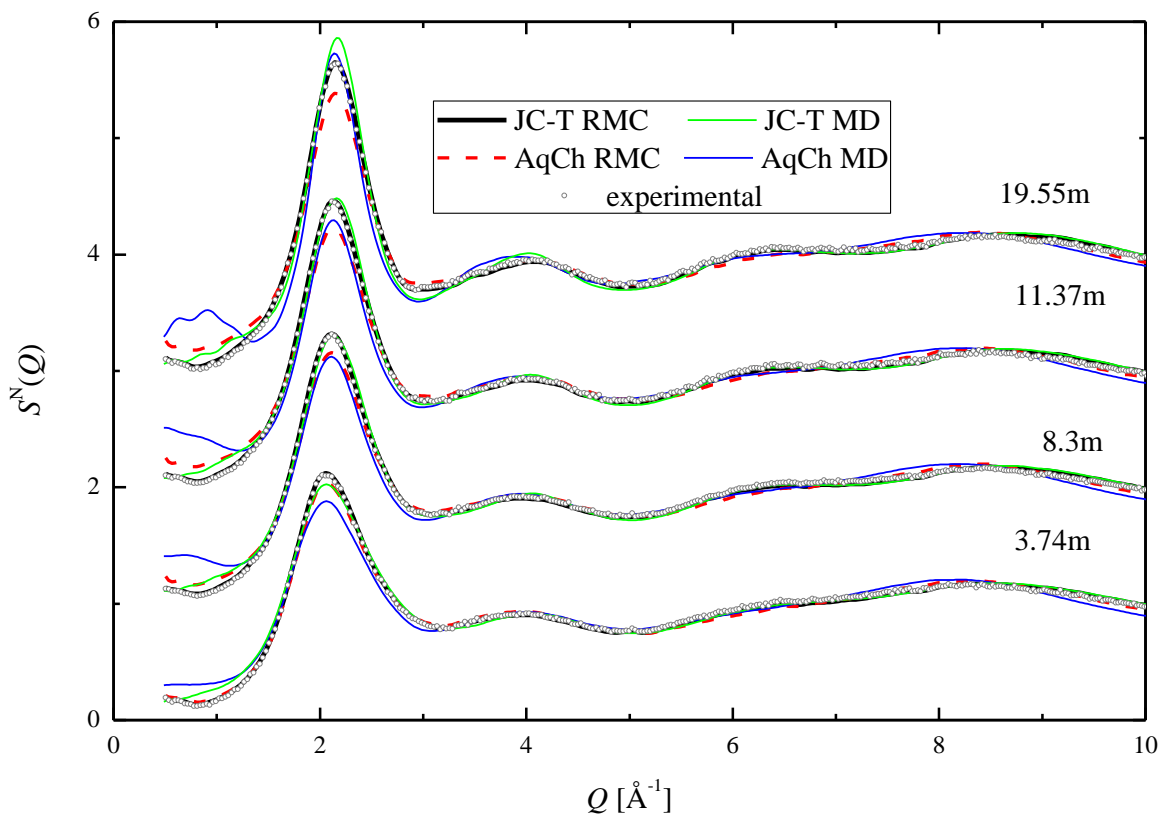

**Figure 19.** Neutron weighted total structure factors from MD simulations (green and blue lines) and after RMC refinement (black and red lines), compared to the experimental curves (symbols) (from Ref. [11]). The qualities of the fits are shown for two models (JC-T (black and green) and AqCh (red and blue)) at four concentrations. (The curves belonging to different concentrations are shifted to improve clarity.)

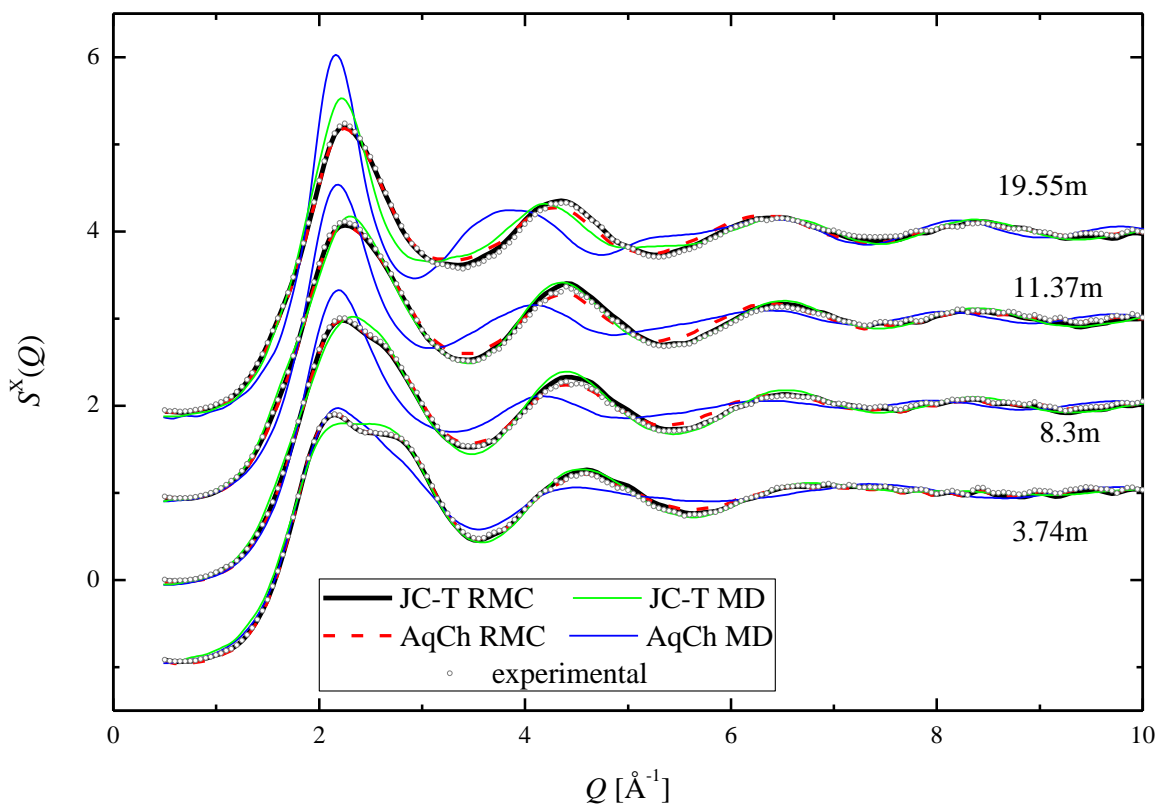

**Figure 20.** X-ray weighted total structure factors from MD simulations (green and blue lines) and after RMC refinement (black and red lines), compared to the experimental curves (symbols) (from Ref. [11]). The qualities of the fits are shown for two models (JC-T (black and green) and AqCh (red and blue)) at four concentrations. (The curves belonging to different concentrations are shifted to improve clarity.)

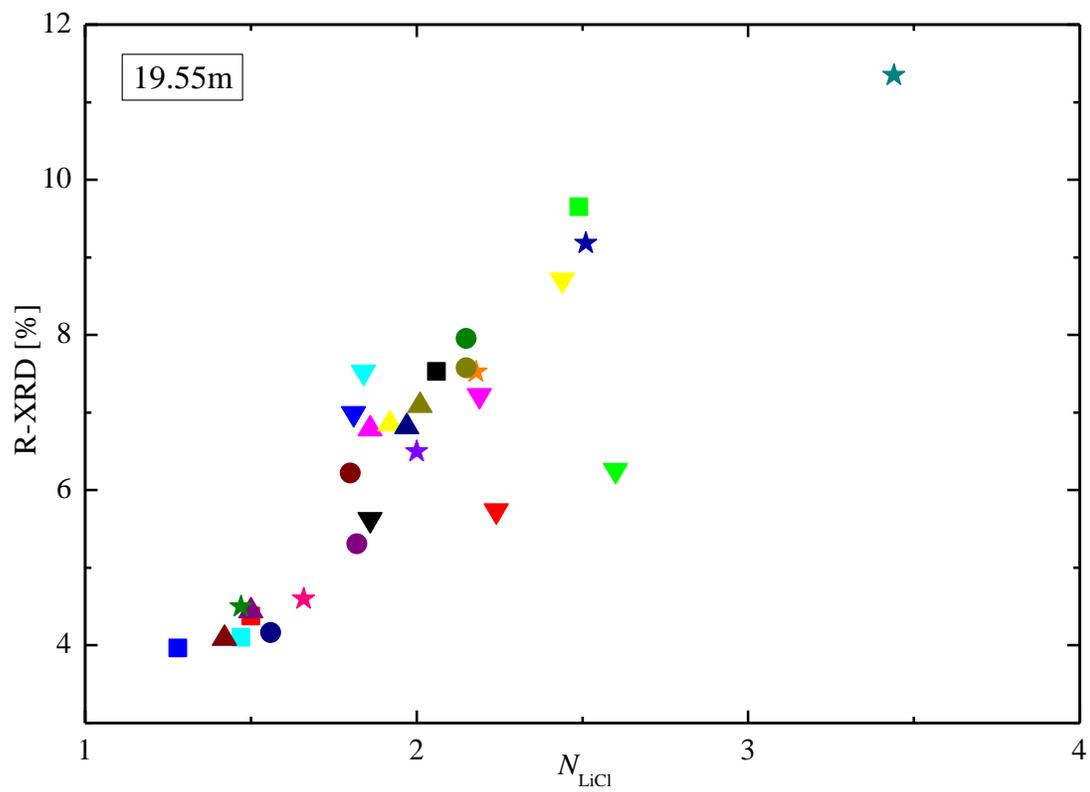

**Figure 21.** *R*-factors of the X-ray structure factor obtained from the RMC simulations as a function of the Li-Cl coordination number. The symbols are shown in the legend in Fig. 4.

Supplementary Material

# The structure of aqueous lithium chloride solutions at high concentrations as revealed by a comparison of classical interatomic potential models

Ildikó Pethes

Wigner Research Centre for Physics, Hungarian Academy of Sciences, H-1525 Budapest, POB 49, Hungary

E-mail address: pethes.ildiko@wigner.mta.hu

**Table S.1**

Positions of the first maximum and the first minimum (following the first peak) of the Li-Cl PPCF obtained from MD simulations (in Å).

|  | position of the first maximum, $r_{max,LiCl}$ [Å] | | | | position of the first minimum after the peak, $r_{min,LiCl}$ [Å] | | | |
| --- | --- | --- | --- | --- | --- | --- | --- | --- |
| forcefield | 3.74m | 8.3m | 11.3 m | 19.55m | 3.74m | 8.3m | 11.37m | 19.55m |
| Ch | 2.22 | 2.2 | 2.24 | 2.24 | 2.8 | 2.8 | 2.8 | 3 |
| DS | 2.44 | 2.44 | 2.44 | 2.44 | 2.9 | 3 | 3 | 3 |
| JJ | 2.38 | 2.4 | 2.4 | 2.42 | 3.1 | 3.1 | 3.1 | 3.2 |
| JC-S | 2.48 | 2.44 | 2.44 | 2.44 | 2.8 | 2.9 | 2.9 | 3 |
| JC-T | 2.38 | 2.38 | 2.38 | 2.36 | 2.8 | 2.9 | 2.9 | 3 |
| HS-g | 2.24 | 2.24 | 2.24 | 2.26 | 2.9 | 2.9 | 2.9 | 3 |
| HM-g | 2.26 | 2.28 | 2.3 | 2.3 | 2.9 | 2.9 | 2.9 | 3 |
| HL-g | 2.28 | 2.3 | 2.3 | 2.32 | 2.9 | 2.9 | 2.9 | 2.9 |
| HS-LB | 2.26 | 2.26 | 2.26 | 2.28 | 2.9 | 2.9 | 2.9 | 3.1 |
| HM-LB | 2.38 | 2.44 | 2.42 | 2.42 | 3 | 3 | 3 | 3 |
| HL-LB | 2.48 | 2.46 | 2.46 | 2.46 | 2.9 | 3 | 3 | 3 |
| Gee | 2.4 | 2.36 | 2.38 | 2.38 | 3 | 2.9 | 3 | 3.1 |
| RH | 2.32 | 2.32 | 2.32 | 2.32 | 3.2 | 3.3 | 3.3 | 3.24 |
| RM | 2.26 | 2.26 | 2.26 | 2.28 | 3 | 2.9 | 2.9 | 3 |
| RL | 2.22 | 2.24 | 2.24 | 2.26 | 2.8 | 2.9 | 2.9 | 3 |
| RL-LB | 2.28 | 2.3 | 2.3 | 2.3 | 2.8 | 2.9 | 2.9 | 3 |
| MP-S | 2.36 | 2.36 | 2.36 | 2.36 | 2.9 | 3 | 3 | 3 |
| MP-T | 2.36 | 2.36 | 2.36 | 2.36 | 3 | 2.9 | 3 | 3 |
| DVH | 2.78 | 2.72 | 2.72 | 2.76 | 3.4 | 3.6 | 3.6 | 3.8 |
| RDVH | 2.92 | 2.9 | 2.92 | 2.9 | 3.6 | 3.8 | 3.8 | 3.84 |
| Li-HFE-S | 2.42 | 2.4 | 2.42 | 2.42 | 3 | 3 | 3 | 3.06 |
| Li-HFE-T | 2.36 | 2.34 | 2.38 | 2.38 | 3 | 3 | 3 | 3.1 |
| Li-IOD-S | 2.4 | 2.38 | 2.38 | 2.38 | 3 | 3 | 3.2 | 3.08 |
| Li-IOD-T | 2.38 | 2.38 | 2.38 | 2.4 | 3 | 3.1 | 3.1 | 3.04 |
| AqCh | 2.3 | 2.32 | 2.34 | 2.32 | 2.9 | 3 | 3 | 3 |
| Pl | 2.34 | 2.34 | 2.34 | 2.36 | 3.1 | 3.2 | 3.16 | 3.1 |
| Ar | 2.34 | 2.3 | 2.32 | 2.32 | 2.8 | 2.9 | 3 | 3 |
| SDG-S | 2.46 | 2.46 | 2.46 | 2.46 | 2.9 | 3.1 | 3 | 3.1 |
| SDG-T | 2.44 | 2.46 | 2.46 | 2.46 | 3 | 3.2 | 3.2 | 3.26 |

**Table S.2**

Positions of the first maximum and the first minimum (following the first peak) of the Li-O PPCF obtained from MD simulations (in Å).

|  | position of the first maximum, $r_{max,LiO}$ [Å] | | | | position of the first minimum after the peak, $r_{min,LiO}$ [Å] | | | |
|---|---|---|---|---|---|---|---|---|
| forcefield | 3.74m | 8.3m | 11.37m | 19.55m | 3.74m | 8.3m | 11.37m | 19.55m |
| Ch | 1.94 | 1.92 | 1.94 | 1.96 | 2.5 | 2.5 | 2.5 | 2.6 |
| DS | 1.96 | 1.96 | 1.96 | 1.96 | 2.6 | 2.7 | 2.6 | 2.56 |
| JJ | 2.06 | 2.06 | 2.06 | 2.06 | 2.8 | 2.8 | 2.7 | 2.8 |
| JC-S | 1.98 | 1.96 | 1.98 | 1.96 | 2.6 | 2.6 | 2.6 | 2.5 |
| JC-T | 1.92 | 1.92 | 1.92 | 1.92 | 2.5 | 2.5 | 2.4 | 2.5 |
| HS-g | 1.94 | 1.94 | 1.94 | 1.94 | 2.46 | 2.5 | 2.5 | 2.6 |
| HM-g | 1.96 | 1.96 | 1.96 | 1.96 | 2.5 | 2.5 | 2.5 | 2.52 |
| HL-g | 1.98 | 1.98 | 1.98 | 1.98 | 2.56 | 2.6 | 2.5 | 2.52 |
| HS-LB | 1.92 | 1.94 | 1.92 | 1.94 | 2.5 | 2.5 | 2.5 | 2.5 |
| HM-LB | 1.94 | 1.94 | 1.94 | 1.94 | 2.5 | 2.6 | 2.6 | 2.5 |
| HL-LB | 1.96 | 1.96 | 1.96 | 1.96 | 2.6 | 2.6 | 2.6 | 2.5 |
| Gee | 1.88 | 1.88 | 1.88 | 1.88 | 2.4 | 2.4 | 2.4 | 2.32 |
| RH | 2.18 | 2.18 | 2.18 | 2.16 | 2.9 | 2.9 | 2.9 | 2.9 |
| RM | 2.08 | 2.06 | 2.08 | 2.08 | 2.6 | 2.6 | 2.6 | 2.64 |
| RL | 2 | 2 | 2.02 | 2 | 2.5 | 2.6 | 2.6 | 2.6 |
| RL-LB | 2 | 2 | 2 | 2 | 2.6 | 2.6 | 2.6 | 2.6 |
| MP-S | 1.96 | 1.96 | 1.96 | 1.96 | 2.5 | 2.6 | 2.6 | 2.54 |
| MP-T | 1.98 | 2 | 2 | 2 | 2.7 | 2.76 | 2.7 | 2.66 |
| DVH | 2.2 | 2.2 | 2.18 | 2.2 | 3 | 3 | 3.1 | 3.12 |
| RDVH | 2.24 | 2.26 | 2.26 | 2.24 | 3 | 3.1 | 3.1 | 3.06 |
| Li-HFE-S | 1.94 | 1.94 | 1.94 | 1.94 | 2.5 | 2.5 | 2.6 | 2.5 |
| Li-HFE-T | 1.9 | 1.9 | 1.9 | 1.9 | 2.4 | 2.6 | 2.5 | 2.44 |
| Li-IOD-S | 2.06 | 2.04 | 2.04 | 2.04 | 2.6 | 2.66 | 2.7 | 2.68 |
| Li-IOD-T | 2.08 | 2.08 | 2.06 | 2.08 | 2.8 | 2.7 | 2.8 | 2.76 |
| AqCh | 2.02 | 2.02 | 2.02 | 2.02 | 2.6 | 2.6 | 2.6 | 2.6 |
| Pl | 1.98 | 1.98 | 1.98 | 1.96 | 2.6 | 2.7 | 2.6 | 2.6 |
| Ar | 1.92 | 1.9 | 1.92 | 1.92 | 2.4 | 2.6 | 2.6 | 2.5 |
| SDG-S | 1.96 | 1.96 | 1.96 | 1.96 | 2.6 | 2.6 | 2.6 | 2.6 |
| SDG-T | 2 | 2 | 2 | 1.98 | 2.7 | 2.7 | 2.8 | 2.76 |

**Table S.3**

Positions of the first maximum and the first minimum (following the first peak) of the Cl-H PPCF obtained from MD simulations (in Å).

| forcefield | position of the first maximum, $r_{max,ClH}$ [Å] | | | | position of the first minimum after the peak, $r_{min,ClH}$ [Å] | | | |
|---|---|---|---|---|---|---|---|---|
| | 3.74m | 8.3m | 11.37m | 19.5 m | 3.74m | 8.3m | 11.37m | 19.55m |
| Ch | 2.28 | 2.26 | 2.24 | 2.24 | 2.9 | 2.9 | 2.9 | 2.9 |
| DS | 2.22 | 2.26 | 2.22 | 2.24 | 3 | 3 | 3 | 2.94 |
| JJ | 2.34 | 2.32 | 2.32 | 2.34 | 2.96 | 2.9 | 2.9 | 2.9 |
| JC-S | 2.16 | 2.12 | 2.14 | 2.14 | 2.96 | 3 | 3 | 2.92 |
| JC-T | 2.22 | 2.2 | 2.2 | 2.2 | 2.96 | 2.94 | 3 | 2.9 |
| HS-g | 2.26 | 2.24 | 2.24 | 2.26 | 3 | 2.96 | 2.94 | 2.94 |
| HM-g | 2.26 | 2.24 | 2.24 | 2.26 | 3.04 | 2.96 | 2.94 | 2.92 |
| HL-g | 2.24 | 2.26 | 2.26 | 2.26 | 3 | 2.96 | 2.96 | 2.92 |
| HS-LB | 2.26 | 2.26 | 2.26 | 2.24 | 3 | 2.96 | 2.94 | 2.92 |
| HM-LB | 2.24 | 2.24 | 2.24 | 2.22 | 3.04 | 3 | 3 | 2.9 |
| HL-LB | 2.22 | 2.22 | 2.22 | 2.22 | 3.04 | 3.04 | 3 | 2.94 |
| Gee | 2.22 | 2.22 | 2.22 | 2.2 | 3 | 3 | 3 | 2.94 |
| RH | 1.98 | 2 | 1.96 | 1.98 | 2.62 | 2.64 | 2.6 | 2.6 |
| RM | 2.08 | 2.06 | 2.06 | 2.06 | 2.8 | 2.74 | 2.74 | 2.74 |
| RL | 2.18 | 2.14 | 2.16 | 2.14 | 2.9 | 2.84 | 2.8 | 2.82 |
| RL-LB | 2.12 | 2.14 | 2.14 | 2.12 | 2.9 | 2.88 | 2.9 | 2.86 |
| MP-S | 2.16 | 2.14 | 2.14 | 2.14 | 3 | 2.94 | 2.94 | 2.9 |
| MP-T | 2.18 | 2.18 | 2.18 | 2.18 | 2.9 | 2.92 | 2.9 | 2.86 |
| DVH | 2.32 | 2.32 | 2.36 | 2.34 | 3.1 | 3.1 | 3.1 | 2.96 |
| RDVH | 2.44 | 2.44 | 2.48 | 2.42 | 3.2 | 3.16 | 3.1 | 2.98 |
| Li-HFE-S | 2.34 | 2.38 | 2.36 | 2.36 | 3.1 | 3.08 | 3 | 3 |
| Li-HFE-T | 2.44 | 2.4 | 2.4 | 2.4 | 3.1 | 3.04 | 3.04 | 2.98 |
| Li-IOD-S | 2.22 | 2.2 | 2.2 | 2.2 | 2.9 | 2.94 | 2.9 | 2.9 |
| Li-IOD-T | 2.26 | 2.24 | 2.26 | 2.24 | 2.9 | 2.86 | 2.86 | 2.84 |
| AqCh | 2.28 | 2.24 | 2.22 | 2.24 | 2.94 | 2.96 | 2.94 | 2.9 |
| Pl | 2.2 | 2.2 | 2.18 | 2.22 | 2.9 | 2.9 | 2.9 | 2.86 |
| Ar | 2.22 | 2.2 | 2.22 | 2.2 | 2.9 | 2.92 | 2.9 | 2.84 |
| SDG-S | 2.28 | 2.32 | 2.3 | 2.3 | 3.04 | 3.04 | 3.06 | 2.96 |
| SDG-T | 2.36 | 2.34 | 2.36 | 2.36 | 3 | 3 | 3 | 2.96 |

**Table S.4**

Positions of the first intermolecular maximum and the first minimum (following the first peak) of the O-H PPCF obtained from MD simulations (in Å).

| forcefield | position of the first intermolecular maximum, $r_{max,OH}$ [Å] | | | | position of the first minimum after the peak, $r_{min,OH}$ [Å] | | | |
|---|---|---|---|---|---|---|---|---|
| | 3.74m | 8.3m | 11.37m | 19.55m | 3.74m | 8.3m | 11.37m | 19.55m |
| Ch | 1.82 | 1.82 | 1.81 | 1.82 | 2.4 | 2.4 | 2.4 | 2.4 |
| DS | 1.77 | 1.76 | 1.76 | 1.76 | 2.4 | 2.3 | 2.3 | 2.26 |
| JJ | 1.82 | 1.82 | 1.82 | 1.81 | 2.38 | 2.4 | 2.36 | 2.36 |
| JC-S | 1.77 | 1.76 | 1.76 | 1.8 | 2.36 | 2.34 | 2.3 | 2.24 |
| JC-T | 1.82 | 1.82 | 1.82 | 1.84 | 2.4 | 2.4 | 2.4 | 2.34 |
| HS-g | 1.77 | 1.76 | 1.76 | 1.76 | 2.4 | 2.34 | 2.34 | 2.3 |
| HM-g | 1.77 | 1.76 | 1.76 | 1.76 | 2.36 | 2.3 | 2.3 | 2.3 |
| HL-g | 1.77 | 1.76 | 1.76 | 1.76 | 2.4 | 2.36 | 2.34 | 2.3 |
| HS-LB | 1.77 | 1.76 | 1.76 | 1.76 | 2.4 | 2.36 | 2.3 | 2.3 |
| HM-LB | 1.77 | 1.76 | 1.76 | 1.76 | 2.4 | 2.3 | 2.3 | 2.22 |
| HL-LB | 1.77 | 1.76 | 1.76 | 1.77 | 2.34 | 2.32 | 2.3 | 2.22 |
| Gee | 1.77 | 1.76 | 1.76 | 1.78 | 2.4 | 2.36 | 2.3 | 2.3 |
| RH | 1.77 | 1.76 | 1.76 | 1.78 | 2.4 | 2.4 | 2.4 | 2.38 |
| RM | 1.77 | 1.76 | 1.76 | 1.77 | 2.4 | 2.38 | 2.36 | 2.36 |
| RL | 1.77 | 1.76 | 1.76 | 1.76 | 2.36 | 2.38 | 2.34 | 2.28 |
| RL-LB | 1.77 | 1.76 | 1.76 | 1.76 | 2.4 | 2.36 | 2.3 | 2.3 |
| MP-S | 1.77 | 1.76 | 1.76 | 1.78 | 2.4 | 2.34 | 2.3 | 2.26 |
| MP-T | 1.82 | 1.82 | 1.82 | 1.83 | 2.4 | 2.38 | 2.36 | 2.34 |
| DVH | 1.77 | 1.76 | 1.78 | 1.79 | 2.34 | 2.32 | 2.3 | 2.24 |
| RDVH | 1.77 | 1.76 | 1.77 | 1.77 | 2.34 | 2.34 | 2.3 | 2.22 |
| Li-HFE-S | 1.77 | 1.76 | 1.75 | 1.76 | 2.36 | 2.34 | 2.3 | 2.22 |
| Li-HFE-T | 1.82 | 1.82 | 1.82 | 1.81 | 2.4 | 2.38 | 2.36 | 2.36 |
| Li-IOD-S | 1.77 | 1.76 | 1.76 | 1.76 | 2.36 | 2.34 | 2.3 | 2.34 |
| Li-IOD-T | 1.82 | 1.82 | 1.82 | 1.82 | 2.4 | 2.4 | 2.4 | 2.36 |
| AqCh | 1.77 | 1.76 | 1.76 | 1.76 | 2.36 | 2.36 | 2.36 | 2.3 |
| Pl | 1.77 | 1.76 | 1.77 | 1.78 | 2.36 | 2.36 | 2.36 | 2.32 |
| Ar | 1.82 | 1.82 | 1.83 | 1.82 | 2.4 | 2.4 | 2.4 | 2.4 |
| SDG-S | 1.77 | 1.76 | 1.76 | 1.76 | 2.4 | 2.34 | 2.3 | 2.26 |
| SDG-T | 1.82 | 1.82 | 1.82 | 1.82 | 2.4 | 2.36 | 2.36 | 2.32 |

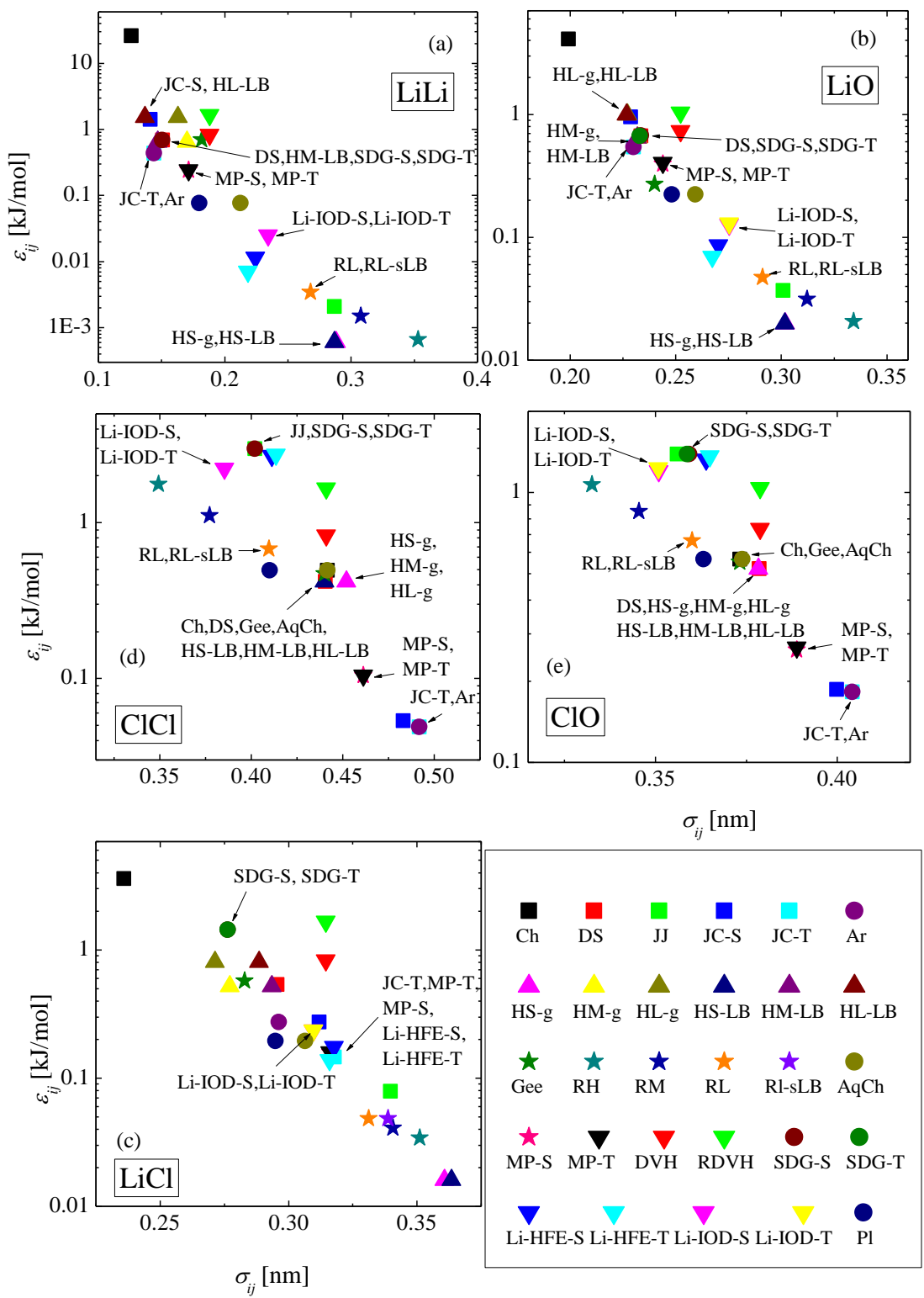

**Figure S.1.** Graphical representation of the LJ ($\varepsilon_{ij}$ and $\sigma_{ij}$) parameters of all investigated models.

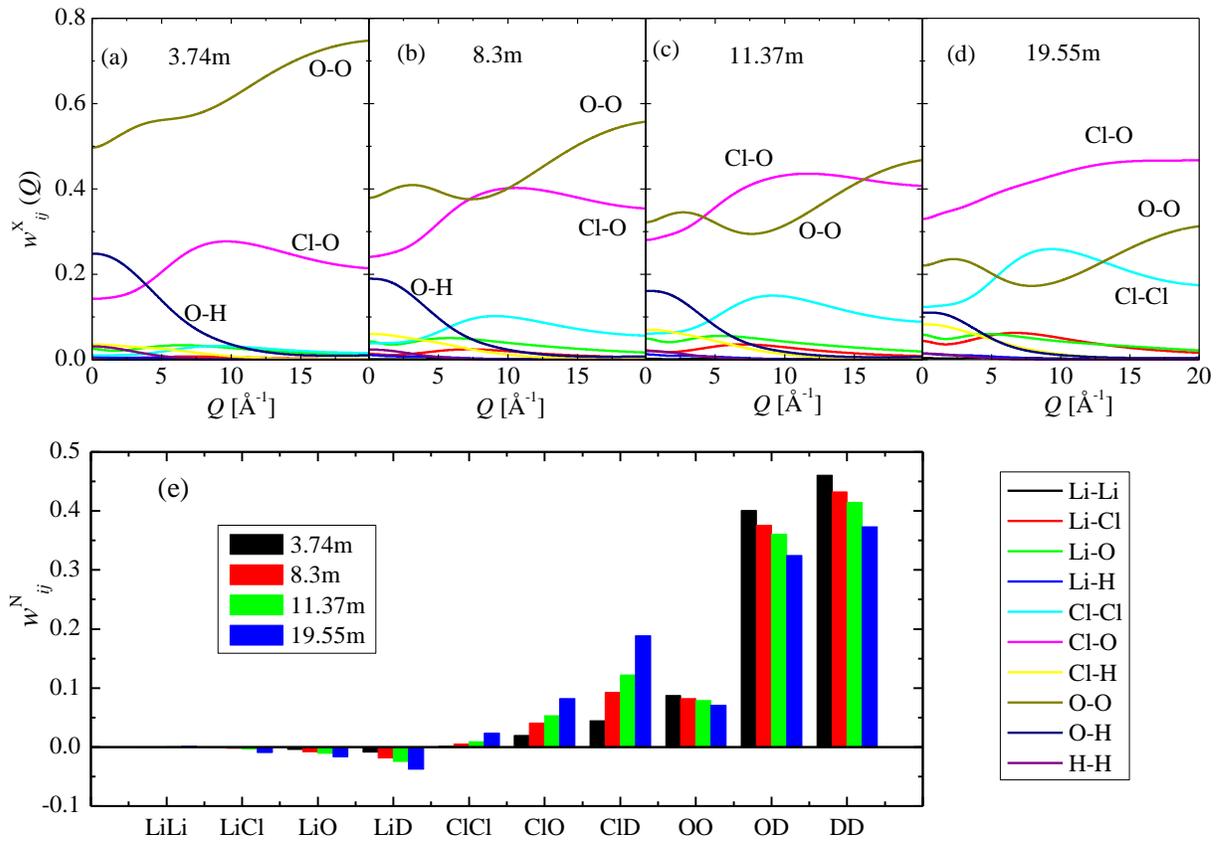

**Figure S.2.** X-ray (a-d) and neutron (e) scattering weights used for the calculations of the X-ray and neutron total structure factors.

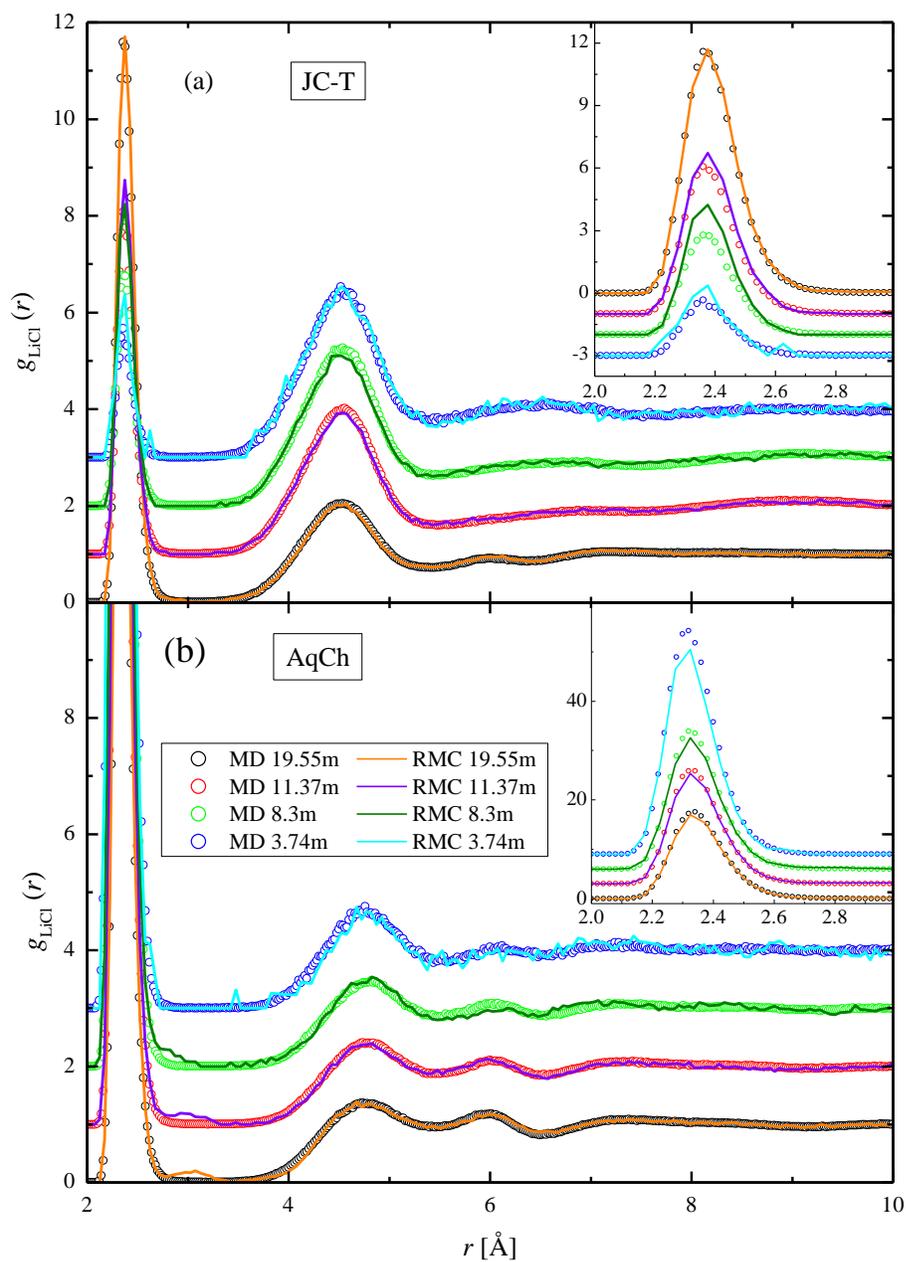

**Figure S.3.** Li-Cl partial pair correlation functions obtained from MD simulations (symbols) and the same after RMC refinements (lines), for the 4 investigated samples: for 3.74m (blue lines and cyan symbols), for 8.3m (green lines and light green symbols), for 11.37m (violet lines and pink symbols), for 19.55m (orange lines and black symbols). The curves are shown for two FFs, with different ion pairing tendency (see text): (a) JC-T (low IPT) and (b) AqCh (high IPT) models. The inset is an enlargement of the first peak.

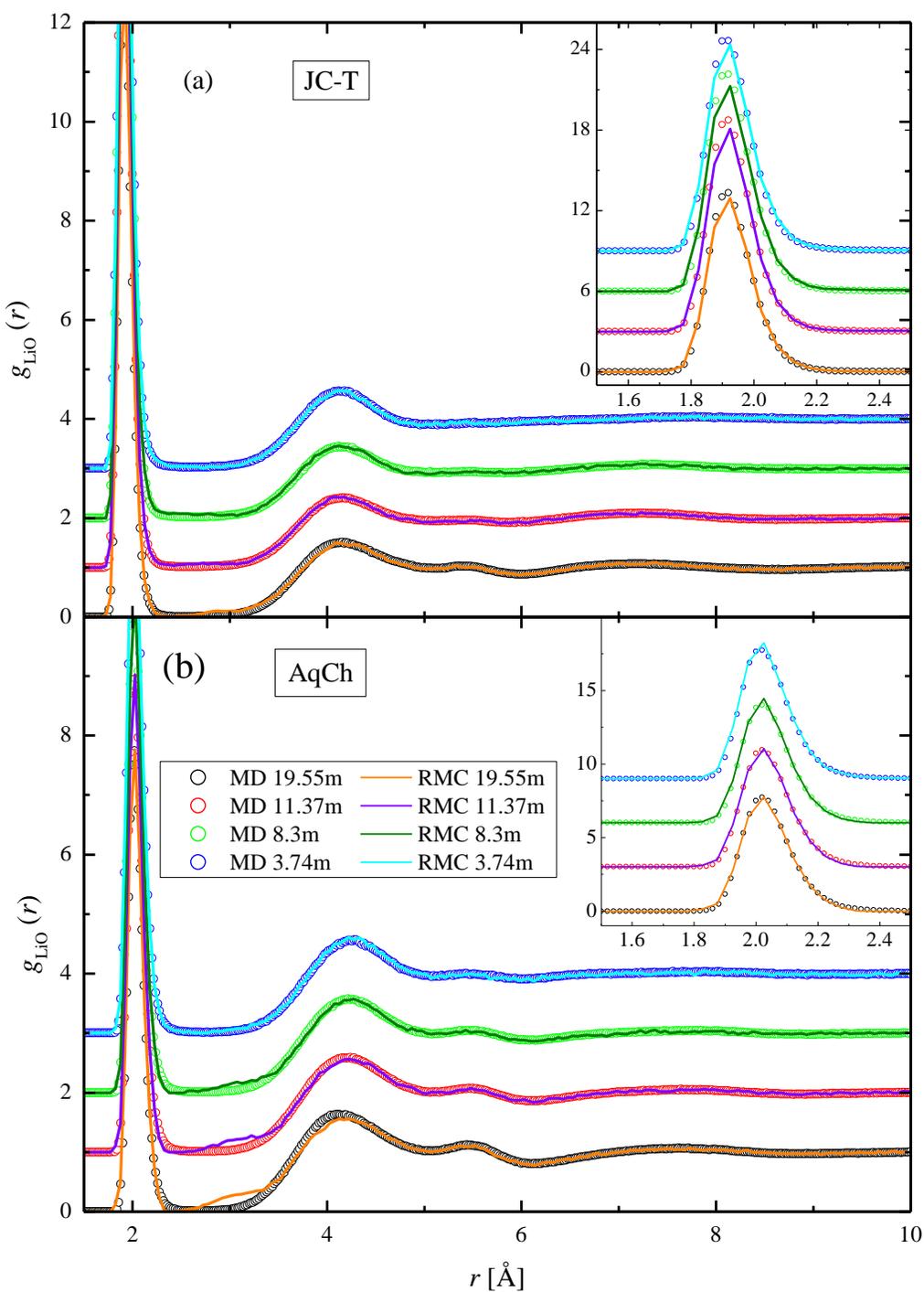

**Figure S.4**. Li-O partial pair correlation functions obtained from MD simulations (symbols) and the same after RMC refinements (lines), for the 4 investigated samples, with (a) JC-T (low IPT) and (b) AqCh (high IPT) models. The inset is an enlargement of the first peak.

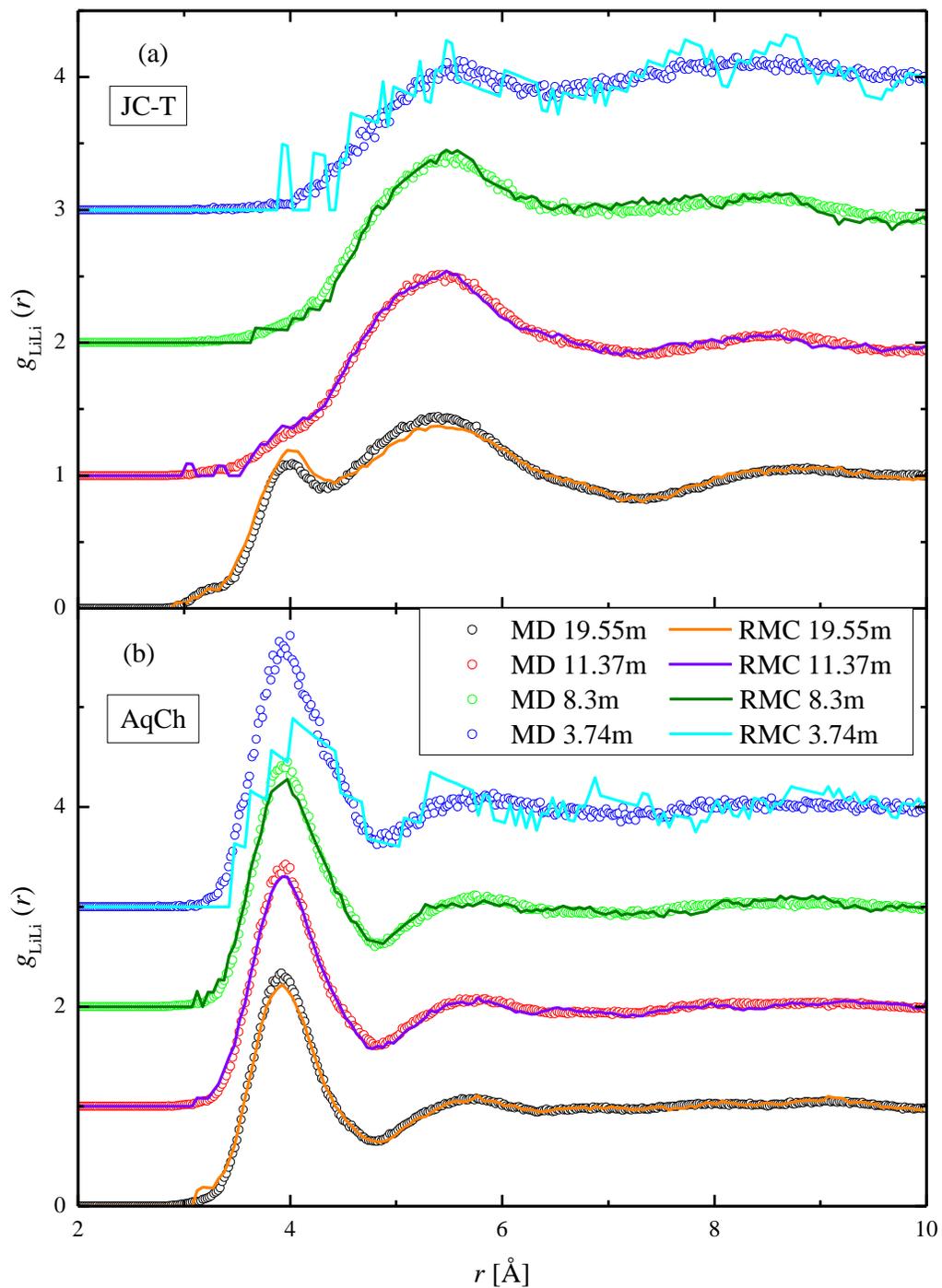

**Figure S.5**. Li-Li partial pair correlation functions obtained from MD simulations (symbols) and the same after RMC refinements (lines), for the 4 investigated samples, with (a) JC-T (low IPT) and (b) AqCh (high IPT) models.

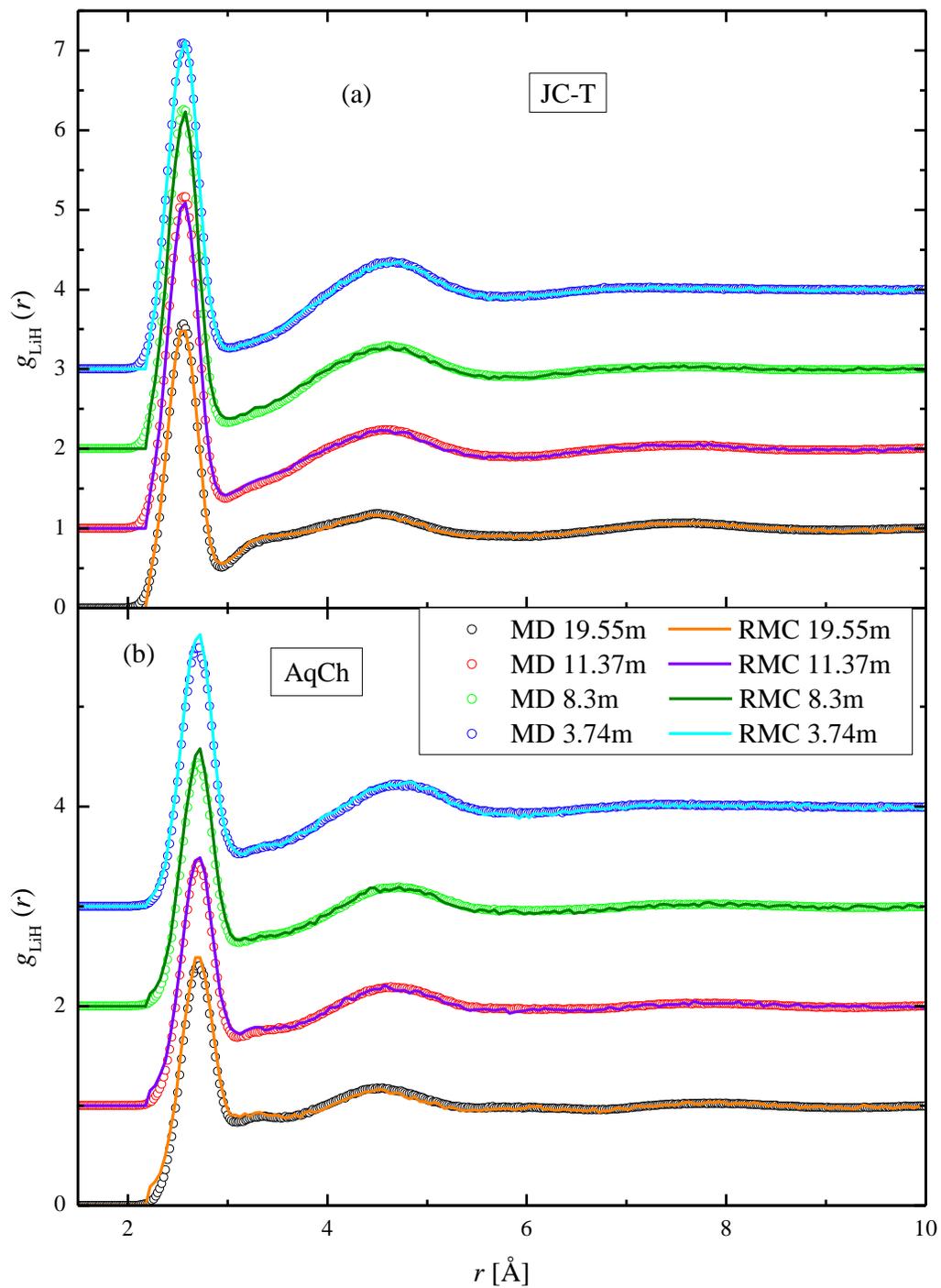

**Figure S.6**. Li-H partial pair correlation functions obtained from MD simulations (symbols) and the same after RMC refinements (lines), for the 4 investigated samples, with (a) JC-T (low IPT) and (b) AqCh (high IPT) models.

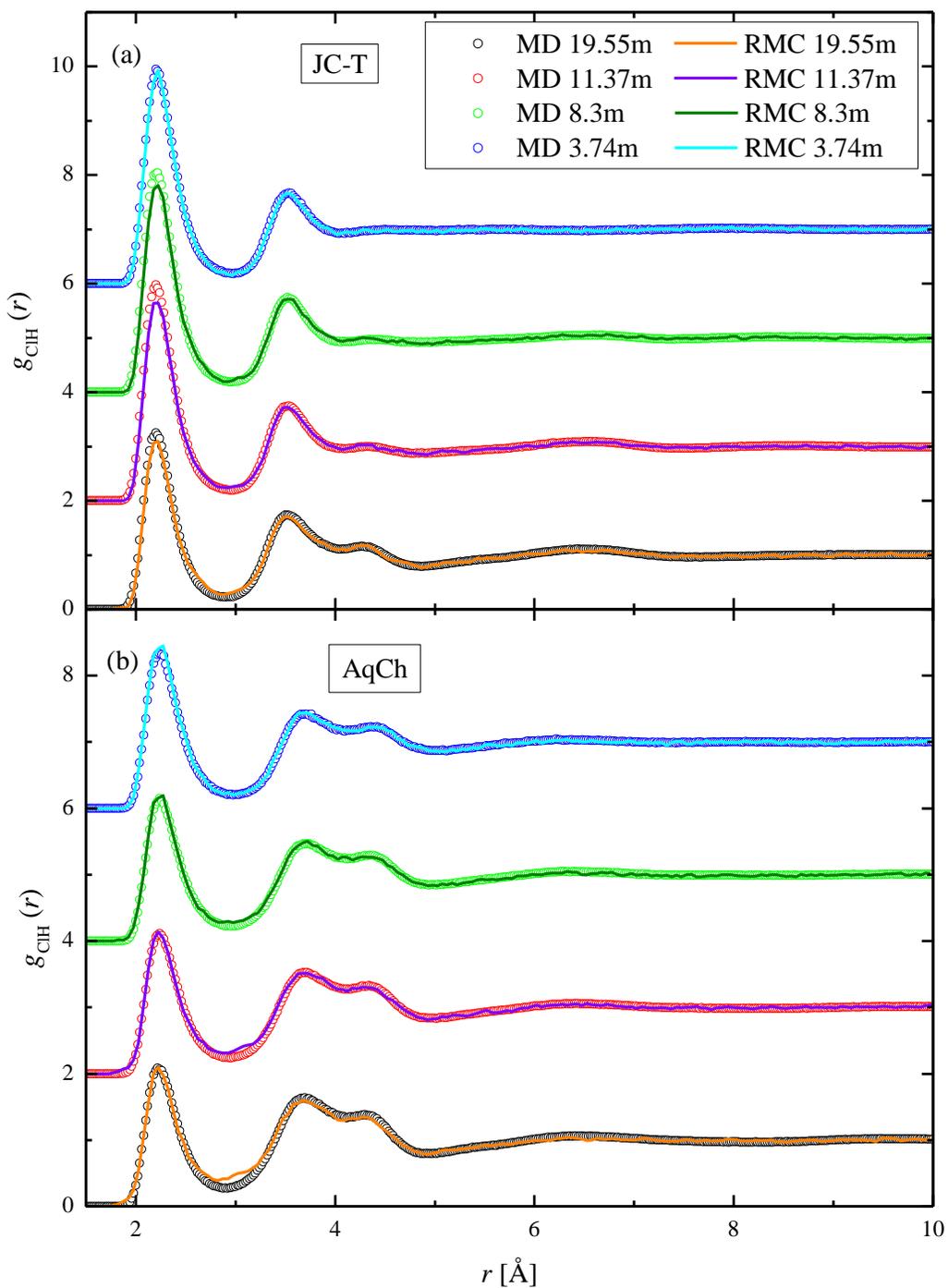

**Figure S.7.** Cl-H partial pair correlation functions obtained from MD simulations (symbols) and the same after RMC refinements (lines), for the 4 investigated samples, with (a) JC-T (low IPT) and (b) AqCh (high IPT) models.

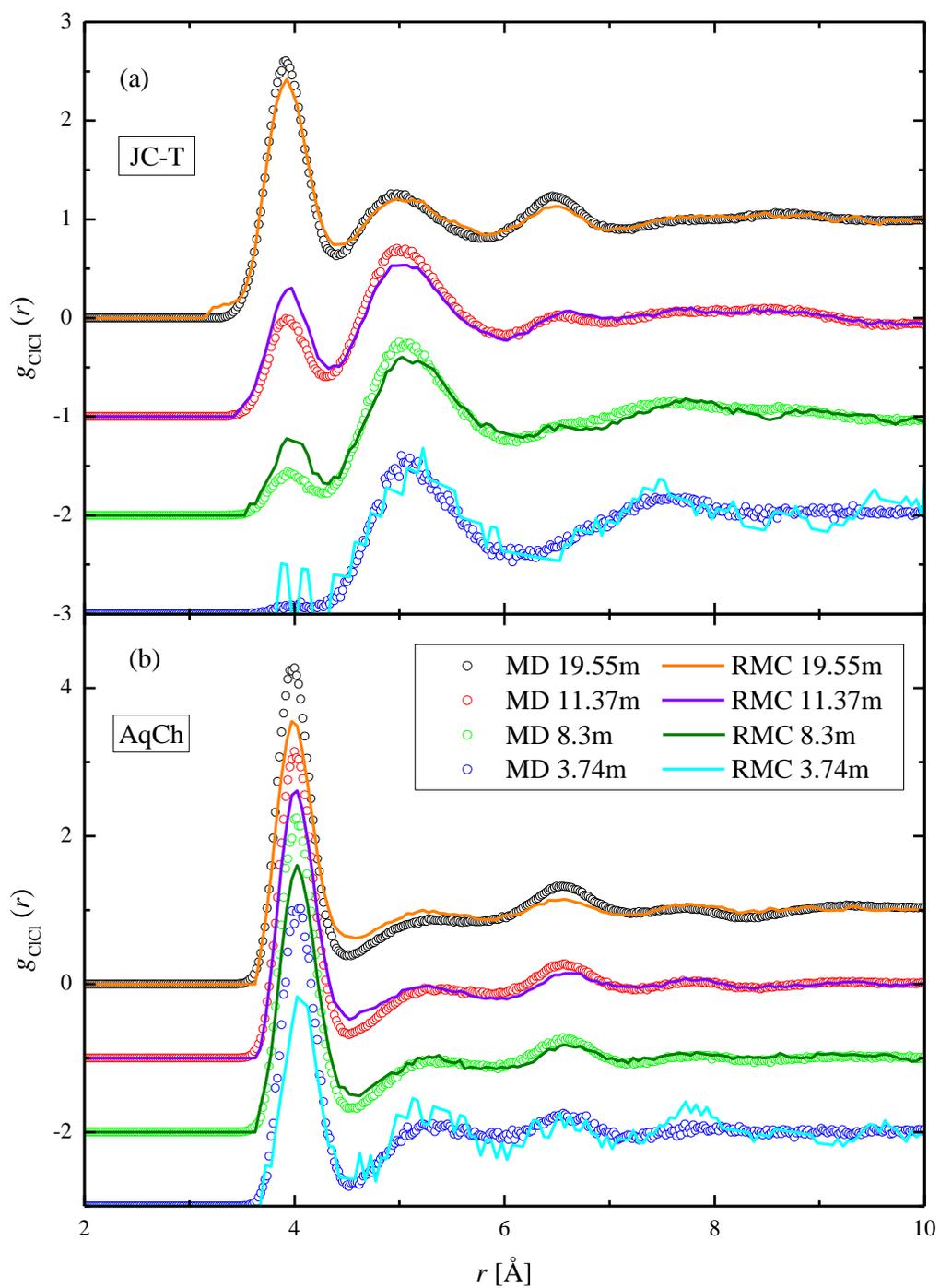

**Figure S.8**. Cl-Cl partial pair correlation functions obtained from MD simulations (symbols) and the same after RMC refinements (lines), for the 4 investigated samples, with (a) JC-T (low IPT) and (b) AqCh (high IPT) models.

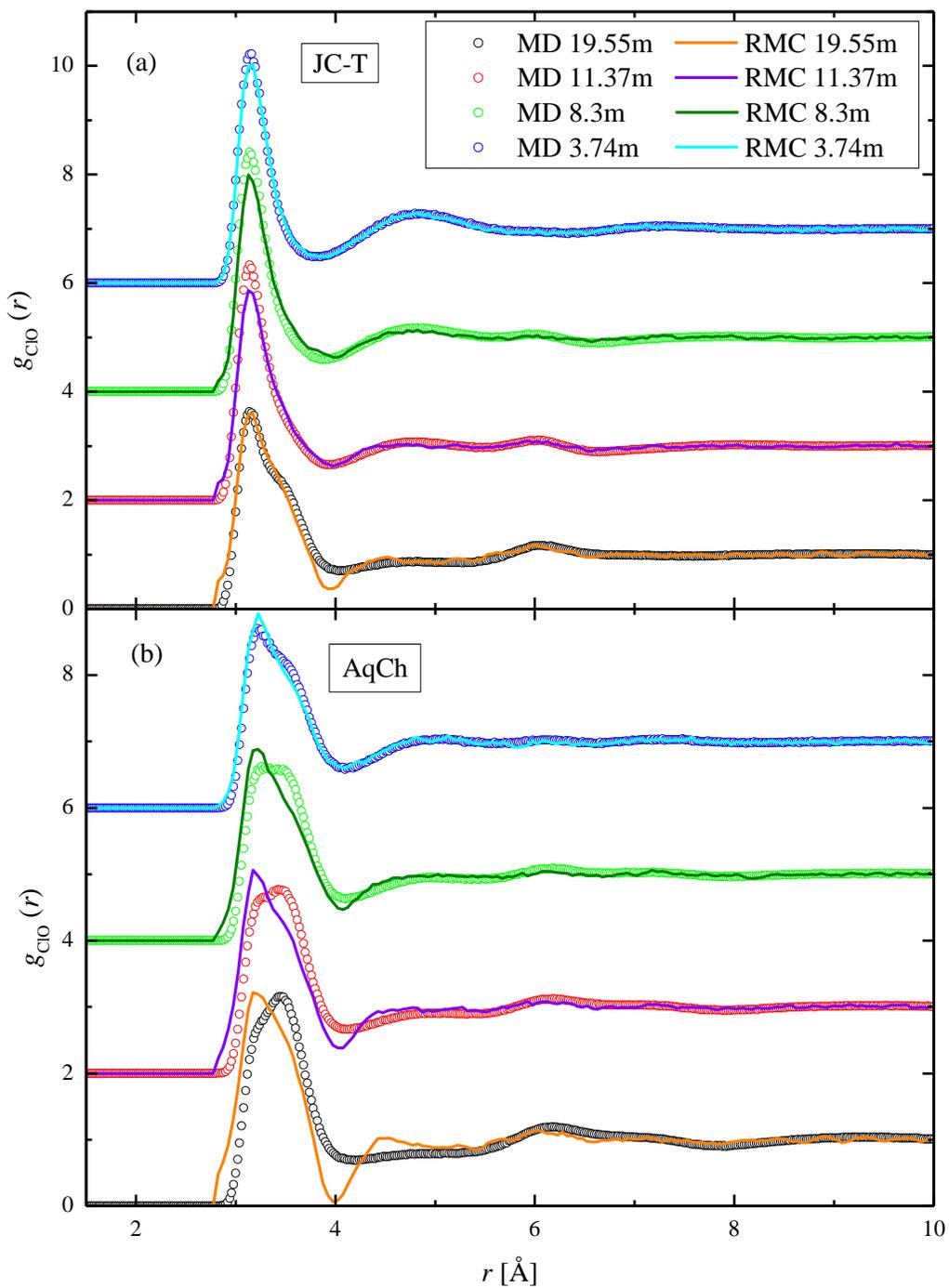

**Figure S.9**. Cl-O partial pair correlation functions obtained from MD simulations (symbols) and the same after RMC refinements (lines), for the 4 investigated samples, with (a) JC-T (low IPT) and (b) AqCh (high IPT) models.

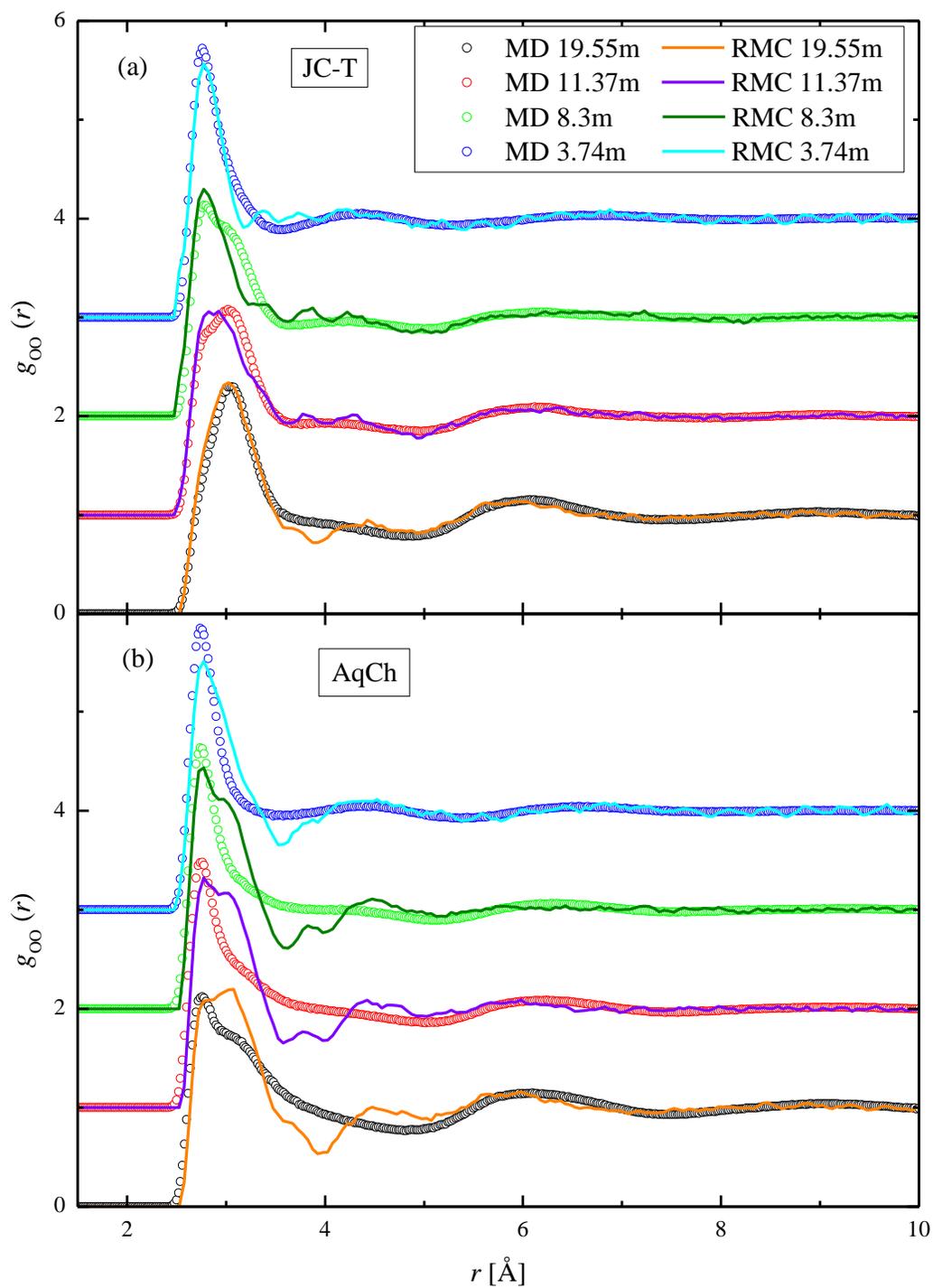

**Figure S.10**. O-O partial pair correlation functions obtained from MD simulations (symbols) and the same after RMC refinements (lines), for the 4 investigated samples, with (a) JC-T (low IPT) and (b) AqCh (high IPT) models.

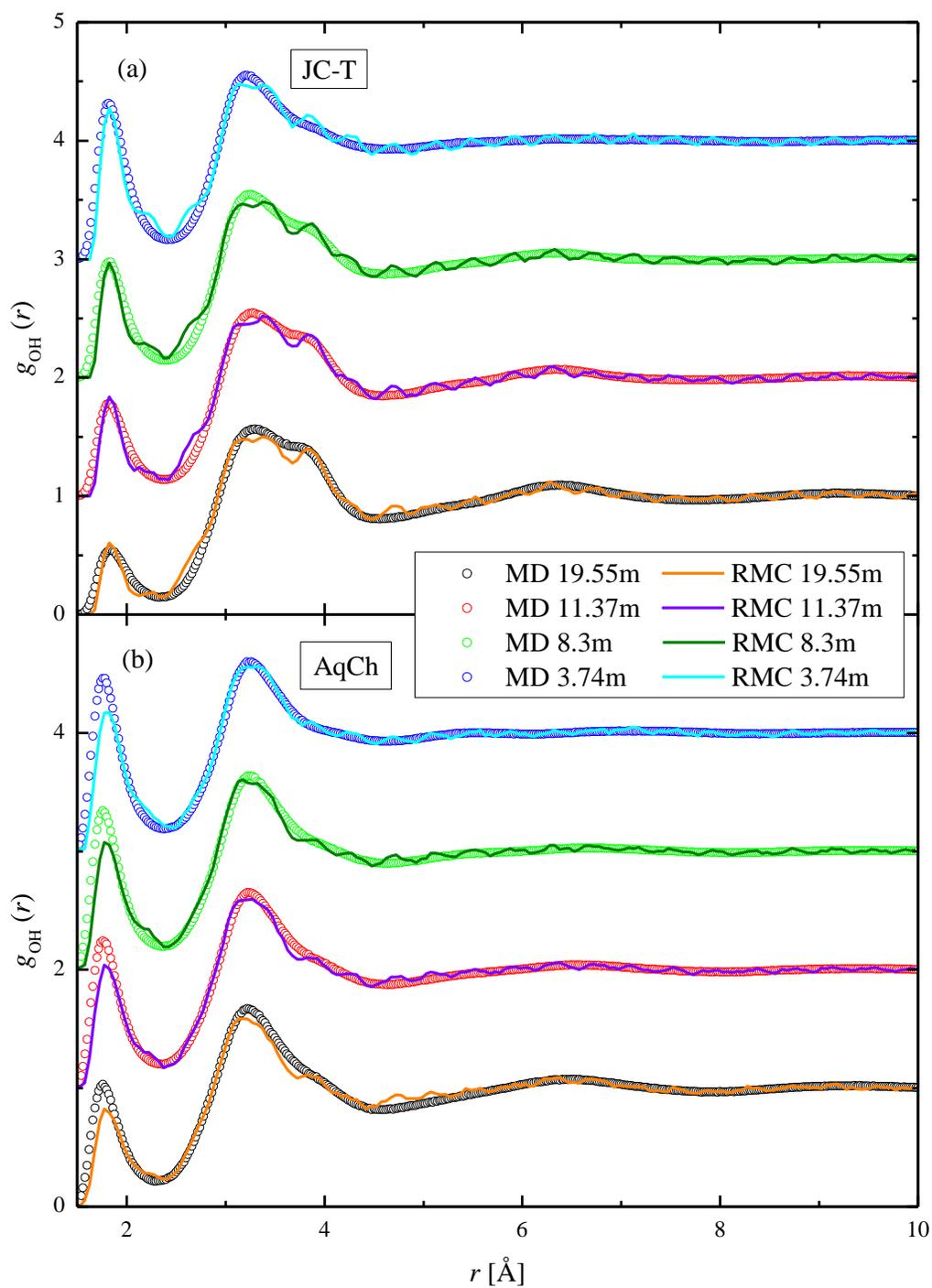

**Figure S.11**. O-H partial pair correlation functions obtained from MD simulations (symbols) and the same after RMC refinements (lines), for the 4 investigated samples, with (a) JC-T (low IPT) and (b) AqCh (high IPT) models.

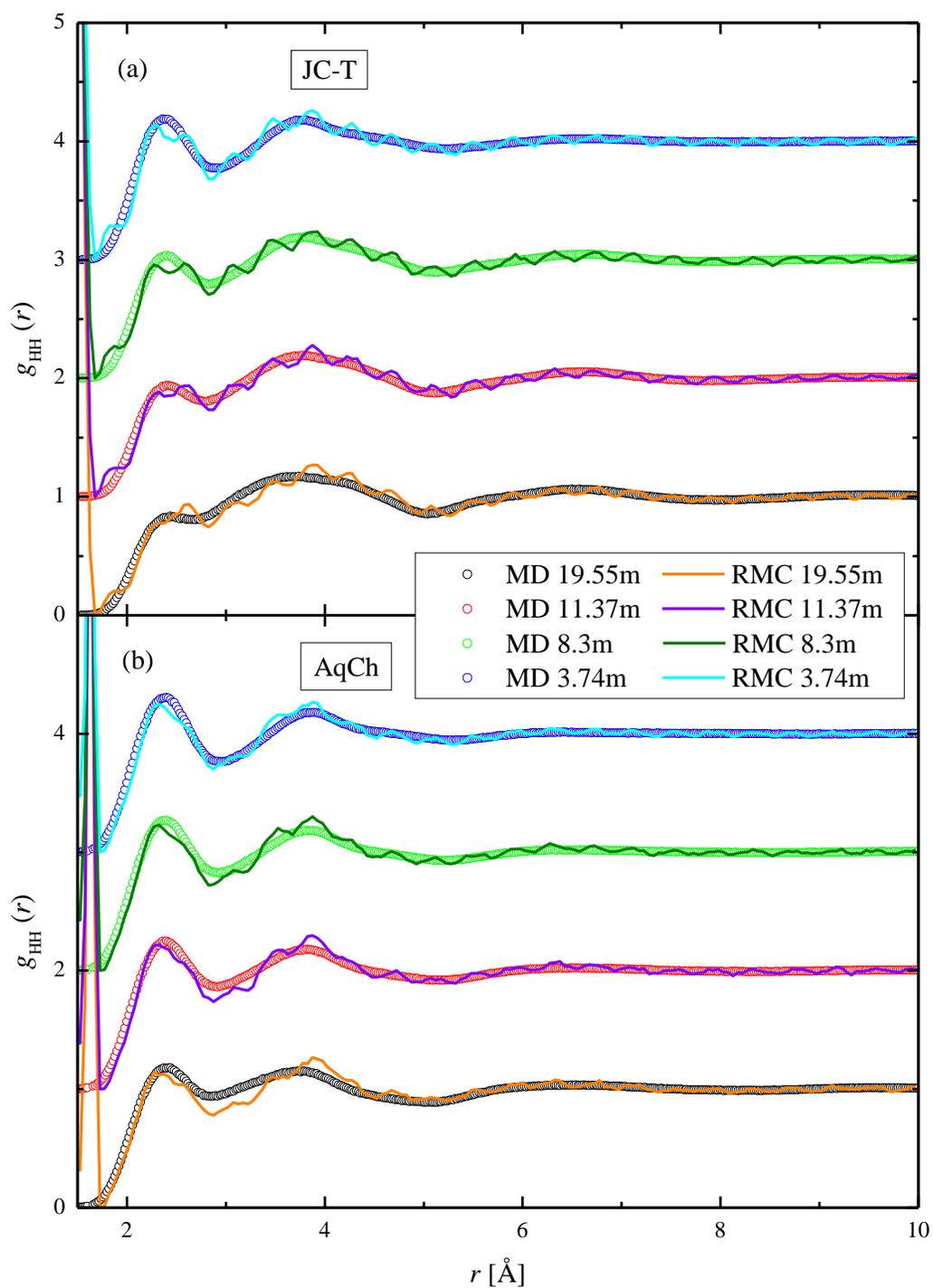

**Figure S.12**. H-H partial pair correlation functions obtained from MD simulations (symbols) and the same after RMC refinements (lines), for the 4 investigated samples, with (a) JC-T (low IPT) and (b) AqCh (high IPT) models.